\documentclass[preprint,12pt,authoryear]{elsarticle}

\usepackage{amssymb}
\usepackage{amsmath}
\usepackage{graphicx}%
\usepackage{multirow}%
\usepackage{amsmath,amssymb,amsfonts}%
\usepackage{amsthm}%
\usepackage{mathrsfs}%
\usepackage[title]{appendix}%
\usepackage{xcolor}%
\usepackage{array}
\usepackage{textcomp}%
\usepackage{manyfoot}%
\usepackage{booktabs}%
\usepackage{algorithm}%
\usepackage{algorithmicx}%
\usepackage{algpseudocode}%
\usepackage{listings}%
\usepackage{mathtools}

\usepackage{amsmath}
\usepackage{ amssymb }
\usepackage{braket}
\usepackage{qcircuit}
\usepackage{soul}
\usepackage{braket}

\journal{Information and Computation}

\begin{document}

\begin{frontmatter}



\title{Variational quantum algorithm for measurement extraction from the Boussniesq-type, Lin-Tsien and Camassa-Holm equations} 


\author{Pete Rigas} 

\affiliation{organization={Applied and Engineering Physics, Cornell University},
            addressline={271 Clark Hall}, 
            city={Ithaca},
            postcode={14850}, 
            state={NY},
            country={United States}}

\begin{abstract}
Classical-quantum hybrid algorithms have recently garnered significant attention, which are characterized by combining quantum and classical computing protocols to obtain readout from quantum circuits of interest. Recent progress due to Lubasch et al in a 2019 paper provides readout for solutions to the Schrodinger and Inviscid Burgers equations, by making use of a new variational quantum algorithm (VQA) which determines the ground state of a cost function expressed with a superposition of expectation values and variational parameters. In the following, we analyze additional computational prospects in which the VQA can reliably produce solutions to other PDEs that are comparable to solutions that have been previously realized classically, which are characterized with noiseless quantum simulations. To determine the range of nonlinearities that the algorithm can process for other IVPs, we study three PDEs, first beginning with the Boussniesq-type equation, and progressing to other equations underlying physical phenomena ranging from those of fluid dynamics, and wave propagation, including the Lin-Tsien and Camassa-Holm equations. To formulate optimization routines that the VQA undergoes for numerical approximations of solutions that are obtained as readout from quantum circuits, cost functions corresponding to each PDE are obtained. (\textbf{MSC Class}: 81P10; 81Q04; 81V08)
\end{abstract}

\begin{graphicalabstract}
\end{graphicalabstract}

\begin{highlights}
\item The following work presents an approach of a recent variational quantum algorithm (VQA), for solving partial differential equations. Given the wide variety of nonlinearities, and widely varying, behaviors that such PDEs can exhibit, it is of great interest to perform time evolution, with polynomial time computational complexity. Underlying the VQA is a quantum-classical feedback loop, in which a suitable parametrisation of a cost function defined over Hilbert space is classical optimized. From a large space of possible optimizers, we extract measurements for approximating the solution to some well posed PDE, ranging from stochastic to deterministic optimizers. 
\item The PDEs that we obtain cost functions for include Boussniesq-type, Lin-Tsien, and Camassa-Holm. Determining whether the solution approximations, which are obtained by generating hundreds of candidate solution wavefunctions to a PDE, are resilient to noise is of great interest. 
\end{highlights}


\end{frontmatter}




\section{Introduction}\label{sec1}

\subsection{Overview}

The variational quantum algorithm (VQA) has attracted wide attention from several computational disciplines in physics, chemistry and materials science {[13]}, with recent studies demonstrating how optimizers for such algorithms can be improved and parametrized for entanglement {[20,21]}, robust encoding in molecules for chemistry computations {[1]}, various implementations for linear algebra problems {[19,23]}, and the construction of quantum Gibbs states {[4]}. To further contribute to rapid developments, Lubasch et al, {[12]}, introduce a novel VQA which can be applied to solve nonlinear problems arising from various PDEs through specifications on quantum circuits, including an optimized quantum nonlinear processing unit (QNPU) for treating nonlinearities of the equation that is being solved for efficiently, in addition to randomly chosen variational parameters $\lambda$, which are randomly initialized with some starting position $\lambda_0$ in the Hilbert space, which parametrizes a cost function $\mathcal{C}$ that is classically optimized. For efficiently performing time evolution, quantum registers prepared in circuits of polynomial depth not only require exponentially fewer parameters, but also provide ripe opportunity for the algorithm to be applied to several other nonlinear problems, some of which are presented in this article, including computational fluid dynamics problems, wave interference and propagation, to provide a few examples. 

In {[12]}, a VQA is applied to obtain high-fidelity solutions to the nonlinear Schrodinger's equation, and to the inviscid Burger's equation, with matrix product state (MPS) ansatzae of polynomial bond dimension {[16,17]}. Such ansatzae are advantageous for quantum computation in their ability to be represented as a product of isometric matrices which can be physically interpreted as the wave function of possible states for the solution to a well posed PDE IVP, after which variational states for the solution are time evolved for determining the value of the PDE solution after the initial state. In comparison to cost functions obtained for NLS and the Burger's equations, cost functions obtained in the Appendix of this work, involve more complicated, and intricate, superpositions of quantum states; for example, the Camassa-Holm cost function includes parameters that are not only directly proportional to the time step magnitude as are cost functions for the Nonlinear Schrodinger's and Burgers equations, but also on several expectation values resulting from quantum states that are acted upon by differential operators which are represented with Adder and Adder dagger operators.

For each of the three PDEs whose solutions can be approximated in polynomial time by the quantum algorithm, we derive the cost function, hence obtaining a superposition of quantum states parameterized in nonlinearities of the equation. After having obtained the cost function, classical, stochastic, gradient-based, and constrained, optimization routines are implemented in order to determine classes of initial data for which the VQA is capable of generating solutions to nonlinear problems, with the Zalka Grover Rudolph-Quantum Fourier Transform (ZGR-QFT) ansatz. For higher-dimensional nonlinearities, simulating time evolution of the three PDEs considered in this article are achieved with QNPUs composed of a significantly higher number of input and output ports, and variational parameters.

To apply such a hybrid quantum algorithm for a broad class of initial data to a nonlinear problem, it is necessary to construct quantum circuits for approximating nonlinearities with expectation values, in addition to variational parameters that are modified throughout the optimization procedure as the algorithm is executed to obtain approximations of analytical solutions for a PDE of interest. From the three PDEs considered in this work, the Camassa-Holm PDE exhibits significant dependency upon initial conditions, in which solution profiles provided in \textit{Figure 5}, and \textit{Figure 6}, are implemented for noiseless quantum simulations consisting of anywhere from $500$ to $20,000$ time steps; noiseless quantum simulations are executed for the Boussniesq-type, and Lin-Tsien, PDEs for $400$, and $500$, time steps of evolution, respectively. Beyond the initial state before performing time evolution in the Hilbert space, several characteristics of ansatzae, ranging from the depth of the circuit to the number of variational parameters that must be specified, are complemented with computation of Fourier coefficients.

In polynomial time, compiling the quantum algorithm with nonlinearities, besides those which are discussed in {[12]} for the nonlinear Schrodinger's and Burgers equations, demonstrates that it is important to not only determine optimal circuit constructions which reduce noise, error mitigation, and error bursts, for potential future applications on near-term hardware {[11,17,26]}, but also the range of time steps that the optimizer must take to arrive to a ground state in order for approximations of the solution converge. From running hundreds of numerical experiments across several PDEs, open-source optimizers, such as the Nevergrad optimizer provided by Facebook Research, fare extremely well in being able to robustly handle wide classes of initial data. To this end, quantum circuits for simulating time evolution of the Boussniesq-type, Lin-Tsien, and Camassa-Holm, PDEs, are of shallow depth, permitting for the extraction of measurements from quantum circuits for classes of nonlinear problems that have previously not been examined with quantum algorithms. Specifically, measurements are extracted from quantum circuits, such as that illustrated in \textit{Figure 4} for input ports $19$-$31$ of the Camassa-Holm PDE, through read out operations.

In comparison to other PDEs that have previously been studied with quantum algorithms, the primary difficulty in applying novel hybrid variational quantum algorithms to solve PDEs, in comparison to nonlinear Schrodinger's and Burger's, equations, is apparent through the QNPU construction for expectation terms from each cost function. Depending on the range of nonlinearities of the PDE, and the corresponding initial condition, whether hybrid quantum algorithms are capable of still providing even a polynomial speedup in runtime is still of great interest to further explore. For time evolution of the Camassa-Holm equation which consists of $20,000$ time steps, the noiseless simulation with the largest number of time steps, classical, deterministic, and stochastic, optimizers alike experience computational difficulties in avoiding spurious minima of optimization landscapes, with the runtime of the optmization rooutine being determined by several budget parameters.

\subsection{Variational quantum algorithmic objects}

\noindent We introduce a few objects central to the VQA approach, which share some similarities and differences with several objects introduced in [3]. That is, for a cost function $C \big( \theta \big)$ for some real parameter $\theta$, the VQA is comprised of a classical-quantum feedback loop, which seeks to compute,

\begin{align*}
  \theta^{*} = \underset{\theta}{\mathrm{arg min}} \text{ }  C \big( \theta \big)   .
\end{align*}

\noindent Cost functions that are considered in VQAs take on a wide variety of forms. For the specific class of nonlinear problems considered in this work, we expand solutions, and possibly derivatives of solutions, in a given PDE. For the following objects, fix some $k \geq 0$. Such cost functions are dependent upon,

\begin{align*}
  f \equiv \textit{Class of suitable functions} , \\ \\ \rho_k \equiv         \textit{Training set data inputs}   , \\ \\ O_k  \equiv  \textit{Observables for each } k        ,   \\ \\  U \big( \theta \big) \equiv \textit{Unitaries dependent upon the parameter } \theta   .
\end{align*}

\noindent where,

\begin{align*}
     C \big( \theta \big)    \equiv     f \big[ \big\{ \rho_k \big\} , \big\{ O_k \big\} , U \big( \theta \big) \big]             .
\end{align*}

\noindent Cost functions, such as the one above that is dependent upon training set data input, admit the decomposition,

\begin{align*}
    C \big( \theta \big)   \propto  \underset{k \geq 0}{\sum}   f_k \big[ \big\{ \rho_k \big\} , \big\{ O_k \big\} , U \big( \theta \big) \big] \equiv   \underset{k \geq 0}{\sum}  f_k \big[ \mathrm{Tr} \big[  O_k U \big( \theta \big) \rho_k U^{\dagger} \big( \theta \big)         \big] \big]    . 
\end{align*}

\noindent In this work, cost functions used within the VQA framework are dependent upon \textit{variational parameters} instead of upon training set input data. Denote,

\begin{align*}
   \lambda_0 \equiv \textit{Variational parameter at the beginning of time evolution}   ,   \\ \\ \lambda   \equiv  \textit{Variational parameter throughout the remainder of time evolution after initialization} .
\end{align*}

\noindent Irrespective of whether one uses training set data input or variational parameters, when beginning time evolution for a Quantum simulation, one makes use of an \textit{ansatz} in Hilbert space. The ansatz considered in this work is dependent upon the computation of Fourier coefficients, in addition to unitary transformations. In previous works, ansatzae $\mathscr{A}$ have taken the form,

\begin{align*}
    \mathscr{A} \equiv        \underset{\theta}{\prod}   \big\{ \textit{Collection of Unitaries } \mathscr{U} \textit{ dependent upon } \theta  \big\}          \equiv   \underset{\theta}{\prod}  \underset{\mathscr{U}}{\prod} \mathscr{U} \big( \theta \big) =     \underset{\theta}{\prod}  \underset{1 \leq i \leq L}{\underset{\mathscr{U}}{\prod}} \mathscr{U}_i \big( \theta \big)             .
\end{align*} 

\noindent Determining whether an ansatz should only consist of unitary transformations or not is highly problem dependent. In the case of nonlinear problems through the PDEs of interest in this work, including Fourier gates, along with corresponding computations of Fourier coefficients, is beneficial. In expansions of admissible ansatze $\mathscr{A}$ provided above, each $\mathscr{U}_i$ admits the decomposition,

\begin{align*}
   \mathscr{U}_i \big( \theta \big) = \underset{m}{\prod}         e^{i \theta_m H_m} W_m        ,
\end{align*}

\noindent for every $i$, with $j < i$, and,

\begin{align*}
 W_m   \equiv   \textit{Unitaries that are not parameterized in } \theta    , \\ \\   H_m \equiv \textit{Hermitian operator}       . 
\end{align*}

\noindent For other comutational purposes, ansatzae have taken the form,

\begin{align*}
   U \big( \gamma , \beta \big) =    \underset{1 \leq l \leq p}{\prod}   \bigg\{     \mathrm{exp} \big[ - i \beta_l H_M \big]  \mathrm{exp} \big[ - \gamma_l H_P \big]     \bigg\}  ,
\end{align*}

\noindent for $\theta \equiv \big( \gamma , \beta \big)$, and,

\begin{align*}
  H_M \equiv \underset{m \in M}{\bigcup} H_m  , \\ \\   H_P \equiv \underset{p \in P}{\bigcup} H_p        , \\ \\   \gamma \equiv \underset{l \in L}{\bigcup} \gamma_l         ,   \\ \\  \beta \equiv \underset{l \in L}{\bigcup} \beta_l      . 
\end{align*}

\noindent The Zalka-Grover-Rudolph (ZGR) Quantum Fourier transform (QFT) ansatz is a beneficial initial encoding for time evolution in VQAs. For the Boussniesq-type, Lin-Tsien, and Camassa Holm PDEs examined in this work, one prepares initial conditions from this ansatz with the following steps, [14]:

\begin{itemize}
    \item[$\bullet$] \textit{Definition of the Quantum Fourier transform}. The QFT, $\ket{r}$, on an $n$-qubit discretized function, $\ket{f^{(n)}}$, is given by,

    \begin{align*}
     \ket{r} \longrightarrow \frac{1}{\sqrt{2^n}}    \underset{0 \leq s \leq 2^n - 1}{\sum} \mathrm{exp} \bigg[ \frac{2 \pi i r s}{2^n} \bigg]    \ket{s}                .
    \end{align*}

    \item[$\bullet$] \textit{Action of the Quantum Fourier transform}. The action associated with the QFT, $\widetilde{\cdot}$, on discretized functions takes the form,

    \begin{align*}
      \ket{\widetilde{f^{(n)}}}      \underset{0 \leq s \leq 2^n - 1}{\sum}        \widetilde{f^{(n)}} \big( p_s \big) \ket{s} \equiv     \widetilde{\mathcal{F}}  \ket{f^{(n)}}                , 
    \end{align*}

    \noindent where,

    \begin{align*}
    \widetilde{\mathcal{F}}  \ket{f^{(n)}}            =    \frac{1}{\sqrt{2^n}} \underset{0 \leq r < s \leq 2^n - 1}{\sum} \mathrm{exp} \bigg[      \frac{2 \pi i s r }{2^n}    \bigg]   f \big( x_r \big) \ket{s}     ,
    \end{align*}

    \noindent for quasimomenta,

   \[ p_s \equiv \frac{2 \pi}{\Delta x^{(n)} 2^n} \left\{\!\begin{array}{ll@{}>{{}}l}                 s \Longleftrightarrow  0 \leq s < 2^n - 1 ,  \\ \\   s - 2^n        \Longleftrightarrow        \textit{otherwise}    . 
\end{array}\right.
\]

    \item[$\bullet$] \textit{Encoding an additional m qubits onto the ansatz}. One can append a strictly positive number, $m$, qubits onto the ansatz, so that,

    \begin{align*}
     s \in \big[ 0 , 2^n -1 \big) \cup \big[ 2^{n+m} - 2^{n-1}  , 2^{n+m} \big)   ,
    \end{align*}

    \noindent with the state,

    \begin{align*}
      \ket{f^{(n+m)}} \equiv    \widetilde{\mathcal{F}}^{-1} \mathscr{U} \bigg[     \ket{0}^{\otimes m}     \otimes     \widetilde{\mathcal{F}} \ket{f^{(n)}}            \bigg]                   .
    \end{align*}

    \item[$\bullet$] \textit{Parameterizing unitaries of the ansatz}. In comparison to a previous parameterization, $W$, which was introduced for ansatzae which do not depend upon QFT, introduce the parameterization,

        \begin{align*}
        \mathscr{W}  \big( \theta \big) \equiv    \underset{0 \leq q \leq n-1}{\prod} R^y_q \big( \theta^{d+1}_q  \big)                 \underset{1 \leq d^{\prime} \leq \textit{depth}}{\prod}  \bigg\{         \underset{0 \leq c \leq n-1}{\prod}       \bigg[   \underset{c<t}{\prod}  \mathrm{CNOT}_{c,t} \bigg] \underset{0 \leq q \leq n-1}{\prod} R^y_q \big( \theta^{d^{\prime}}_q \big)   \bigg\}     \\ \equiv    \underset{0 \leq q \leq n-1}{\prod} R^y_q \big( \theta^{\textit{depth}+1}_q  \big)            \underset{1 \leq d^{\prime} \leq \textit{depth}}{\prod}  \bigg\{         \underset{0 \leq c \leq n-1}{\prod}       \bigg[   \underset{c<t}{\prod}  \mathrm{CNOT}_{c,t} \bigg] \underset{0 \leq q \leq n-1}{\prod} R^y_q \big( \theta^{d^{\prime}}_q \big)   \bigg\}      , 
        \end{align*}

    \noindent for unitaries appearing in the ZGR-QFT ansatz, where,

    \begin{align*}
        R^y_q \big( \theta^{d+1}_q  \big)  \equiv      \mathrm{exp} \bigg[ -     i \frac{\theta \sigma_y}{2}     \bigg]    \textit{ many rotations }   ,
    \end{align*} 

\noindent and $\mathrm{CNOT}_{c,t}$ denotes CNOT gates which depend upon each $c,t$.
    
\end{itemize}

\noindent The Quantum Fourier Transform, which we denote QFT for short, satisfies, [19]:

\begin{itemize}
\item[$\bullet$] \textit{Ket representation}: Denote a Quantum state over Hilbert space with $\ket{\psi}$. One has that, given $\alpha_k >0$ for all $k$, that,

\begin{align*}
 \ket{\psi} = \underset{0 \leq k \leq 2^n - 1}{\sum} \alpha_k \ket{k}   ,
\end{align*}

\noindent where,

\begin{align*}
   \alpha_k = \frac{1}{\sqrt{2^n}} \underset{0 \leq j \leq 2^n - 1}{\sum} a_j \omega^{k_j} = \frac{1}{\sqrt{2^n}} \underset{0 \leq j \leq 2^n - 1}{\sum} a_j \mathrm{exp} \bigg[     \frac{\pi i}{2^n - 1}     \bigg] . 
\end{align*}

\item[$\bullet$] \textit{Linearity of the QFT}. The transforming, being linear, satisfies,

\begin{align*}
  \mathrm{QFT}  \ket{\psi} =             \underset{0 \leq j \leq 2^n - 1}{\sum} \alpha_j     \mathrm{QFT}   \ket{j} . 
\end{align*}

\item[$\bullet$] \textit{Embedded ket representation}. Besides the representation provided in the previous item, the transform can also be represented as,

\begin{align*}
      \mathrm{QFT} \ket{\psi} = \underset{ 0 \leq j \leq 2^n - 1}{\sum} \alpha_j \bigg[               \frac{1}{\sqrt{2^n}} \underset{0 \leq k \leq 2^n - 1}{\sum} \omega^{k_j} \ket{k}       \bigg]    . 
\end{align*}

\item[$\bullet$] \textit{Alternative factorization from the roots of unity}. As a superposition over the roots of unity, one has that,

\begin{align*}
   \mathrm{QFT} \ket{j}  =  \frac{1}{\sqrt{2^2}} \bigg[  \ket{00} +         \omega^j \ket{01} + \omega^{2j} \ket{10} + \omega^{3j} \ket{11}    \bigg]    ,
\end{align*}

\noindent for roots of unity, $\omega^j$.

\item[$\bullet$] \textit{Alternative factorization depending upon the first application of the QFT}. Straightforwardly, the alternative factorization presented in the previous item above can be expressed as,

\begin{align*}
\frac{1}{\sqrt{2^2}} \bigg[  \ket{00} +         \omega^j \ket{01} + \omega^{2j} \ket{10} + \omega^{3j} \ket{11}    \bigg]  = \frac{1}{\sqrt{2^2}} \big[ \ket{0} +  \omega^{2j} \ket{1} \big] \big[ \ket{0} +  \omega^j \ket{1} \big]       \\   \equiv   \frac{1}{\sqrt{2^2}} \big[ \ket{0} +  \omega^{2j} \ket{1} \big] \mathscr{Q}\mathscr{F}\mathscr{T}_1  , 
\end{align*}

\noindent for the first application of the QFT, $\mathscr{Q}\mathscr{F}\mathscr{T}_1$.

\item[$\bullet$] \textit{Quantum circuit basis}. Fix some $\varphi \in \big[ 0 , 2 \pi \big]$. The states,

\begin{align*}
   \frac{1}{\sqrt{2}} \big[ \ket{0} + \mathrm{exp} \big[  i \varphi \big] \ket{1} \big] , 
\end{align*}

\noindent for the phase shift gate, $P \big( \varphi \big)$, with,

\begin{align*}
 P \big( \varphi \big) = \begin{bmatrix} 1 & 0 \\ 0 & \mathrm{exp} \big[ i \varphi \big] \end{bmatrix}  , 
\end{align*}

\noindent and the Hadamard test,

\begin{align*}
 H = \frac{1}{\sqrt{2}} \begin{bmatrix}   1 & 1 \\ 1 & -1    \end{bmatrix}   , 
\end{align*}

\noindent are used to construct the QFT.

\end{itemize}

\noindent ZGR-QFT ansatzae, in comparison to variational Hamiltonian, mixed state, unitary coupled, and alternating operator, ansatzae, have the following similarities and differences:

\begin{itemize}
\item[$\bullet$] \textit{The ZGR-QFT ansatz, as does the alternating operator ansatz, depends upon trotterization}. The ZGR-QFT and alternating operator ansatzae each depend upon trotterizing, which is widely used within the field of Quantum Computing for preparing initial states of time evolution.

\item[$\bullet$] \textit{The ZGR-QFT ansatz, in comparison to the variational Hamiltonian ansatz, does not depend upon an explicit formulation of an energy functional}. The variational Hamiltonian ansatz depends upon parametrizations of unitary transformation through a Hamiltonian. Unitaries for the ZGR-QFT ansatz are not defined in terms of a Hamiltonian energy functional.

\item[$\bullet$] \textit{The ZGR-QFT ansatz, in comparison to the unitary coupled ansatz, does not depend upon the excitation of Hartree-Fock states}. Similar to the previous difference mentioned above, the ZGR-QFT does not depend upon excitations, and consequently, to annihilation and creation operators that are used to define the Hartree-Fock Hamiltonian. 

\item[$\bullet$] \textit{The ZGR-QFT ansatz, in comparison to the mixed state ansatz, does not depend upon the computation of eigenvalues}. The mixed state ansatz is dependent upon the computation of eigenvalues, as well as upon subsequent training of a probability distribution $p_i$, where the index $i$ varies over the number of inputs. While one could straightforwardly introduce training set parameters into the ZGR-QFT ansatz, in this work we choose not to.

\item[$\bullet$] \textit{The ZGR-QFT ansatz, in comparison to the variational Hamiltonian, mixed state, unitary coupled, and alternating operator, ansatze, depends upon the computation of Fourier coefficients}. The ZGR-QFT ansatz, as previously described, depends upon the QFT and the accompanying computation of Fourier coefficients.

\item[$\bullet$] \textit{The ZGR-QFT ansatz, as does the variational Hamiltonian, mixed state, unitary coupled, and alternating operator, ansatzae, depends upon a computationl quantum-classical feedback loop}. The ZGR-QFT ansatz, variational Hamiltonian, mixed state, unitary coupled, and alternating operator, ansatzae, alike depend upon constrainted optimization problems. Namely, the quantum-classical feedback loop significantly depends upon the optimization of the three cost functions obtained in the appendix, given an encoding for the initial state through the ansatz.

\item[$\bullet$] \textit{The ZGR-QFT ansatz, in comparison to the variational Hamiltonian, mixed state, unitary coupled, and alternating  operator, ansatze, provides a polynomial speedup for some nonlinear problems of interest}. As demonstrated through Quantum simulation for time evolution in this work, the ZGR-QFT ansatz provides, in some cases, polynomial speedup. It is of great interest to determine whether other ansatzae, rather than the ZGR-QFT ansatz, can provide similar speedups to algorithmic runtime. 

\end{itemize}

\subsection{PDE nonlineaerity library}

\noindent The cost functions obtained for the PDEs of interest in this work take the form,

\begin{align*}
    \mathcal{C}_{\mathrm{BT}} \big( \vec{\lambda}_{\mathrm{BT}} \big)  \equiv \mathcal{C}_{\mathrm{BT}} \big( \lambda_{0,\mathrm{BT}} , \lambda_{\mathrm{BT}} \big)    \propto  \underset{k \geq 0}{\sum}   f_{\mathrm{BT},k} \big[ \big\{ \lambda_k \big\} , \big\{ O_k \big\} , U \big( \theta \big) \big] \equiv   \underset{k \geq 0}{\sum} f_{\mathrm{BT},k}  \big[ \mathrm{Tr} \big[  O_k U \big( \theta \big)\lambda_k U^{\dagger} \big( \theta \big)         \big] \big]  , \\ \\ \mathcal{C}_{\mathrm{LT}} \big( \vec{\lambda}_{\mathrm{LT}} \big)  \equiv \mathcal{C}_{\mathrm{BT}} \big( \lambda_{0,\mathrm{LT}} , \lambda_{\mathrm{LT}} \big)    \propto  \underset{k \geq 0}{\sum}   f_{\mathrm{LT},k} \big[ \big\{ \lambda_k \big\} , \big\{ O_k \big\} , U \big( \theta \big) \big] \equiv   \underset{k \geq 0}{\sum} f_{\mathrm{LT},k}  \big[ \mathrm{Tr} \big[  O_k U \big( \theta \big)\lambda_k U^{\dagger} \big( \theta \big)         \big] \big]  , \\ \\ \mathcal{C}_{\mathrm{CH}} \big( \vec{\lambda}_{\mathrm{CH}} \big)  \equiv \mathcal{C}_{\mathrm{CH}} \big( \lambda_{0,\mathrm{CH}} , \lambda_{\mathrm{CH}} \big)    \propto  \underset{k \geq 0}{\sum}   f_{\mathrm{CH},k} \big[ \big\{ \lambda_k \big\} , \big\{ O_k \big\} , U \big( \theta \big) \big] \equiv   \underset{k \geq 0}{\sum} f_{\mathrm{CH},k}  \big[ \mathrm{Tr} \big[  O_k U \big( \theta \big)\lambda_k U^{\dagger} \big( \theta \big)         \big] \big]  , 
\end{align*}

\noindent for,

\[ \left\{\!\begin{array}{ll@{}>{{}}l} \lambda_{0,\mathrm{BT}} \equiv \textit{Variational parameter for beginning of time evolution of solution state} \\ \textit{ wavefunctions for the Boussniesq-type PDE} , \\ \\  \lambda_{0,\mathrm{LT}} \equiv \textit{Variational parameter for beginning of time evolution of solution state} \\ \textit{ wavefunctions for the Lin Tsien PDE}   ,  \\ \\ \lambda_{0,\mathrm{CH}} \equiv \textit{Variational parameter for beginning of time evolution of solution state} \\ \textit{ wavefunctions for the Camassa Holm PDE}   , 
\end{array}\right.
\]

\noindent corresponding to instances of initial conditions through variational parameters $\lambda_0$,

\[ \left\{\!\begin{array}{ll@{}>{{}}l}                  \Lambda_{0,\mathrm{BT}} \equiv    \underset{\lambda \in \Lambda}{\bigcup}  \lambda_{0,\mathrm{BT}}      , \\ \\     \Lambda_{0,\mathrm{LT}} \equiv    \underset{\lambda \in \Lambda}{\bigcup}  \lambda_{0,\mathrm{LT}}      ,  \\ \\    \Lambda_{0,\mathrm{CH}} \equiv    \underset{\lambda \in \Lambda}{\bigcup}  \lambda_{0,\mathrm{CH}}      , 
\end{array}\right.
\]

\noindent corresponding to the set of all possible $\lambda_0$ for each PDE,

\[ \left\{\!\begin{array}{ll@{}>{{}}l}                \lambda_{\mathrm{BT}} \equiv \textit{Variational parameter after the  beginning of time evolution of solution state} \\ \textit{ wavefunctions for the Boussniesq-type PDE} , \\ \\  \lambda_{\mathrm{LT}} \equiv \textit{Variational parameter after the beginning of time evolution of solution state} \\ \textit{ wavefunctions for the Lin Tsien PDE}   ,  \\ \\ \lambda_{\mathrm{CH}} \equiv \textit{Variational parameter after the beginning of time evolution of solution state} \\ \textit{ wavefunctions for the Camassa Holm PDE}   , 
\end{array}\right.
\]

\noindent corresponding to the set of all possible $\lambda$ after the initial point of time evolution, and,

\[ \left\{\!\begin{array}{ll@{}>{{}}l}                  \Lambda_{\mathrm{BT}} \equiv    \underset{\lambda \in \Lambda}{\bigcup}  \lambda_{\mathrm{BT}}      , \\ \\     \Lambda_{\mathrm{LT}} \equiv    \underset{\lambda \in \Lambda}{\bigcup}  \lambda_{\mathrm{LT}}      ,  \\ \\    \Lambda_{\mathrm{CH}} \equiv    \underset{\lambda \in \Lambda}{\bigcup}  \lambda_{\mathrm{CH}}      , 
\end{array}\right.
\]

\noindent corresponding to the set of all possible $\lambda$ for each PDE. The observables $O_k$ for each cost function take on a wide variety of intriguing forms, which are ultimately dependent upon the intrinsic nonlinearities of each PDE. For example, observables,

\begin{align*}
    O_k \equiv O_{\mathrm{BT}_k} , 
\end{align*}

\noindent corresponding to terms appearing in the Boussniesq-type cost function can be expressed with,

\begin{align*}
   \mathrm{span} \bigg\{         \big(  \lambda^B_0 \big)^2  ,  A_x , A^{\dagger}_x ,  2   \big( \frac{\beta}{(\Delta x)^2} \big)^2    ,   \bra{u} , \ket{u} ,  \bra{u} \ket{u} ,         \bra{\widetilde{u}}   , \bra{\widetilde{\widetilde{u}}}   , \ket{\widetilde{u}}   , \ket{\widetilde{\widetilde{u}}}  ,   \bra{\widetilde{u}}   \ket{\widetilde{\widetilde{u}}}    , \bra{\widetilde{\widetilde{u}}}   \ket{\widetilde{\widetilde{u}}}                                                             \bigg\}                .
\end{align*}

\noindent Straightforwardly, one can formulate the spanning sets,

\begin{align*}
  O_{\mathrm{LT}_k} \equiv \mathrm{span}  \big\{ \textit{LT Operators} \big\}   , \\ \\ O_{\mathrm{CH}_k} \equiv \mathrm{span}  \big\{ \textit{CH Operators} \big\}  , 
\end{align*}

\noindent for the two remaining PDEs using similar expressions, which can be read off from the cost functions provided in the Appendix.

\subsection{Paper organization}

Beyond the initial discussion and overview of motivation for simulating time evolution, novel applications of hybrid quantum algorithms are presented. After having obtained a suitably defined cost function for each PDE, variatioonal states corresponding to the initial condition are further characterized. Within the framework of {[12]}, we formulate applications to various nonlinear problems of interest, which could be simulated on next-generation near term devices. From several characteristics fo solutions that can be classically predicted, including the curvature, initial conditions, and well-posedness, we identify nonlinear problems of interest that can be efficiently studied with recent variants of the variational quantum eigensolver. From well-established representations of quantum networks as sequences of unitary transformations that are applied to qubits initialized in the $0$ state at the beginning of time evolution, we obtain readout from quantum circuits through adaptations of the VQA of {[12]} with ZGR-QFT anstzae that are classically simulable in polynomial time.

From QNPUs provided in {[12]} for approximating solutions to the nonlinear Schrodinger's, and Burgers, equations with a few qubits encooded in the ansatz, the total number of qubits which must be time evolved in a quantum circuit is indicative of the computational complexity of each time step in the evolution. To account for the ways in which computational complexity posed by the cost function impacts the efficieny of the hybrid quantum algoorithm, we efficiently address computational difficulties associated with vanishing gradients of the cost function to ansatz construction {[12]}, expressibility of parametrizations for quantum circuits {[20]}, qubit quality and stable quantum representations of solutions to PDE IVPs, data compression in machine learning for classification tasks, error mitigation, error correcting codes, and noise and complexity growth {[2,7]}. To systematically characterize the coverage of the Hilbert space that we expect from such quantum simulations, we sample entries from a state wavefunction, from which comparisons with classical solution approximations can be established. To obtain the final wavefunction that will be used to construct the solution approximation, we perform time evolution for however many time steps there are in the noiseless simulation. Despite having great promise for being able to accommodate noise in the future, currently the VQA extensively studied in this work would require significantly more computational power that could potentially make the runtime larger than that of classical solvers. Nevertheless, determining whether error rates can be mitigated, while simultaneously preserving advantages in runtime, remains of interest to further explore.

To ensure that the VQA robustly performs across different parameter initialization schemes that can be used for encoding initial states of the time evolution in quantum circuits, we generate a broad ensemble of quantum states for direct comparison with classically known exact solutions to each PDE.

\begin{figure}
\begin{align*}
\includegraphics[width=0.43\columnwidth]{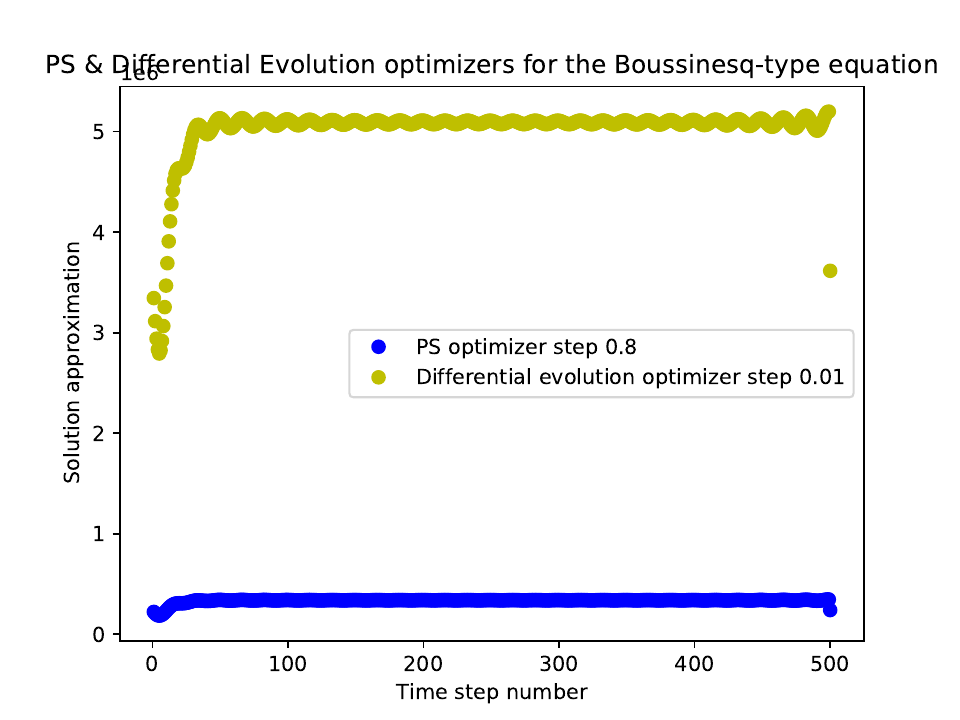} \\ \includegraphics[width=0.43\columnwidth]{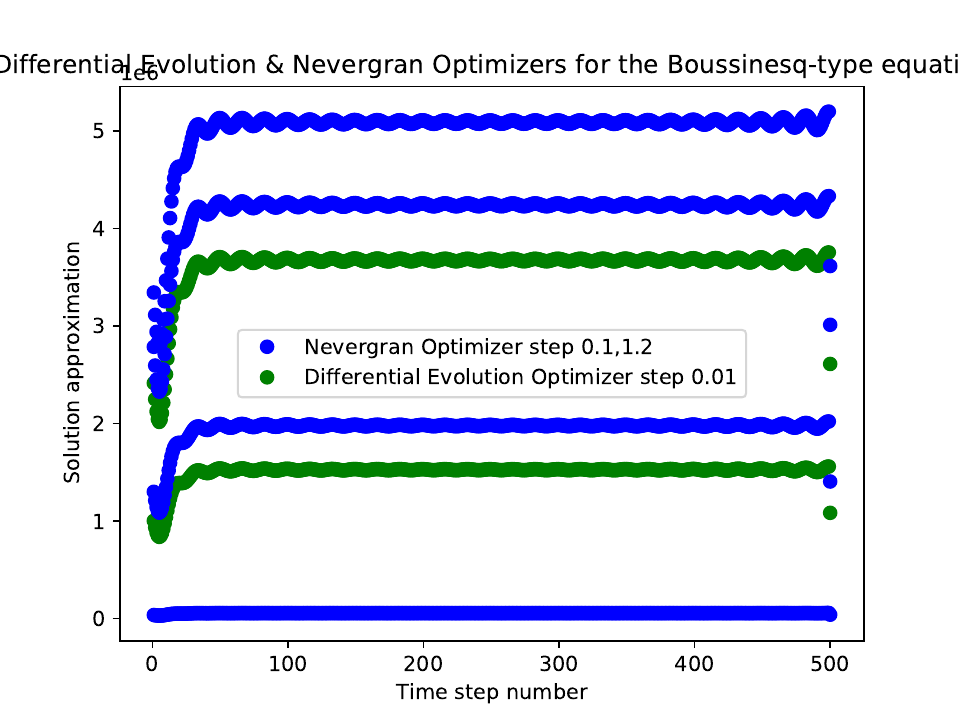}\\\includegraphics[width=0.43\columnwidth]{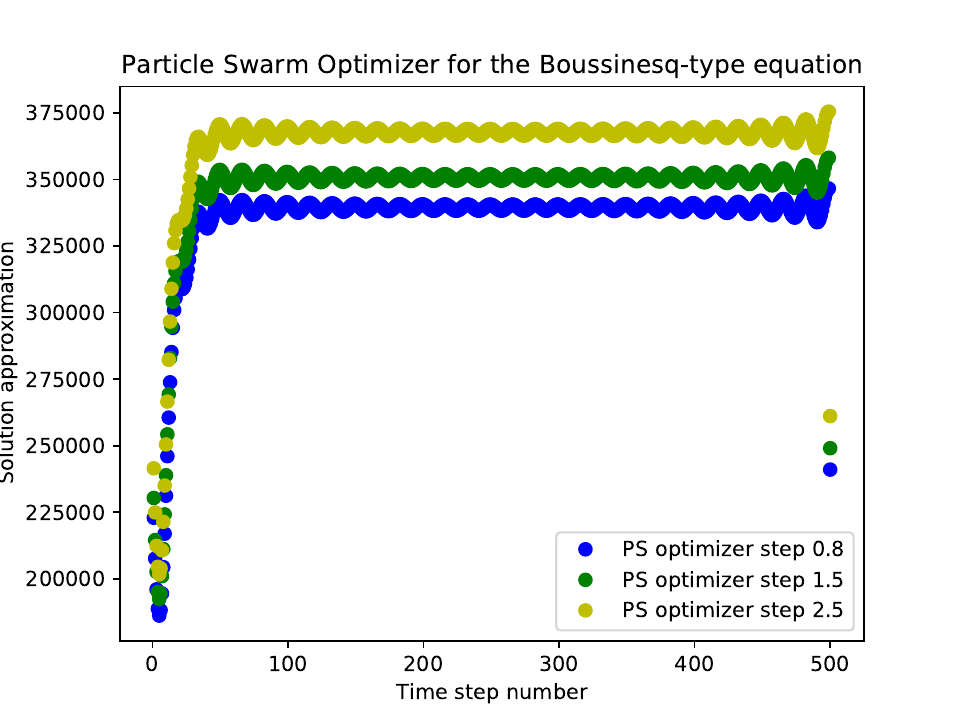}\\
\end{align*}
\caption{\textit{Extracting measurements from the Boussinesq-type equation with $500$ time steps of evolution.}}
\end{figure}

 \begin{figure}
\begin{align*}
\includegraphics[width=0.65\columnwidth]{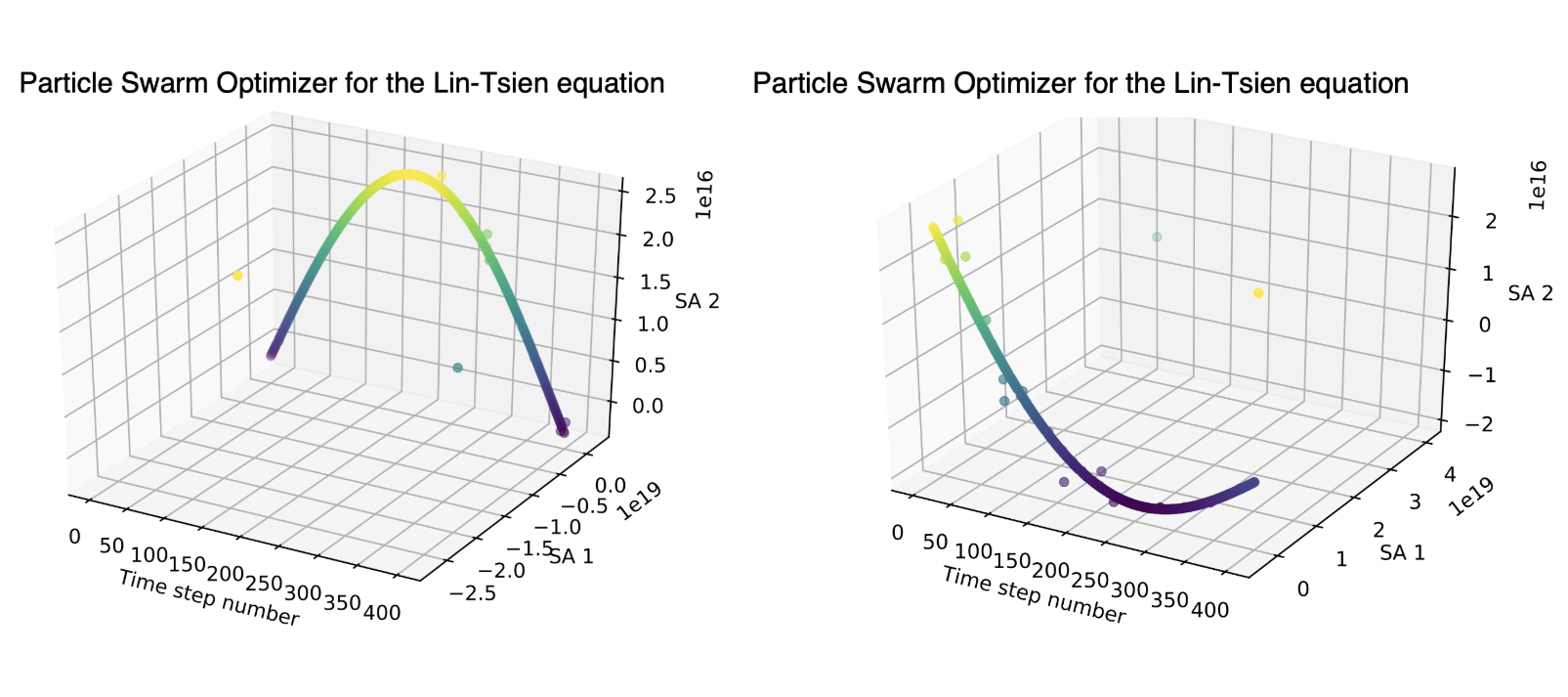}\\
\includegraphics[width=0.65\columnwidth]{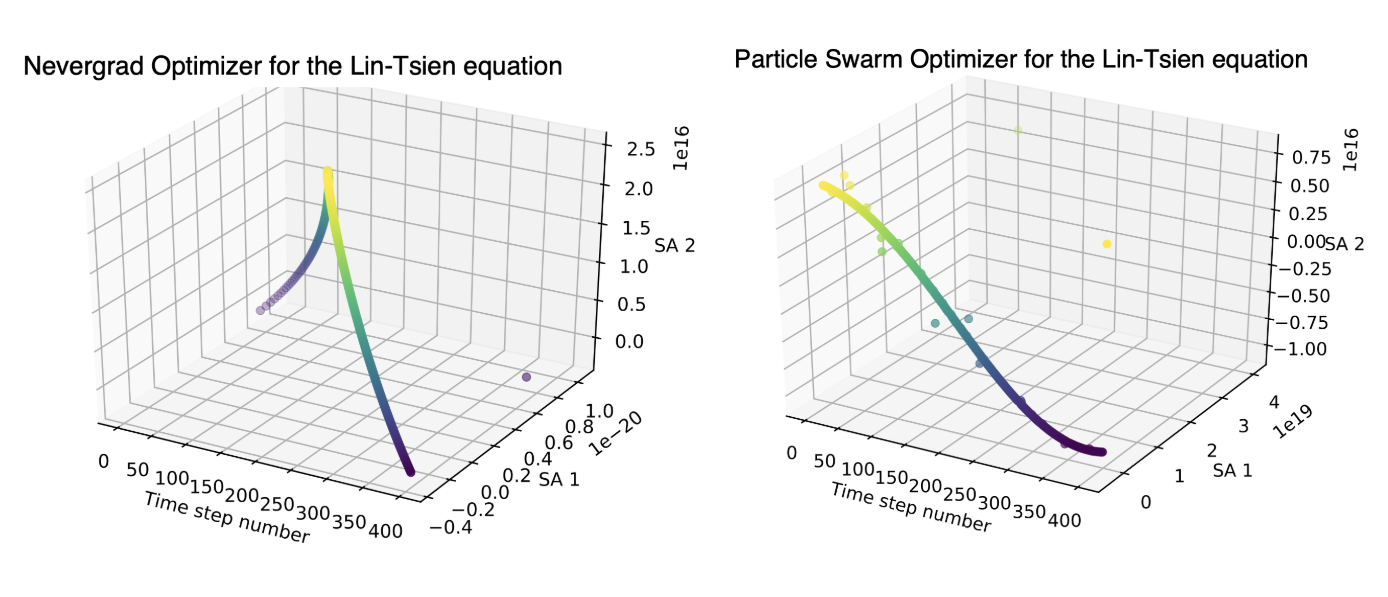}\\\includegraphics[width=0.65\columnwidth]{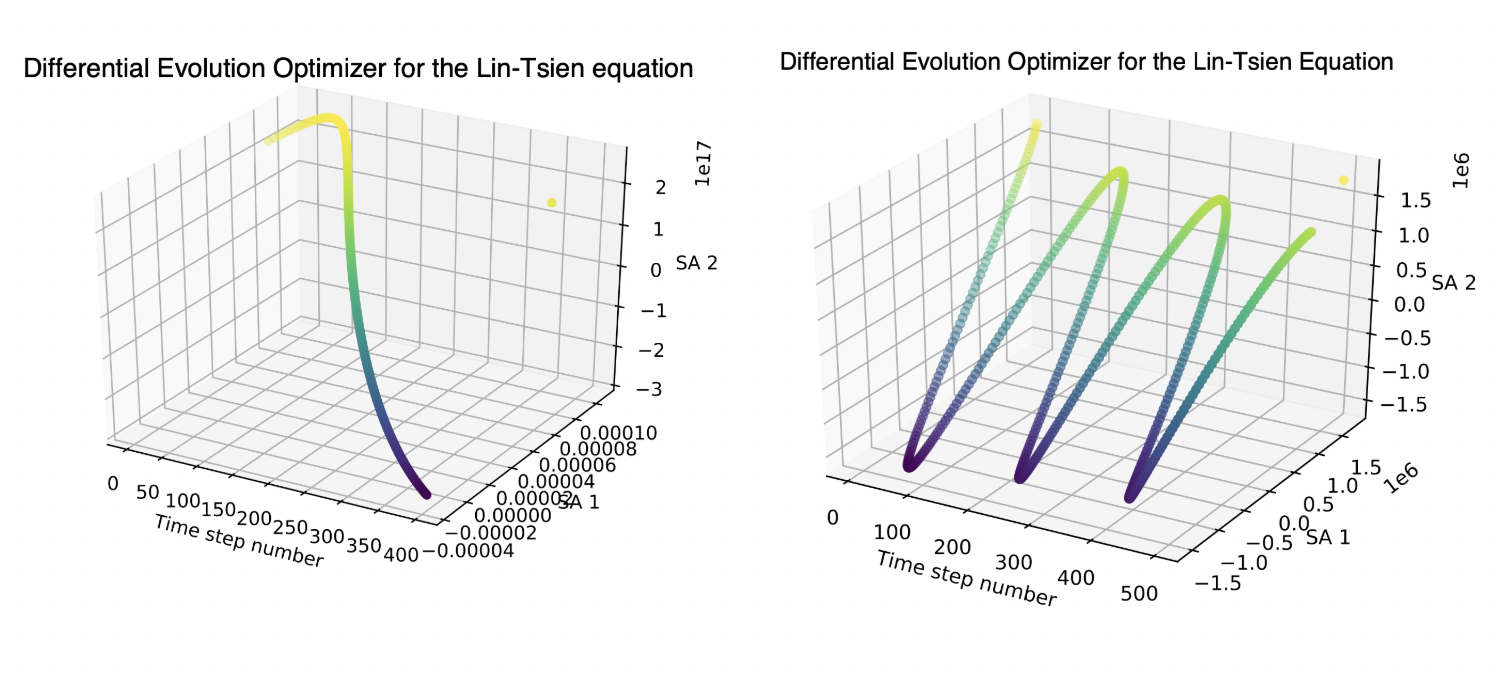}\\\includegraphics[width=0.65\columnwidth]{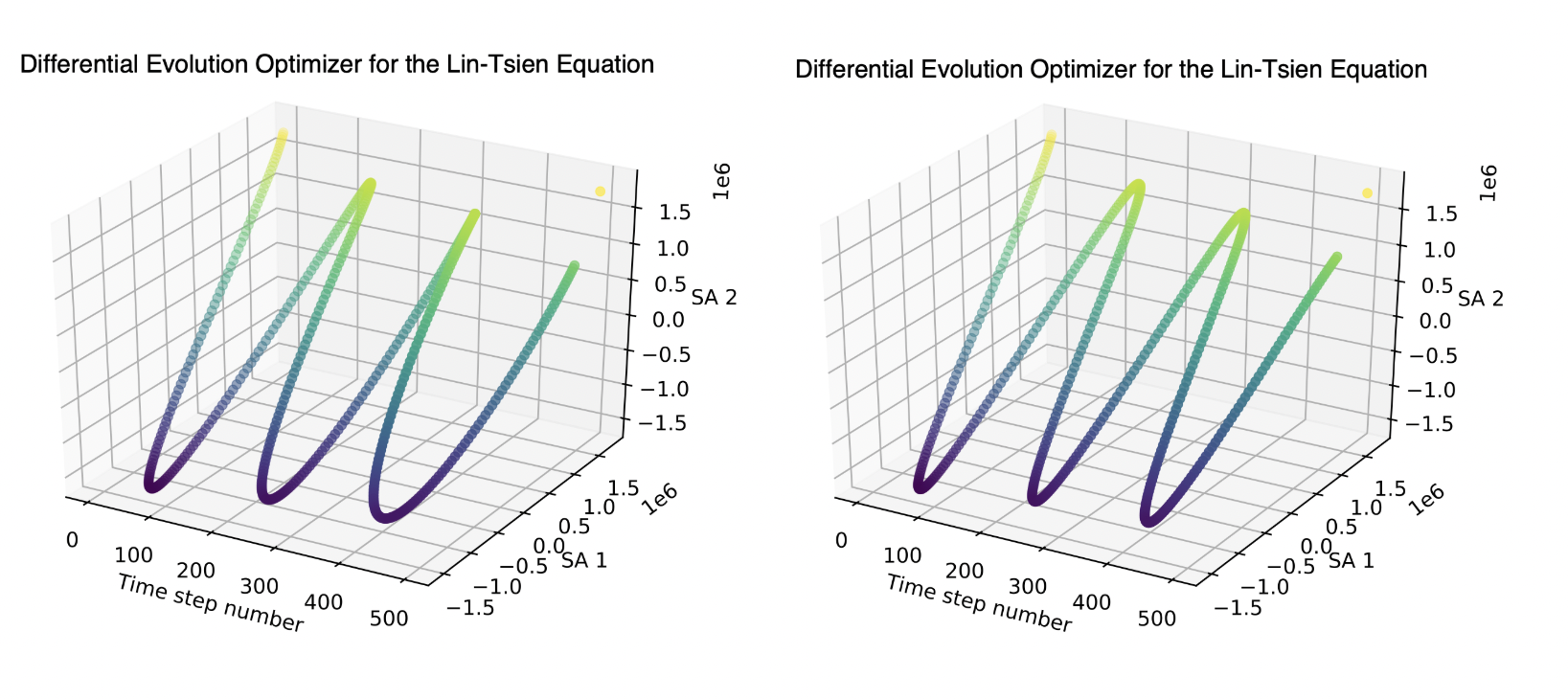}\\
\end{align*}
\caption{\textit{Extracting measurements from the Lin-Tsien equation with $400$ time steps of evolution}.}
\end{figure}

\begin{figure}
\begin{align*}
   \includegraphics[width=0.7\columnwidth]{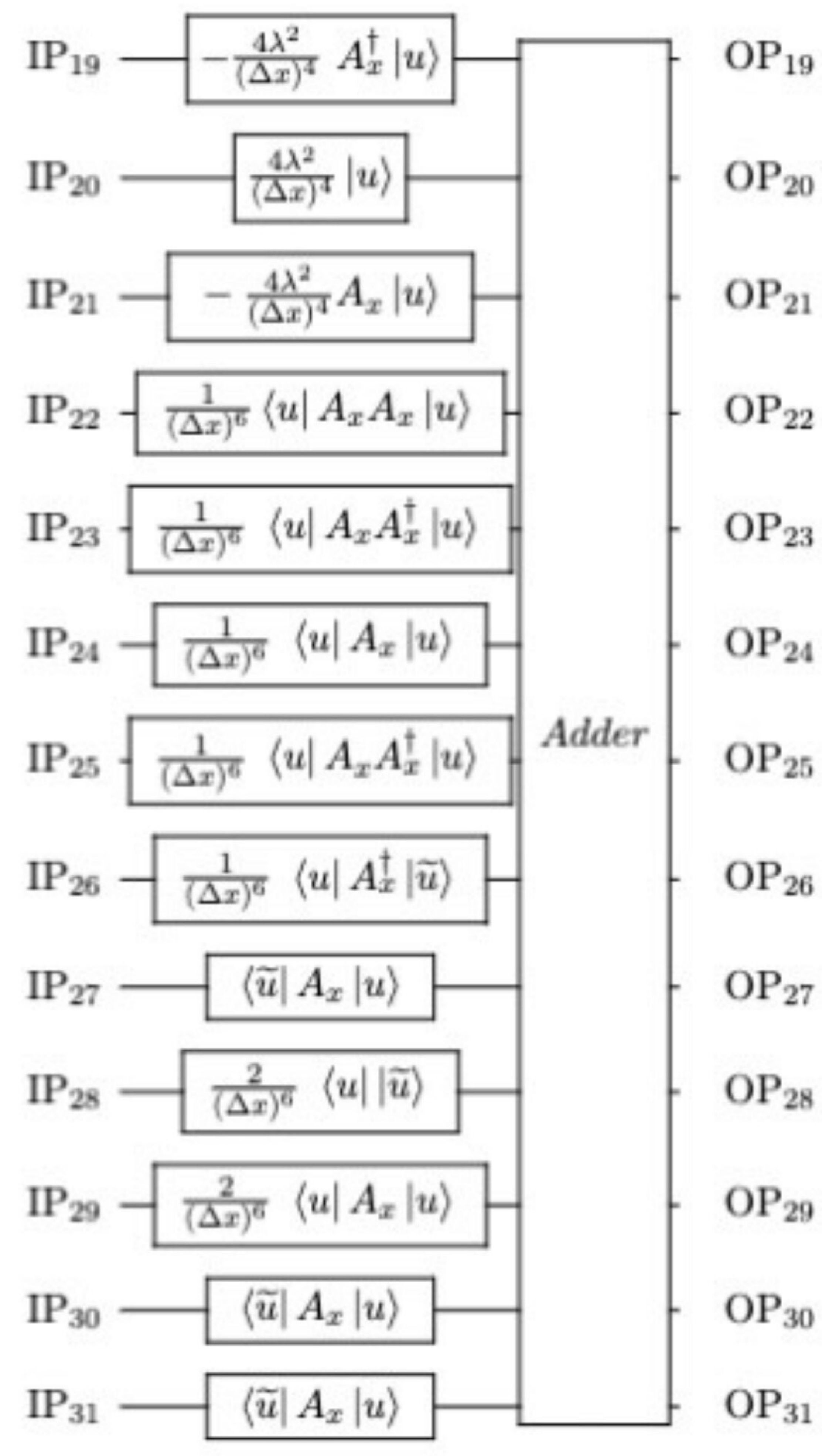}\\
\end{align*}
\caption{\textit{A portion of the QNPU in quantum circuits for running the VQA  for the Camassa-Holm PDE}.}
\end{figure}

\begin{figure}
\begin{align*}
\includegraphics[width=0.96\columnwidth]{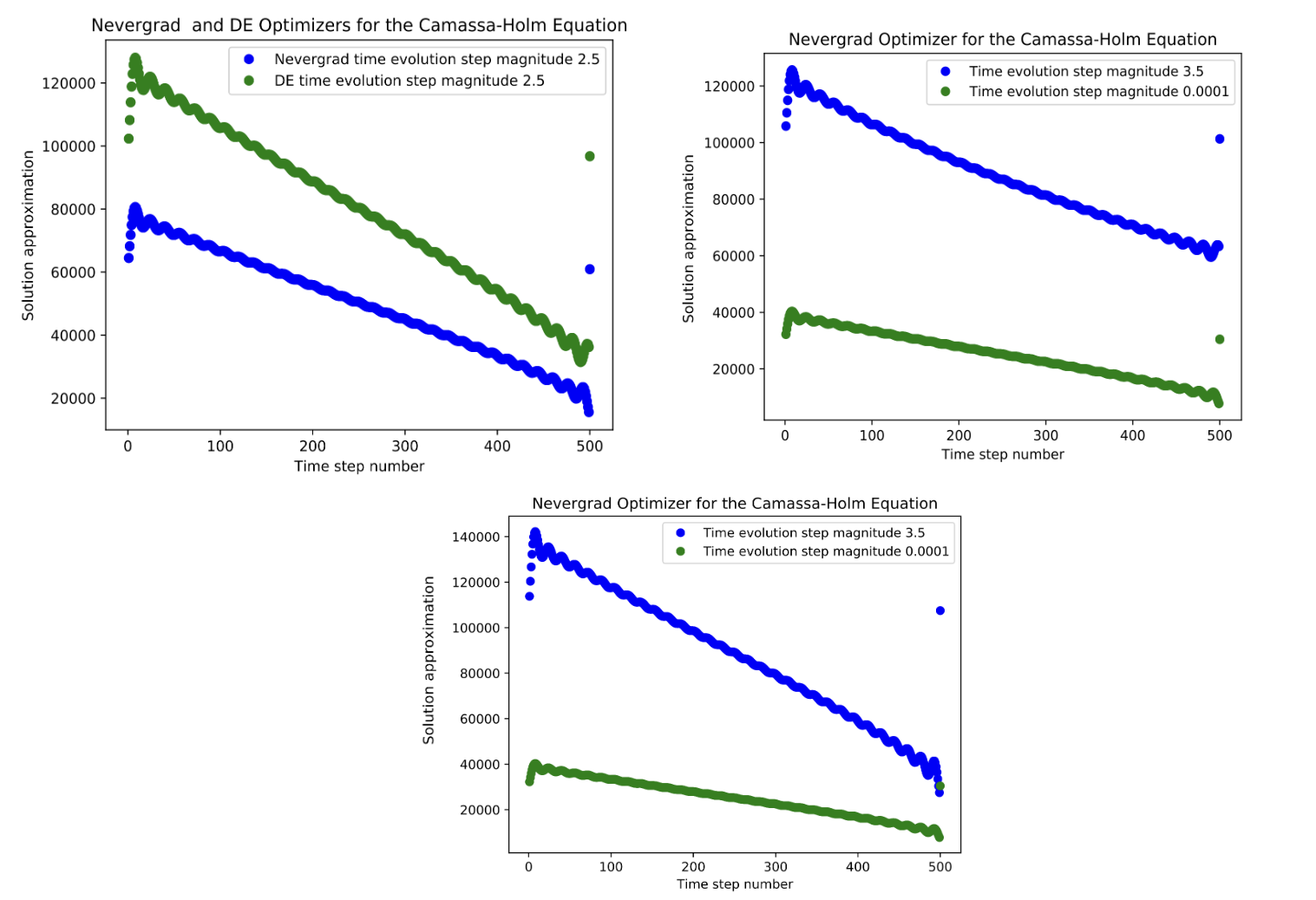}\\\includegraphics[width=0.9\columnwidth]{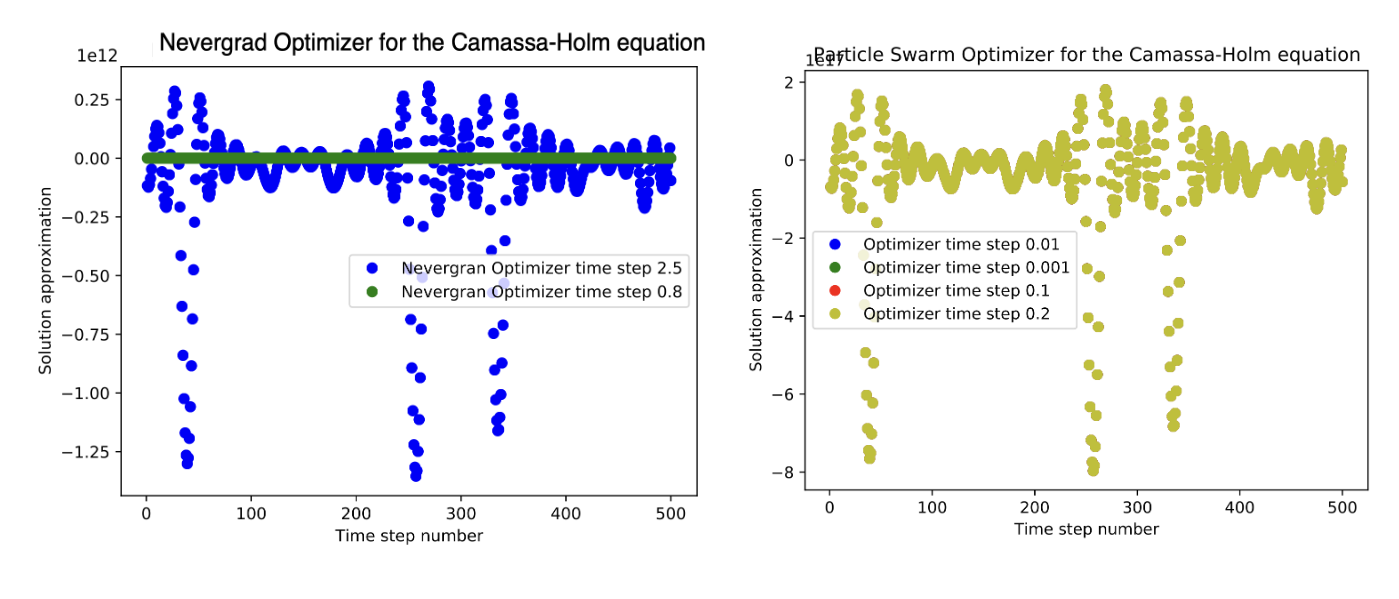}
\end{align*}
\caption{\textit{Extracting measurements from the Camassa-Holm equation with $8500$ time steps}. Implementing the variational quantum algorithm for $8500$ time steps yields sinusoidal profiles of motion.}
\end{figure}

  \begin{figure}
\begin{align*}
\includegraphics[width=0.8\columnwidth]{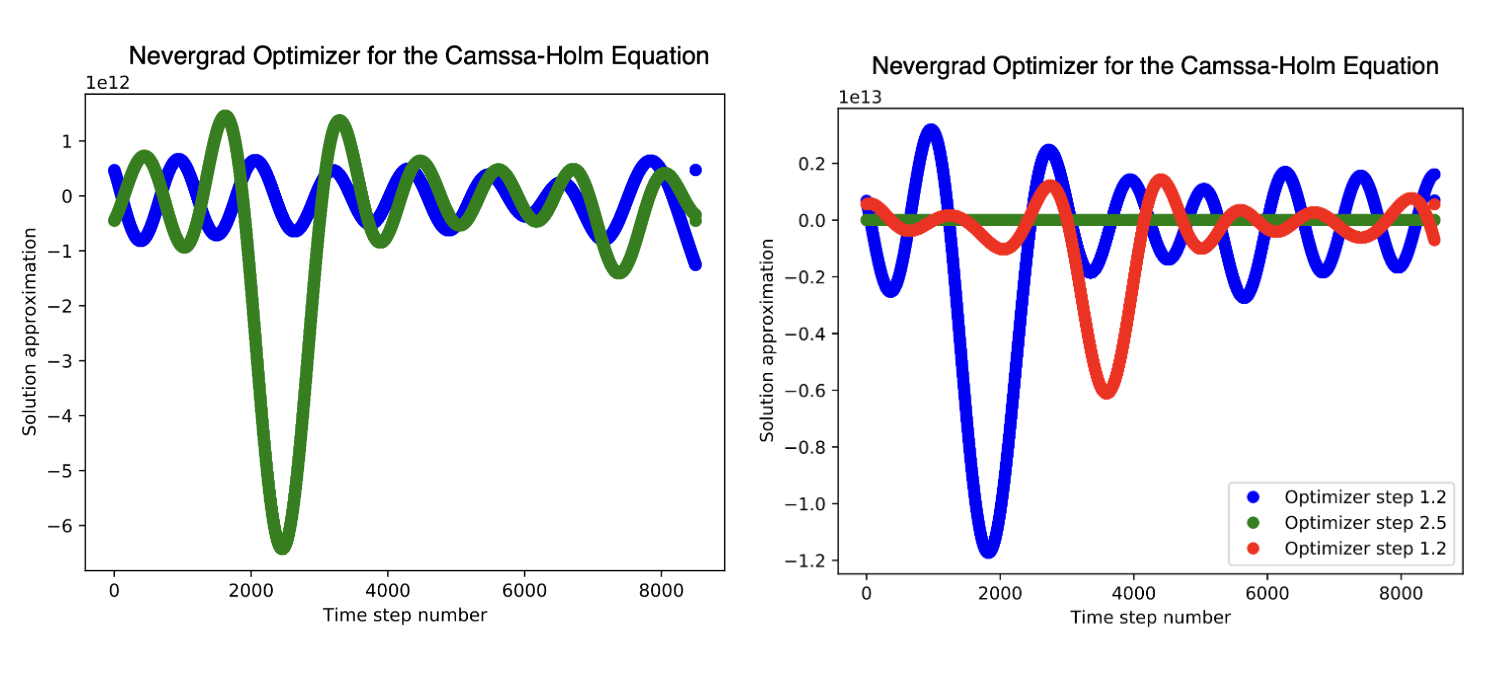}\\
\includegraphics[width=0.8\columnwidth]{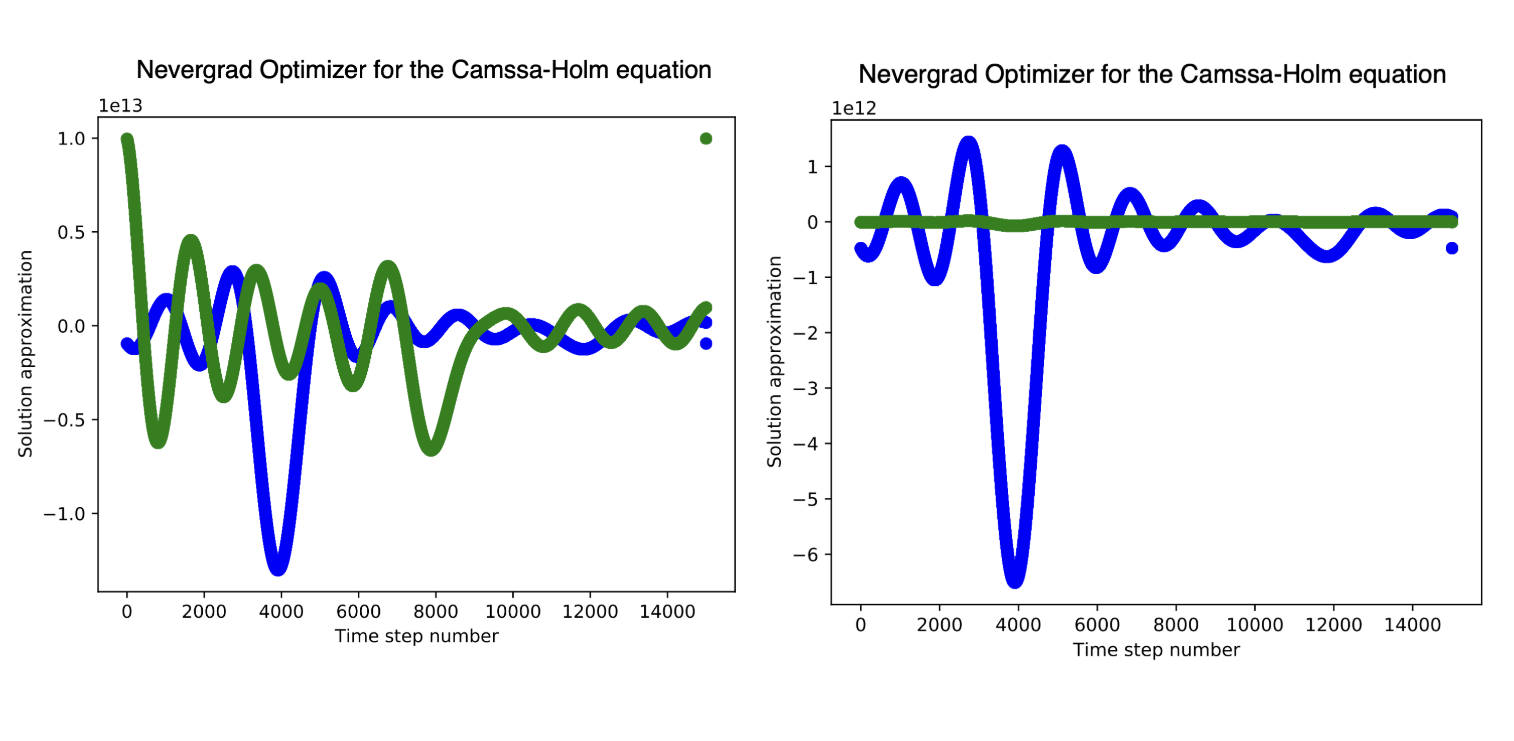}\\\includegraphics[width=0.8\columnwidth]{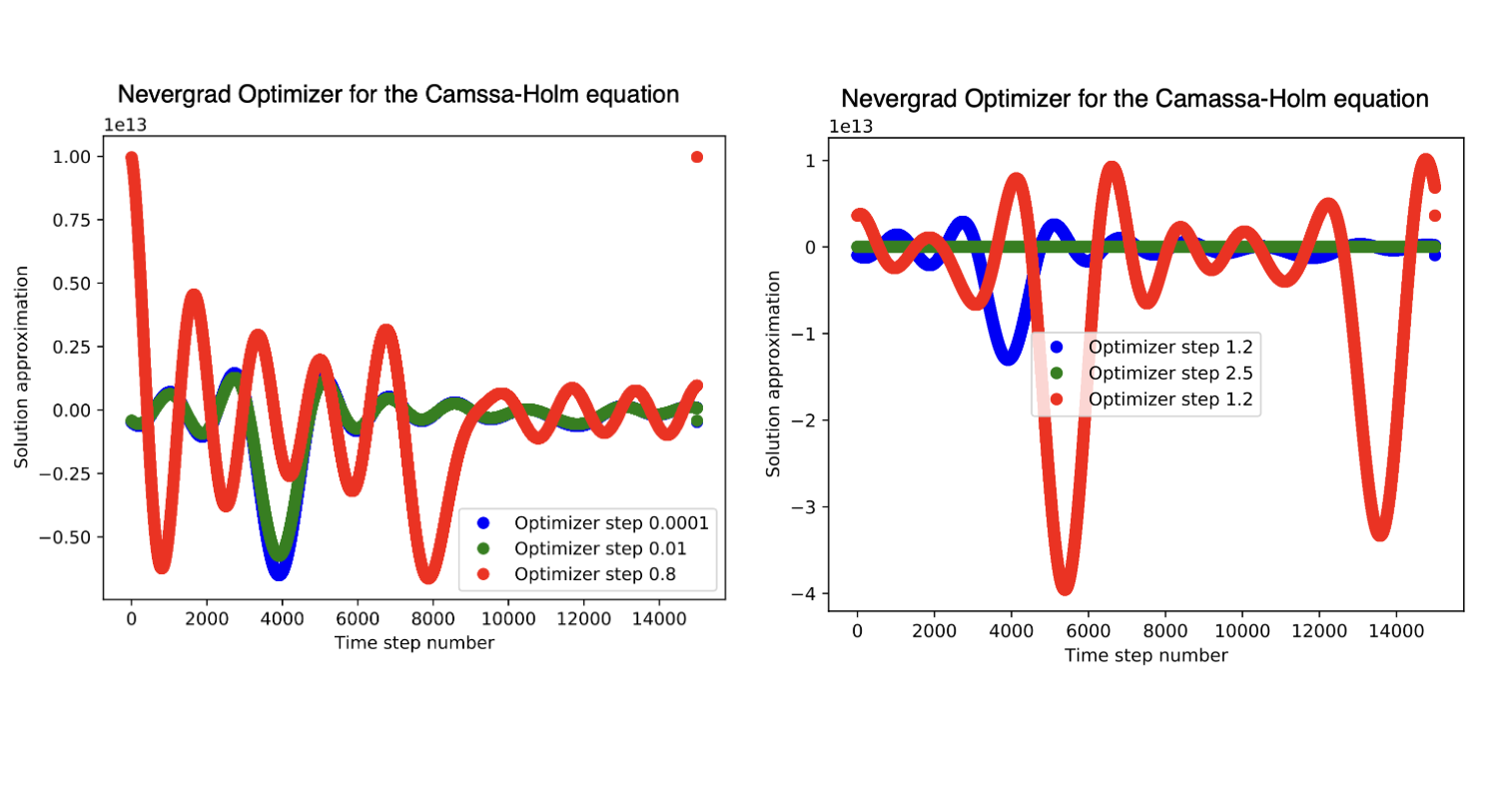}
\end{align*}
\caption{\textit{Extracting measurements from the Camassa-Holm equation}. Implementing the variational quantum algorithm for $4500$ time steps yields sinusoidal profiles of motion.}
\end{figure}

Throughout the time evolution, at each step the VQA extracts measurements from the solution state wavefunction.

With several qubits typically being used for encoding the ansatz for the solution state wavefunction, with rapid developments to quantum hardware with new prototypes emerging, running variants of such hybrid algorithms, or novel ones altogether, continues to remain a valuable pursuit.






\newpage 

\section{Cost function derivations}

\subsection{Boussinesq type equation}

\subsubsection{Statement}
            
                   \begin{align*}
                  \underline{ \mathcal{C}^B(\lambda,\lambda^B_0)}  =    \text{ }\big(  \lambda^B_0 \big)^2 +    \mathrm{Re}  \bigg\{     2   \big( \frac{\beta}{(\Delta x)^2} \big)^2    \bra{\widetilde{u}}   A^{\dagger}_x \ket{\widetilde{\widetilde{u}}}                - 2  \big( \frac{\beta}{(\Delta x)^2} \big)^2 \frac{(\Delta x)^2}{\beta}   \bra{\widetilde{u}}  \ket{\widetilde{\widetilde{u}}}             - 2    \big( \frac{\beta}{(\Delta x)^2} \big)^2 \end{align*}
                  
                  \begin{align*} \times      \bra{\widetilde{u}} A_x A^{\dagger}_x \ket{\widetilde{u}}  +     2   \big( \frac{\beta}{(\Delta x)^2} \big)^2    \big( 2 + \frac{\beta}{(\Delta x)^2} \big)   \bra{\widetilde{u}} A_x \ket{\widetilde{\widetilde{u}}}  \text{        }  +   4          \big( \frac{\beta}{(\Delta x)^2} \big)^2    \bra{\widetilde{u}} A_x  A_x \ket{\widetilde{\widetilde{u}}}   \\ -    4  \big( \frac{\beta}{(\Delta x)^2} \big)^2   \big( 2 + \frac{\beta}{(\Delta x)^2} \big)   \bra{\widetilde{u}} \ket{\widetilde{\widetilde{u}}}  +  2  \big( \frac{\beta}{(\Delta x)^2} \big)^2   \bra{\widetilde{u}} A_x A_x \ket{\widetilde{\widetilde{u}}}  +   8  \big( \frac{\beta}{(\Delta x)^2} \big)^2        \bra{\widetilde{u}} A^{\dagger}_x \ket{\widetilde{u}}   \\ -  
                   \big( \frac{\beta}{(\Delta x)^2} \big)^2  \bra{\widetilde{\widetilde{u}}}     A^{\dagger}_x A^{\dagger}_x      \ket{\widetilde{\widetilde{u}}}    
                        \text{        }   + 2  \big( \frac{\beta}{(\Delta x)^2} \big)^2             \bra{\widetilde{\widetilde{u}}} A^{\dagger}_x \ket{\widetilde{\widetilde{u}}}    +  \big( \frac{\beta}{(\Delta x)^2} \big)^2          \frac{(\Delta x)^2}{\beta}  \bra{\widetilde{\widetilde{u}}} A^{\dagger}_x \ket{\widetilde{\widetilde{u}}}  \\ - 
 \big( \frac{\beta}{(\Delta x)^2} \big)^2  \bra{\widetilde{\widetilde{u}}} A^{\dagger}_x A_x \ket{\widetilde{\widetilde{u}}}   -         \big( \frac{\beta}{(\Delta x)^2} \big)^2         \bra{\widetilde{\widetilde{u}}}   A_x A_x       \ket{\widetilde{\widetilde{u}}}       +  2  \big( \frac{\beta}{(\Delta x)^2} \big)^2             A_x \ket{\widetilde{\widetilde{u}}}   \\ + 
      \big( \frac{\beta}{(\Delta x)^2} \big)^2           \frac{(\Delta x)^2}{\beta}  \bra{\widetilde{\widetilde{u}}} A_x \ket{\widetilde{\widetilde{u}}}    -   2  \big( \frac{\beta}{(\Delta x)^2} \big)^2                  \bra{\widetilde{\widetilde{u}}} A^{\dagger}_x \ket{\widetilde{\widetilde{u}}}      -  \big( \frac{\beta}{(\Delta x)^2} \big)^3   \bra{\widetilde{\widetilde{u}}} A^{\dagger}_x \ket{\widetilde{\widetilde{u}}}   \\ + 2   \big( \frac{\beta}{(\Delta x)^2} \big)^2   \bra{\widetilde{\widetilde{u}}} A_x \ket{\widetilde{\widetilde{u}}}           \text{          }     - 2   \big( \frac{\beta}{(\Delta x)^2} \big)^3        \bra{\widetilde{\widetilde{u}}} \ket{\widetilde{\widetilde{u}}}    +  \big( \frac{\beta}{(\Delta x)^2} \big)  \bra{\widetilde{\widetilde{u}}} A_x \ket{\widetilde{\widetilde{u}}} \\ +   \big( \frac{\beta}{(\Delta x)^2} \big)  \bra{\widetilde{\widetilde{u}}} A^{\dagger}_x \ket{\widetilde{\widetilde{u}}} +  \big( \frac{\beta}{(\Delta x)^2} \big)^2  \bra{\widetilde{\widetilde{u}}} A^{\dagger}_x \ket{\widetilde{\widetilde{u}}}     +   4  \big( \frac{\beta}{(\Delta x)^2} \big)^2       \big( \bra{\widetilde{\widetilde{u}}}  A_x     A^{\dagger}_x \ket{\widetilde{\widetilde{u}}}     A^{\dagger}_x \ket{\widetilde{\widetilde{u}}}   \\ -  2   \big( \frac{\beta}{(\Delta x)^2} \big)                           \bra{\widetilde{\widetilde{u}}}  \ket{\widetilde{u}}     
       + 2   \big( \frac{\beta}{(\Delta x)^2} \big)^2              \bra{\widetilde{\widetilde{u}}} A^{\dagger}_x \ket{\widetilde{\widetilde{u}}}  +  4  \big( \frac{\beta}{(\Delta x)^2} \big)^2              \bra{\widetilde{\widetilde{u}}} A_x \ket{\widetilde{\widetilde{u}}}   \\ -     \big( \frac{\beta}{(\Delta x)^2} \big)              \bra{\widetilde{\widetilde{u}}} A_x \ket{\widetilde{\widetilde{u}}}    \text{          }    +   \big( \frac{\beta}{(\Delta x)^2} \big)            \bra{\widetilde{\widetilde{u}}} A^{\dagger}_x \ket{\widetilde{\widetilde{u}}}            - 2  \big( \frac{\beta}{(\Delta x)^2} \big)           \bra{\widetilde{\widetilde{u}}} A_x \ket{\widetilde{\widetilde{u}}}    \\     +   \big( \frac{2 \beta^3 }{(\Delta x)^6} \big)              \bra{u} A^{\dagger}_x A^{\dagger}_x \ket{u}   -     \alpha   \big( \frac{\beta^2}{(\Delta x)^6} \big)     A^{\dagger}_x \ket{u} \big( \bra{u} A_x \ket{u} \big)  \\    - \tau^2   \big( \frac{\beta^2}{(\Delta x)^4} \big)                   \bra{u} A^{\dagger}_x A^{\dagger}_x \ket{u}                    -  2  \tau^2  \big( \frac{\beta^2}{(\Delta x)^4} \big)       \bra{u} A^{\dagger}_x A^{\dagger}_x \ket{u}                  \\          + 4 \tau^2 \alpha    \big( \frac{\beta}{(\Delta x)^2} \big)^2   \bra{u}  A^{\dagger}_x A_x \ket{u}            +     \tau^4   \big( \frac{\beta}{(\Delta x)^2} \big)^2  \bra{u} A^{\dagger}_x A^{\dagger}_x \ket{u}   \bigg\}       \text{ . }    
\end{align*}

\subsubsection{Derivation}

\noindent The equation with smooth solutions $u$ is,

\begin{align*}
        u_{tt} - u_{xx} - 2 \alpha \big( u u_x \big)_x - \beta u_{xxtt} = 0        \\
               \text{              } \Updownarrow      \text{                       }    \\ 
        \frac{1}{\tau^2}     \text{         }      \ket{u(t+2\tau)} -  \frac{1}{\tau^2}  \ket{u(t+\tau)}             \text{         }    - \big\{ A^{\dagger}_x - 2 \textbf{1} + A_x       \big\} \ket{u}  - 2 \alpha \bigg\{       \ket{u} \times \bigg\{  \frac{A_x - \textbf{1}}{\Delta x}     \bigg\} \ket{u}   \bigg\}_x \\ -         \text{          }     \frac{\beta}{(\Delta x)^2} \bigg\{        \big\{ A^{\dagger}_x - 2 \textbf{1} + A_x \big\} \ket{u}      \text{        } \bigg\}_{tt}  = 0 \text{ . } 
                      \end{align*}
                      
                      \noindent Differentiating the final term with respect to time yields,
                      
                      \begin{align*}
     \frac{\beta}{(\Delta x)^2} \text{             } \bigg\{    \frac{1}{\tau} \big(    \text{    }       \big\{ A^{\dagger}_x - 2 \textbf{1} + A_x \text{        } \big\} \text{             } \ket{u(t + \tau)}- \big\{ A^{\dagger}_x - 2 \textbf{1} + A_x \big\} \ket{u(t)} \big) \text{             } \bigg\}_t  \text{, }\end{align*}

     \noindent which is equivalent to the superposition,
     
     \begin{align*} \frac{\beta}{(\Delta x)^2 \tau} \bigg\{ \frac{1}{\tau}     \bigg[   \bigg\{ \big\{ A^{\dagger}_x - 2 \textbf{1} + A_x \big\} \ket{u(t+ 2 \tau)} -  \big\{ A^{\dagger}_x - 2 \textbf{1} +  A_x \big\} \ket{u(t+\tau)} \bigg\}  -  \bigg\{ \text{     }       \big\{ A^{\dagger}_x - 2 \textbf{1} +A_x \big\} \\ \times \ket{u(t+\tau)}   - 
    \big\{ A^{\dagger}_x - 2 \textbf{1} + A_x \big\} \ket{u(t)}  \bigg\}\bigg]  \bigg\}    \text{.}
    \end{align*}

  \noindent Rearranging the state before differentiation with respect to space yields the expression, 
  
                        \begin{align*}
                           2 \alpha \bigg\{ \ket{u} \times \bigg\{ \frac{A_x - \textbf{1}}{\Delta  x } \bigg\} \ket{u} \text{    }   \bigg\}_x   \Longleftrightarrow  \text{                                                   }      2 \alpha \text{            } \bigg\{ \frac{ A_x - \textbf{1}}{\Delta x} \bigg\} \bigg\{ \text{            } \ket{u}  \bigg\{ \frac{A_x - \textbf{1}}{\Delta x }    \bigg\} \text{  } \ket{u} \text{     }         \bigg\} \\ \Longleftrightarrow   2 \alpha \text{      } \bigg\{ \text{             } \frac{ A_x \ket{u} - \ket{u}  }{\Delta x}   \times \text{              }         \frac{A_x \ket{u} - \ket{u}}{\Delta x} \bigg\}     \Longleftrightarrow    \frac{2 \alpha}{(\Delta x)^2} \text{            } \bigg\{   \text{       }     A_x \ket{u} \bra{u} A^{\dagger}_x - 2 A_x \ket{u} \ket{u}    + \ket{u} \bra{u}         \text{          }         \bigg\}     \text{           }  \\ \Longleftrightarrow \frac{4 \text{        } \alpha}{(\Delta x)^2} \bigg\{    \lambda^{2}    \text{         } -  A_x \ket{u}   \bigg\}   \text{.}\end{align*}
                             
                \noindent The time evolved variational state is,
                                      
                             \begin{align*}
                               \frac{1}{\tau^2} \text{      } \ket{u(t+\tau)} \text{       }    = \text{        } \frac{1}{\tau^2} \ket{u(t+2\tau)}  \text{       }   - \text{           } A^{\dagger}_x \ket{u} + 2 \ket{u} + A_x \ket{u} \text{      }       - \frac{4 \alpha}{(\Delta x)^2} \bigg\{    \lambda^{2}    \text{         }  -  A_x \ket{u}   \bigg\}      \text{        }  \\ - 
                                    \frac{\beta}{\tau^{ 2} (\Delta x)^2 } \text{      }          \bigg\{ \text{           }  A^{\dagger}_x \ket{u(t+2\tau)} - 2 \ket{u(t+2\tau)}    + A_x \ket{u(t+2\tau)} \text{         }      - A^{\dagger}_x \ket{u(t+\tau)} + 2 \ket{u(t+\tau)}  \\  +  A_x \ket{u(t+\tau)} - 
                               \text{       } A^{\dagger}_x \ket{u(t+\tau)} + 2 \ket{u(t+\tau)}      -  A_x \ket{u(t+\tau)} \text{            }          - A^{\dagger}_x \ket{u(t)} \\  + 2 \ket{u(t)} - A_x \ket{u(t)}  \text{           }    \bigg\}     \text{       .   }  \\
                               \end{align*}
                       \noindent Rescaling by $\tau^2$ from the time evolved variational state from a previous second application of the time derivative gives,         
                               \begin{align*}
 \ket{u(t+\tau)} \text{       }    = \text{        }  \ket{u(t+2\tau)}  \text{       } - \bigg\{ \big(  \text{    }  \frac{\beta}{(\Delta x)^2} \text{     }     - \tau^2  \big) A^{\dagger}_x +  \big(  2 \text{   } \frac{\beta}{(\Delta x)^2 } - \frac{4 \text{      }  \bra{u}  \alpha}{(\Delta x)^2 } - \tau^2 \big) A_x  + 2   \big(  \frac{\beta}{(\Delta x)^2}    \\ - \tau^2 \big) \text{      } \bigg\}  \ket{u} -   \frac{4 \alpha}{(\Delta x)^2}  \lambda^{2}    \text{        }  -  
                                   \frac{\beta}{(\Delta x)^2} \text{      }          \bigg\{  \big( A^{\dagger}_x  - 2 + A_x  \text{         }   \big)\text{    } \ket{u(t+2\tau)}   -  \big( 2 A^{\dagger}_x - 4  \text{      }    \big) 
                        \ket{u(t+\tau)}      \bigg\}     \text{.          }  
\end{align*}    

\noindent Expanding the cost function, $\mathcal{C}^B(\lambda,\lambda_0) = \text{       }     \big|\big|     \ket{u(t+\tau)}  -        \ket{u(t)}  \big|\big|^{2}$, yields,

\begin{align*}
          \text{        }        \bigg|\bigg|    -  \bigg\{  \big( \frac{\beta}{(\Delta x)^2} - \tau^2 \big) A^{\dagger}_x + \big( 2 \frac{\beta}{(\Delta x)^2} - 4 \frac{\bra{u}  \alpha}{(\Delta x)^2} \text{         }  - \tau^2 \big)   \text{        } A_x +  2 \big( \frac{\beta}{(\Delta x)^2} - \tau^2 - \frac{1}{2} \big) \bigg\} \text{    } \ket{u} \\ -   \frac{\beta}{(\Delta x)^2} \bigg\{ \big( A^{\dagger}_x  - 2 - \frac{(\Delta x)^2}{\beta} + A_x \big) \text{        } \ket{u(t+2\tau)}  -  \big( 2A^{\dagger}_x - 4 \big) \ket{u(t+\tau)}  \bigg\} \text{        }   \bigg| \bigg|^{2}    \text{, }
\end{align*}

\noindent with, 

\begin{align*}
       \text{      }      \bigg\{ \big( \frac{\beta}{(\Delta x)^2} - \tau^2 \big) A^{\dagger}_x + \big( 2 \frac{\beta}{(\Delta x)^2} - 4 \textbf{ } \frac{\bra{u} \alpha}{(\Delta x)^2 } - \tau^2 \big) A_x +  \text{        }   2 \text{      } \big( \text{       }      \frac{\beta}{(\Delta x)^2} - \tau^2 - \frac{1}{2} \big) \bigg\} \ket{u}  \bigg\{  \big( \frac{\beta}{(\Delta x)^2} \\ - \tau^2 \big) A^{\dagger}_x
     \text{              }  \big( 2 \frac{\beta}{(\Delta x)^2} - 4 \frac{\bra{u} \alpha}{(\Delta x)^2} \text{       } - \tau^2     \big) A_x  +  2 \big( \frac{\beta}{(\Delta x)^2} - \tau^2 - \frac{1}{2} \big)  \text{        } \bigg\} \ket{u}   \text{        }    \text{   ,   }  \\
    \end{align*}

    \noindent corresponding to the first term of the expansion,

    \begin{align*}
\text{            }             2 \text{        }         \bigg\{ \big( \frac{\beta}{(\Delta x)^2} - \tau^2 \big) A^{\dagger}_x + \big( \text{         } 2 \frac{\beta}{(\Delta x)^2 } - 4 \frac{\bra{u} \alpha}{(\Delta x)^2} \text{       } - \tau^2 \big) A_x  +  2 \big( \frac{\beta}{(\Delta x)^2} - \tau^2 - \frac{1}{2} \big)  \bigg\} \text{      }                     \ket{u}      
    \text{   }       \frac{\beta}{(\Delta x)^2 }    \\ \times    \bigg\{        \text{            }      \big( A^{\dagger}_x - 2  - \frac{(\Delta x)^2}{\beta} + A_x \big) \ket{u(t+2\tau)}     - \big(   2 A^{\dagger}_x - 4         \big)         \ket{u(t+\tau)} \text{       } \bigg\}     \text{  ,         } \\
    \end{align*}

\noindent corresponding to the second term of the expansion, and the superposition,

\begin{align*}
       \frac{\beta}{(\Delta x)^2} \bigg\{ \text{    } \big( A^{\dagger}_x - 2 - \frac{(\Delta x)^2}{\beta} + A_x  \big) \text{       }      \ket{u(t+2\tau)} \text{      } - \big( 2A^{\dagger}_x - 4 \big) \ket{u(t+\tau)} \text{       }   \bigg\}   \text{      }  \bigg\{            \big( A^{\dagger}_x - 2 \\ - \frac{(\Delta x)^2}{\beta}        + A_x \big) \\ \times \ket{u(t+2\tau)}         - \big( 2 A^{\dagger}_x - 4 \big) \ket{u(t+\tau)}      \bigg\}   \text{         }      \bigg\{ \big( \frac{\beta}{(\Delta x)^2} - \tau^2 \big) A^{\dagger}_x  \ket{u}  + \big( 2 \frac{\beta}{(\Delta x)^2} \text{           } \\ - 4 \frac{\bra{u} \alpha}{(\Delta x)^2} \text{                  }                     - \tau^2 \big) \\ \times  A_x \ket{u}   \text{        }     2 \big( \text{         }    \frac{\beta}{(\Delta x)^2 } - \tau^2 - \frac{1}{2} \text{       } \big) \ket{u}  \text{       } \bigg\}    \bigg\{       \text{          }            \big( \frac{\beta}{(\Delta x)^2} - \tau^2 \big) A^{\dagger}_x  \ket{u}    + 
\big( 2 \frac{\beta}{(\Delta x)^2} \text{           }   - 4 \frac{\bra{u} \alpha}{(\Delta x)^2} \text{                  }                     - \tau^2 \big) \\ \times  A_x \ket{u}    +      \text{        }     2 \big( \text{         }    \frac{\beta}{(\Delta x)^2 } - \tau^2 - \frac{1}{2} \text{       } \big) \ket{u}        \text{          }       \bigg\}     \text{ ,   }      
     \end{align*}

\noindent corresponding to the last term. Term by term, we consolidate each prefactor for all quantum states in the superposition, in which,

\begin{align*}
       \text{                }      \big( \text{   } \frac{\beta}{(\Delta x)^2} - \tau^2 \big) \text{      }              A^{\dagger}_x \text{        }  \ket{u}   \big( \frac{\beta}{(\Delta x)^2} - \tau^2 \big)    A^{\dagger}_x \ket{u} \text{         } 
          +   \big(   \big(  \text{          }     \frac{\beta}{(\Delta x)^2} - \tau^2 \big) A^{\dagger}_x \text{      }  \ket{u}               \text{          }            \big)    \text{          }       \big( 2 \frac{\beta}{(\Delta x)^2}  - 4  \frac{\bra{u} \alpha}{(\Delta x)^2} - \tau^2 \big) \\ \times A_x \ket{u}  
          + 
                  \big( \frac{\beta}{(\Delta x)^2} - \tau^2 \big) A^{\dagger}_x \ket{u} \times 2 \big(    \frac{\beta}{(\Delta x)^2} - \frac{1}{2}      \big)          \text{          }      \ket{u}    + \big(   2 \frac{\beta}{(\Delta x)^2}  - 4 \frac{\bra{u} \alpha}{(\Delta x)^2} - \tau^2               \big) \\ \times  A_x \ket{u} \big( \frac{\beta}{(\Delta x)^2} - \tau^2 \big) A^{\dagger}_x \ket{u}  + 
                   \big( \text{        }   2 \frac{\beta}{(\Delta x)^2} - 4 \frac{\bra{u} \alpha}{(\Delta x)^2} - \tau^2        \big)   A_x \ket{u}    \big( \text{        }   2 \frac{\beta}{(\Delta x)^2} -        4 \frac{\bra{u} \alpha}{(\Delta x)^2} - \tau^2        \big)   A_x \ket{u} \end{align*}

                   \begin{align*} + 
                \big( 2 \frac{\beta}{(\Delta x)^2} - 4 \frac{\bra{u} \alpha}{(\Delta x)^2} - \tau^2 \big)  A_x \ket{u}  2 \big( \frac{\beta}{(\Delta x)^2} - \tau^2 - \frac{1}{2} \big) \text{           } \ket{u}   \text{           }        
          +   2 \big( \frac{\beta}{(\Delta x)^2} - \tau^2 \\ - \frac{1}{2} \big)           \ket{u}  \big( \frac{\beta}{(\Delta x)^2} - \tau^2 \big)  A^{\dagger}_x \ket{u}  + 2 \big( \frac{\beta}{(\Delta x)^2} - \tau^2 - \frac{1}{2} \big) \ket{u} 
               \big( 2 \frac{\beta}{(\Delta x)^2} - 4 \frac{\bra{u} \alpha}{(\Delta x)^2} - \tau^2 \big) A_x \ket{ u} \\ +  \text{             } 
                     2 \big( \frac{\beta}{(\Delta x)^2} - \tau^2 - \frac{1}{2} \big) \ket{u}    2 \big( \frac{\beta}{(\Delta x)^2} - \tau^2 - \frac{1}{2} \big) \ket{u}  \text{      }   \text{, }
                     \end{align*}
                     
           \noindent for the four terms previously obtained. We further rearrangements by grouping together terms depending on whether a scalar prefactor is projected onto a quantum state for the solution $u$,          
                 \begin{align*}
                  \big( \frac{\beta}{(\Delta x)^2} - \tau^2 \text{         } \big)^2 \text{           }  \lambda^{2} \text{    }   + \text{    }  \big( \frac{\beta}{(\Delta x)^2} A^{\dagger}_x \ket{u}  \text{      } - \tau^2 A^{\dagger}_x \ket{u}  \big) \text{      } \big( \text{                 } \frac{2 \beta}{(\Delta x)^2} A_x \ket{u} \text{          }    -  \frac{4 \bra{u} \alpha}{(\Delta x)^2 }  A_x \ket{u} \text{          }     - \tau^2 A_x \ket{u} \text{           } \big)                                            \\ + 
      \big( \frac{\beta}{(\Delta x)^2} A^{\dagger}_x\ket{u}  - \tau^2 A^{\dagger}_x \ket{u} \text{            } \big) \text{      }   \big( \text{            } \frac{ 2 \beta}{(\Delta x)^2} \text{            }      \ket{u} \text{       }  - \ket{u} \text{            }    \big) + 
              \big( 2 \frac{\beta}{(\Delta x)^2} A_x \ket{u} \text{        } - 4 \frac{\bra{u} \alpha}{(\Delta x)^2} A_x \ket{u} \text{                  }    \\  - \tau^2 A_x \ket{u} \text{            }    \big) \text{      } \big( \text{         } \frac{\beta}{(\Delta x)^2} \text{      }  A_x \ket{u}   - \tau^2 A_x \ket{u} \text{                } \big)   +  \lambda^{2} \text{     } \big(    2 \frac{\beta}{(\Delta x)^2} - 4 \frac{\bra{u} \alpha}{(\Delta x)^2} - \tau^2 \big) \\ \times   \text{          } \big(   \text{         }   2 \frac{\beta}{(\Delta x)^2} - 4 \frac{\bra{u} \alpha}{(\Delta x)^2} - \tau^2    \big)  \\ 
    \Updownarrow  \\  
    \text{             }                     \big( \frac{\beta}{(\Delta x)^2} - \tau^2 \big)^2 \lambda^{2} +    \frac{\beta}{(\Delta x)^2} A^{\dagger}_x \ket{u} \frac{2 \beta}{(\Delta x)^2} A_x \ket{u} - \frac{\beta}{(\Delta x)^2} A^{\dagger}_x \ket{u} \frac{4 \bra{u} \alpha}{(\Delta x)^2} A_x \ket{u}          \text{             }     \\
       - \frac{ \beta}{(\Delta x )^2}    A^{\dagger}_x  \ket{u} \text{            } \tau^2 A_x \ket{u} \text{         } - \tau^2 A^{\dagger}_x \ket{u} \text{      } \frac{2 \beta}{(\Delta x )^2} A_x \ket{u} + \tau^2 A^{\dagger}_x \ket{u} \frac{4 \bra{u} \alpha}{(\Delta x)^2} A_x \ket{u} + 2  \tau^4 \text{  } \bra{u} A^{\dagger}_x A^{\dagger}_x \ket{u}  \text{       }   \\
                  +    \text{            }             2 \text{        }         \bigg\{ \big( \frac{\beta}{(\Delta x)^2} - \tau^2 \big) A^{\dagger}_x + \big( \text{         } 2 \frac{\beta}{(\Delta x)^2 } - 4 \frac{\bra{u} \alpha}{(\Delta x)^2} \text{       } - \tau^2 \big) A_x + 2 \big( \frac{\beta}{(\Delta x)^2} - \tau^2 - \frac{1}{2} \big)  \bigg\} \text{      }                     \ket{u}      \text{            }     \\ \times 
    \text{            }       \frac{\beta}{(\Delta x)^2 }       \bigg\{             \big( A^{\dagger}_x - 2 - \frac{(\Delta x)^2}{\beta} + A_x \big) \ket{u(t+2\tau)}     - \big(   2 A^{\dagger}_x - 4         \big)         \ket{u(t+\tau)} \bigg\} \text{ . }              \end{align*}   
        
   \noindent Grouping together expectation terms into several quantum superposition takes the form,      
        \begin{align*}
        \text{              }       2\bigg\{                      \frac{\beta}{(\Delta x)^2} A^{\dagger}_x \ket{u} - \tau^2 A^{\dagger}_x \ket{u} \text{  } + 2 \frac{\beta}{(\Delta x)^2} A_x \ket{u} - 4 \frac{\bra{u} \alpha}{(\Delta x)^2} A_x \ket{u} - \tau^2 A_x \ket{u}  +  2 \frac{\beta}{(\Delta x)^2} \ket{u} \\ - \tau^2 \ket{u}  - \ket{u}    \bigg\} 
     \bigg\{   \frac{\beta}{(\Delta x)^2} \text{        }  A^{\dagger}_x \ket{u(t+2\tau)} -  \text{      } \frac{2\beta}{(\Delta x)^2} \ket{u(t+2\tau)}  -  
     \ket{u(t+2\tau)} \\ + \frac{\beta}{(\Delta x)^2} A_x \ket{u(t+2\tau)}   - \frac{2 \beta}{(\Delta x)^2} \ket{u(t+\tau)} \text{         }  +  \frac{4 \beta}{(\Delta x)^2} \text{      } \ket{u(t+\tau)} \text{      }    \bigg\}  
     \\ \times \frac{\beta}{(\Delta x)^2} \text{       } \bigg\{ \text{            } \big( A^{\dagger}_x - 2 - \frac{(\Delta x)^2}{\beta }   + A_x \big) \text{                } \ket{u(t+ 2\tau)} \text{         } - \big( 2 A^{\dagger}_x - 4 \big) \ket{u( t+ \tau)} \bigg\} \\ \times 
 \frac{\beta}{(\Delta x)^2} \bigg\{ A^{\dagger}_x \ket{u(t+2 \tau)}  - 2      \ket{u(t+ 2 \tau)} \text{       }  +  A_x \ket{u(t + 2 \tau)} \text{            }      - \ket{u(t+ 2 \tau)} \text{         }   - 2 A^{\dagger}_x \ket{u(t+\tau)} \text{            }   \\ + 4 \ket{u(t+\tau)} \bigg\}
         \text{ . }  
             \end{align*}

   \noindent   For fixed $\beta$, given the spatial increment associated with the second partial derivative with respect to space, the superposition takes the form,        
             
             \begin{align*}
             \big( \frac{\beta}{(\Delta x)^2}          \big)^2 \text{           }       \bigg\{ \text{      }                  \big( A^{\dagger}_x - 2 - \frac{(\Delta x)^2}{\beta} +A_x \big) \text{          }     \ket{u(t+ 2 \tau)}        \big( A^{\dagger}_x \text{        } - 2 - \frac{(\Delta x)^2}{\beta} \text{            }   
             + A_x \big) \\ \times  \ket{u(t+2 \tau)}   - 2  
           \big( A^{\dagger}_x - 2  - \frac{(\Delta x)^2}{\beta}   + A_x \big) \ket{u(t+ 2 \tau)} \times \big( 2 A^{\dagger}_x - 4 \big)  \\ \times \ket{u(t+\tau)} \text{                 }    + 
              \big( 2 A^{\dagger}_x - 4 \big) \ket{u(t+\tau)}  \big( 2 A^{\dagger}_x - 4 \big) \ket{u(t+\tau)}   \bigg\}  \text{, }  
              \end{align*}
              
        \noindent which is equivalent to,

              \begin{align*}
              \big( \frac{\beta}{(\Delta x)^2} \big)^2  \bigg\{ \big( A^{\dagger}_x - 2 - \frac{(\Delta x)^2}{\beta} \big) \ket{u(t+ 2 \tau)} \big( A^{\dagger}_x - 2 - \frac{\beta}{(\Delta x)^2} + A_x \big) \ket{u(t + 2 \tau)} + A_x \ket{u(t+2 \tau)} \\ \times  \big( A^{\dagger}_x - 2 - \frac{(\Delta x)^2}{\beta} + A_x  \big) \ket{u(t+2 \tau)}  -   A^{\dagger}_x \ket{u(t+\tau)} \big( \text{           } 2 A^{\dagger}_x - 4 \big) \ket{u(t+ \tau)} \\ - \big( 2 + \frac{\beta}{(\Delta x)^2} \big) \ket{u(t+ 2 \tau)} \big( 2 A^{\dagger}_x - 4 \big) \ket{u(t+\tau)} + A_x \ket{u(t+ 2 \tau)} \big( 2 A^{\dagger}_x - 4\big) \ket{u(t+\tau)} \\ +  \big( \text{     } 2 A^{\dagger}_x \ket{u(t+\tau)} - 4 \ket{u(t+\tau)} \text{               } \big) \text{               } \big( 2 A^{\dagger}_x \ket{u(t+\tau)} - 4 \ket{u(t+\tau)} \big) \bigg\} \text{  .  } 
              \end{align*}
              
      \noindent From the superposition state, expectation terms of time evolved variational states can be multiplied onto other time evolved variational states, as,

            \begin{align*}      
              \text{        }    \big( \frac{\beta}{(\Delta x)^2} \big)^2     \text{        }       \bigg\{     \text{      }        \big(      A^{\dagger}_x \ket{u(t+2\tau)} - 2 \ket{u(t+2\tau)} - \frac{(\Delta x)^2}{\beta} \ket{u(t+2\tau)}  \text{     }   \big) \\ \times 
           \big(   \text{       }     A^{\dagger}_x \ket{u(t+2\tau)} - 2 \ket{u(t+2\tau)} - \frac{\beta}{(\Delta x)^2} \ket{u(t+2\tau)}     \text{          }           + A_x \ket{u(t+2\tau)}    \text{       }  \big) \\ + 
              \big(   \text{       }      A_x \ket{u(t+2\tau)} A^{\dagger}_x \ket{u(t+2\tau)} - 2 A_x \ket{u(t+2\tau)} \text{      }      - A_x \ket{u(t+2\tau)} \frac{(\Delta x)^2}{\beta}    \ket{u(t+2\tau)}             \big)  \\ -  
              A^{\dagger}_x \ket{u(t+\tau)} \big( 2 A^{\dagger}_x \ket{u(t+\tau)} - 4 \ket{u(t+\tau)} \text{         }       \big) -          \big( 2 + \frac{\beta}{(\Delta x)^2} \text{      } \big) \ket{u(t+2\tau)} \big( 2 A^{\dagger}_x \ket{u(t+\tau)} \text{             }     \\   -  4 \ket{u(t+\tau)} \big)   +  A_x \ket{u(t+2\tau)} 2 A^{\dagger}_x \ket{u(t+\tau)} \text{       }       - A_x \ket{u(t+2\tau)} 4 \ket{u(t+\tau)}\\ +   4  | \lambda |^2 \text{        }   \text{             }           - 2 A^{\dagger}_x \ket{u(t+\tau)} 4 \ket{u(t+\tau)} \text{      }          -  4 \ket{u(t+\tau)} 2 A^{\dagger}_x \ket{u(t+\tau)} \text{      }     - 16 \lambda^2  \bigg\}   \text{ . }   
                          \end{align*}

     \noindent The resultant superposition is,         
                     \begin{align*}
                 \text{            }        \big( \frac{\beta}{(\Delta x)^2} \big)^2  \text{           }          \bigg\{             \text{              }    \big(    \text{              }      A^{\dagger}_x \ket{u(t+2\tau)} - 2 \ket{u(t+2\tau)} - \frac{(\Delta x)^2}{\beta} \ket{u(t+2\tau)} \text{            }    \big)  \\ \times \big(      A^{\dagger}_x \ket{u(t+2\tau)}  - 2 \ket{u(t+2\tau)}  - 
                     \frac{\beta}{(\Delta x)^2} \ket{u(t+2\tau)} \text{           }                 \big)  + 
                         \text{          } \big(        A^{\dagger}_x \ket{u(t+2\tau)} \\ - 2 \ket{u(t+2\tau)} - \frac{(\Delta x)^2}{\beta} \ket{u(t+2\tau)} \text{       } \big)        \text{            }     A_x \ket{u(t+2\tau)} \text{           }  \\ +    \big(        \text{                }          A_x \ket{u(t+2\tau)} A^{\dagger}_x - 2 A_x - A_x \ket{u(t+2\tau)} \frac{(\Delta x)^2}{\beta} \big)   \text{                }   \ket{u(t+2\tau)}           \text{                }   \\  - \big(            \text{          }                           A^{\dagger}_x \ket{u(t+\tau)} 2 A^{\dagger}_x - 4 - \big( 2 + \frac{\beta}{(\Delta x)^2  } \text{            }  + 
           4   A_x    \big) \ket{u(t+2\tau)} \text{        } 2 A^{\dagger}_x \text{        }         \\  + \big( 2 + \frac{\beta}{(\Delta x)^2 \text{       } }   \big) \text{           }   4 \ket{u(t+2\tau)}   \text{             }  - 
              \text{             }   A_x \ket{u(t+2\tau)} 2 A^{\dagger}_x \text{              }       \text{             }      \\  - 8 A^{\dagger}_x \ket{u(t+\tau)} \text{          }      \text{             }  \big)    \ket{u(t+\tau)} \text{            }            - \text{            }      12 \lambda^2      \text{                }       \bigg\}        \text{. }
                \end{align*}
        
   \noindent The desired superposition now takes the form,

              \begin{align*}
                     \text{                }        \big( \text{       } \frac{\beta}{(\Delta x)^2} \text{       } \big)^2 \bigg\{            \text{  } \big( A^{\dagger}_x \ket{u(t+2\tau)} - 2 \ket{u(t+2\tau)} \text{         }       \big)            \text{      } \big( A^{\dagger}_x \ket{u(t+2\tau)} - 2 \ket{u(t+2\tau)} \\ -  \frac{\beta}{(\Delta x)^2 } \ket{u(t+2\tau)} \text{          }      \big)  -    \frac{(\Delta x)^2}{\beta} \ket{u(t+2\tau)} \text{             }                \big( A^{\dagger}_x \ket{u(t+2\tau)} - 2 \ket{u(t+2\tau)} \\ - \frac{\beta}{(\Delta x)^2} \ket{u(t+2\tau)} \text{             }           \big)  \text{      }  
                     + \big( A^{\dagger}_x \ket{u(t+2\tau)} A_x  -  2 \ket{u(t+2\tau)} A_x \\ -  \frac{(\Delta x)^2}{\beta} \ket{u(t+2\tau)}  A_x +  A_x \ket{u(t+2\tau)} A^{\dagger}_x \text{      } \\ - 2 A_x  -    A_x \ket{u(t+2\tau)} \frac{(\Delta x)^2}{\beta} \big)     \text{    } \ket{u(t+2\tau)}     -   12  \lambda^2                 \bigg\} \text{, }  
              \end{align*}  
              
       \noindent leading to the state,       
                        
          \begin{align*}        \big( \frac{\beta}{(\Delta x)^2} \big)^2 \bigg\{      \lambda^2 \text{            }       -   A^{\dagger}_x \ket{u(t+2\tau} \text{          }  2 \ket{u(t+2\tau)}  -  A^{\dagger}_x \ket{u(t+2\tau)} \frac{\beta}{(\Delta x)^2} \ket{u(t+2\tau)} \\  - 2 \ket{u(t+2\tau)}       A^{\dagger}_x \ket{u(t+2\tau)} \text{             }             +         
                        4  \lambda^2 \text{           }                     + 2 \ket{u(t+2\tau)} \frac{\beta}{(\Delta x)^2} \ket{u(t+2\tau)} \text{              }  \\ -          \frac{(\Delta x)^2}{\beta} \ket{u(t+2\tau)} A^{\dagger}_x \ket{u(t+2\tau)}  - 2 \frac{(\Delta x)^2}{\beta}  \lambda^2 \text{           }  -     11   \lambda^2           \bigg\}   \text{. }
               \end{align*}

      \noindent The superposition above is equivalent to,                  
                \begin{align*}        \text{      }  \big( \text{        }    \frac{\beta}{(\Delta x)^2} \text{        }   \big)^2  \bigg\{        \text{          }    - \big( 6 + 2 \frac{(\Delta x)^2}{\beta}  \text{          } \big)  \lambda^{2}    + \big( A^{\dagger}_x \ket{u(t+2\tau)} \text{         }     A_x     - 2 \ket{u(t+2\tau)} A_x \\ - \frac{(\Delta x)^2}{\beta} \ket{u(t+2\tau)} A_x  +  A_x \ket{u(t+2\tau)} A^{\dagger}_x  - 
                        2 A_x - A_x \ket{u(t + 2\tau)} \frac{(\Delta x)^2}{\beta} \text{       }      - \text{      } 2 A^{\dagger}_x       \ket{u(t+2\tau)} \\  - A^{\dagger}_x \frac{\beta}{(\Delta x)^2 \text{      } } \ket{u(t+2\tau)}     \text{      }          - 
2 \ket{u(t+2\tau)} \text{        }   A^{\dagger}_x + 2 \frac{\beta}{(\Delta x)^2} \text{      } \ket{u(t+2\tau)} \\   -  \text{          }      \frac{(\Delta x)^2}{\beta} \ket{u(t+2\tau)} A^{\dagger}_x  \text{          }                   \text{          }         \big) \ket{u(t+2\tau)}   -  \big( A^{\dagger}_x \ket{u(t+\tau)} 2 A^{\dagger}_x - 4 - \big( 2 + \frac{\beta}{(\Delta x)^2} \\ + 4 A_x \big)       \ket{u(t+2\tau)} \text{             }       2 A^{\dagger}_x  +  \big( 2 + \frac{\beta}{(\Delta x)^2} \text{         } \big) 4 \ket{u(t+2\tau)} \text{       }    \\   - A_x \ket{u(t+2\tau)} 2 A^{\dagger}_x  -      8 A^{\dagger}_x \ket{u(t+\tau)}  \text{            }   \big)  \text{            }   \ket{u(t+\tau)}    \text{            }    \bigg\}  \text{. }
                        \end{align*}
          \noindent Rearrangements of the last term of the expansion imply that,              
                        \begin{align*}
   \frac{\beta}{(\Delta x)^2} \bigg\{\big( A^{\dagger}_x - 2 - \frac{(\Delta x)^2}{\beta} + A_x  \big) \text{       }      \ket{u(t+2\tau)} \text{      }   \big( 2A^{\dagger}_x - 4 \big) \ket{u(t+\tau)} \text{       }   \bigg\} \bigg\{            \big( A^{\dagger}_x  - 2 - \frac{(\Delta x)^2}{\beta}        + A_x \big) \\ \times \ket{u(t+2\tau)}       - \big( 2 A^{\dagger}_x - 4 \big) \ket{u(t+\tau)}      \bigg\}   \text{, }   \end{align*}
   
   \noindent can be expressed as,

   \begin{align*} \frac{\beta}{(\Delta x)^2} \text{       }           \bigg\{ \text{       }       A^{\dagger}_x \ket{u(t+2\tau)} - 2 \ket{u(t+2\tau)}  - \frac{(\Delta x)^2}{\beta} \ket{u(t+2\tau)} +  A_x \ket{u(t+2\tau)} \text{      }     \bigg\}   \bigg\{       \text{          }     A^{\dagger}_x \ket{u(t+2\tau)} \\ - 2 \ket{u(t+2\tau)} - \frac{(\Delta x)^2}{\beta} \ket{u(t+2\tau)} + A_x \ket{u(t+2\tau)}  -  2 A^{\dagger}_x \ket{u(t+\tau)} \text{         } + 4 \ket{u(t+\tau)}        \bigg\} \text{,     }  \end{align*}

\noindent which can be further rearranged as,

   \begin{align*}
     \frac{\beta}{(\Delta x)^2}   \bigg\{       \big( A^{\dagger}_x \ket{u(t+2\tau)}   2 \ket{u(t+2\tau)} \text{         }      - \frac{(\Delta x)^2}{\beta} \text{      }       \ket{u(t+2\tau)} \text{       }              \big)  \bigg\{    \text{         }        A^{\dagger}_x \ket{u(t+2\tau)} \text{            } -  2 \ket{u(t+\tau)}         \\                            - \frac{(\Delta x)^2}{\beta }      \text{          } \ket{u(t+2 \tau)} \text{               } + A_x \ket{u(t+2 \tau)} \text{                   } - 2 A^{\dagger}_x \ket{u(t+ 2 \tau)} \text{            }               + 4 \ket{u(t+\tau)}     \bigg\}  \\ +    A_x \ket{u(t+2\tau)} \text{       }           \bigg\{ \text{       } A^{\dagger}_x \ket{u(t+2\tau)} - 2 \ket{u(t+2\tau)} - \frac{(\Delta x)^2}{\beta} \ket{u(t+2\tau} \\ +   A_x \ket{u(t+2\tau)} \text{      }          - 2 A^{\dagger}_x \ket{u(t+\tau)} + 4 \ket{u(t+\tau)}     \bigg\} \bigg\}\text{. } 
              \end{align*}
              
              \noindent Further distributing terms,
              
              \begin{align*}
              \frac{\beta}{(\Delta x)^2} \text{       }   \bigg\{  \big( A^{\dagger}_x \ket{u(t+2\tau)} - 2 \ket{u(t+2\tau)} \text{   }   \text{       }           \big) \bigg\{ \text{            }   A^{\dagger}_x \ket{u(t+2\tau)} - 2 \ket{u(t+\tau)} \text{       } \\   - \frac{(\Delta x)^2}{\beta} \ket{u(t+2\tau)} \text{      }        A_x \ket{u(t+2\tau)} \text{          }           - 2 A^{\dagger}_x \ket{u(t+2\tau)} +  4 \ket{u(t+\tau)} \text{       }         \bigg\} \\  -                  \frac{(\Delta x)^2}{\beta} \ket{u(t+2\tau)} \text{          }           \bigg\{ A^{\dagger}_x \ket{u(t+2\tau)} - 2 \ket{u(t+\tau)}   - \frac{(\Delta x)^2}{\beta} \text{      }       \ket{u(t+2\tau)}    + A_x \ket{u(t+2\tau)} \end{align*}
              
              \begin{align*} - 2 A^{\dagger}_x \ket{u(t+2\tau)} + 4 \ket{u(t+\tau)}    \text{          }        \bigg\}  + 
             A_x \ket{u(t+2\tau)} A^{\dagger}_x \ket{u(t+2\tau)} - A_x \ket{u(t+2\tau)} \\ \times 2 \ket{u(t+2\tau)} -   A_x \ket{u(t+2\tau)}    \frac{(\Delta x)^2}{\beta} \ket{u(t+2\tau)} \text{           }           + \lambda^2  -  A_x \ket{u(t+2\tau)} \\ \times  2 A^{\dagger}_x \ket{u(t+\tau)}  +  A_x \ket{u(t+ 2 \tau)} \text{        }     4 \ket{u(t+\tau)}    \bigg\} \text{.}
             \end{align*}

     \noindent We group together several product states of variational states for the solution time evolved forwards one or two time steps,

            \begin{align*} \frac{\beta}{(\Delta x)^2} \text{                    }    \bigg\{  \lambda^{2} -  A^{\dagger}_x \ket{u(t+ 2 \tau)}  2 \ket{u(t+\tau)} \text{                   }   - A^{\dagger}_x \ket{u(t+2 \tau)}     \frac{(\Delta x)^2}{\beta}  \text{         }    \ket{u(t+2\tau)}  +  A^{\dagger}_x \\ \times  \ket{u(t+2 \tau)} A_x \ket{u(t+2\tau)}  - A^{\dagger}_x \ket{u(t+ 2\tau)}  2 A^{\dagger}_x \ket{u(t+2 \tau)}  + A^{\dagger}_x \ket{u(t+2\tau)} 4 \ket{u(t+\tau)}  \\ - 2 \ket{u(t+2\tau)} ^{\dagger}_x \ket{u(t+2\tau)}  + 4 \lambda^2  + 
                   2 \ket{u(t+\tau)} \text{       }      \frac{(\Delta x)^2}{\beta} \ket{u(t+2\tau)}  - 2 \ket{u(t + 2 \tau)} A_x \ket{u(t+ 2 \tau)} \\ + 2 \ket{u(t + 2 \tau)} 2 A^{\dagger}_x \ket{u(t + 2 \tau)} - 8 \lambda^2 \text{       }   -    \frac{(\Delta x)^2}{\beta}   \ket{u(t+2\tau)} A^{\dagger}_x \ket{u(t+2\tau)}   \end{align*}

                   \begin{align*} - \frac{(\Delta x)^2}{\beta}  \ket{u(t+2\tau)} \text{           }       2 \ket{u(t+\tau)}         - \big( \frac{(\Delta x)^2}{\beta}   \big)^2 \text{         }\lambda^2   + 
                   \frac{(\Delta x)^2}{\beta}     \ket{u(t+2\tau)} A_x \ket{u(t+2\tau)} \\  - \frac{(\Delta x)^2}{\beta }   \ket{u(t+2\tau)}      2 A^{\dagger}_x \ket{u(t+2\tau)}  - \frac{(\Delta x)^2}{\beta}   \ket{u(t + 2\tau)} 4  \ket{u(t+\tau)}          \bigg\}  \text{, } \end{align*}

                   \noindent which is equivalent to,

                   \begin{align*}
                   \frac{\beta}{(\Delta x)^2} \text{       }      \bigg\{ \lambda^2 -        \text{      }  \big(     \text{        }   A^{\dagger}_x \ket{u(t+2\tau)} 2 - A^{\dagger}_x \ket{u(t+2\tau)} 4 + \frac{(\Delta x)^2}{\beta} \ket{u(t+2\tau)} 2  \text{           }      \big) \text{         } \ket{u(t+\tau)}    \\ -    \big(      A^{\dagger}_x \ket{u(t+2\tau)}  \frac{(\Delta x)^2}{\beta} - \text{        } A^{\dagger}_x \ket{u(t+2\tau)}   \text{        }       - A^{\dagger}_x \ket{u(t+2\tau)} 2 A^{\dagger}_x \text{      }            2 \ket{u(t+2\tau)} \text{           }   A^{\dagger}_x \\  +            2 \ket{u(t+\tau)} \frac{(\Delta x)^2}{\beta}  - 2 \ket{u(t+2\tau)} A_x \text{      }        - 2 \ket{u(t+2\tau)} 2 A^{\dagger}_x \text{          }           + \frac{(\Delta x)^2}{\beta} \ket{u(t+2\tau)} A^{\dagger}_x \text{                 }    \\   -           \frac{(\Delta x)^2}{\beta} \ket{u(t+2\tau)} A_x  + \frac{(\Delta x)^2}{\beta}   \ket{u(t+2\tau)} \text{      }      2 A^{\dagger}_x \text{       }                         \big) \text{           } \ket{u(t+2\tau)}   \bigg\}   \text{.}
                   \end{align*}


\subsection{Lin-Tsien equation}

\subsubsection{Statement}

    \begin{align*}
                  \underline{   \mathcal{C}^{\mathrm{L-T}}(\lambda,\lambda^{\mathrm{L-T}}_0) }  =  \big( \lambda^{\mathrm{L-T}}_0\big)^{2} +         \mathrm{Re }   \Bigg\{       \big(   \frac{\tau}{(\Delta x)^2} \big)^2   \text{    }  \bra{u}   A^{\dagger}_x A^{\dagger}_x       \ket{\widetilde{u}}    \text{    }    - 2 \big(   \text{ }\frac{\tau}{(\Delta x)^2} \big)^2 \bra{u}  A^{\dagger}_x A^{\dagger}_x A^{\dagger}_x       \ket{u}  \end{align*}

                  \begin{align*} -     2  \big( \frac{\tau}{(\Delta x)^2} \big) ^2    \bra{\widetilde{u}} A_x \ket{u}         +               6   \lambda^2   \bra{u}    A_x A_x   A_x   \ket{u}  - 2  \lambda^2 
   \bra{u}A_x A_x A^{\dagger}_x  \ket{u}    
  \\ - 4 \lambda^2 \bra{u} A_x A^{\dagger}_x \ket{u}  + \bra{u} A_x A_x   \ket{u}  \text{           }   + \bra{u} A_x A^{\dagger}_x \ket{u} +  \bra{u} A^{\dagger}_x A_x   \ket{u}    \\ +  \bra{u} A^{\dagger}_x A^{\dagger}_x \ket{u}    -     A^{\dagger}_x  \ket{u}  +   \frac{ 1}{(\Delta x)^2} A^{\dagger}_y A_x   \ket{u}     +   \frac{1}{(\Delta y)^2} A^{\dagger}_y A^{\dagger}_x \ket{u}  + \frac{ 1}{(\Delta y)^4} \bra{u} \ket{u}       \\ +     \frac{2}{(\Delta y)^4}   A^{\dagger}_x    \ket{u} +  3  \bra{u} A_x  \ket{u}  + 3 \bra{u} A^{\dagger}_x \ket{u}     \text{    }  + \frac{1}{(\Delta y)^2} A^{\dagger}_y  \ket{u}    + 2       \ket{u}  \text{  } +    \frac{1}{(\Delta y)^2} A^{\dagger}_x   \ket{u}   \\ +         3   \bra{u}   A^{\dagger}_x  \ket{u}   +    \frac{1}{(\Delta y)^2} A^{\dagger}_y \ket{u}  -   \frac{2}{(\Delta y)^2}  \ket{u}   +    \frac{1}{(\Delta y)^2} A_y    \ket{u}  +   \frac{1}{(\Delta y)^2} A^{\dagger}_x A^{\dagger}_y   \ket{u}  \\ +  2         \frac{1}{(\Delta y)^2}    A^{\dagger}_x  \ket{u}       +  \frac{1}{(\Delta y)^2}  A^{\dagger}_x A_y \ket{u}  +  \frac{1}{(\Delta y)^2}  A_x A^{\dagger}_y   \ket{u}  +     2  \frac{1}{(\Delta y)^2} A_x     \ket{u}  \\ +  \frac{1}{(\Delta x)^2} \bra{u} A^{\dagger}_x A_x \ket{u} A^{\dagger}_x  \ket{u}   -     \frac{ 2 \tau }{(\Delta x)^2} \bra{u} A^{\dagger}_x A_x \ket{u}      + \frac{\tau}{(\Delta x)^2 (\Delta y)^2} \bra{u} A_x A_x \ket{u}     \\ + 
             \text{  } \bra{u}  A_x \ket{u}     +   \text{  }  2 \frac{\tau}{(\Delta x)^2} +   \frac{\tau}{(\Delta x)^2} \bra{\widetilde{u}}A^{\dagger}_x A_x \ket{\widetilde{u}}    
          \Bigg\}     \text{. }       \end{align*}

\subsubsection{Derivation}

\noindent The equation is,

\begin{align*}
        2 u_{tx} + u_x u_{xx} - u_{yy} = 0       \\
               \text{              } \Updownarrow      \text{                       }        \\
            \frac{2}{\tau \Delta x} \big(    A_x - \textbf{1}     \big)   \ket{u(x,y,t+\tau)} + \big(    \frac{A_x - \textbf{1}}{\Delta x}        \big)    \ket{u(x,y,t)}  \big(    
            \frac{A^{\dagger}_x - 2\textbf{1}  + A_x}{(\Delta x)^2 }  \big) \ket{u(x,y,t)}\\ -    \Bigg\{ \big( \frac{A^{\dagger}_y - 2 \textbf{1} + A_y}{(\Delta y)^2} \big)  \ket{u(x,y,t)}   \Bigg\} = 0   \text{. }
            \end{align*}
      
      \noindent After representing all spatial and time derivatives with the corresponding quantum operators, the PDE takes the form,

            \begin{align*}
              \frac{2}{\tau \Delta x} \big( A_x \ket{u(x,y,t+\tau) } -     \ket{u(x,y,t+\tau) } \big) + \frac{1}{(\Delta x)^3} \big( A_x \ket{u(x,y,t)} -     \ket{u(x,y,t)}  \big)   \\ \times  \big(     A^{\dagger}_x \ket{u(x,y,t)} - 2 \ket{u(x,y,t)}    + A_x  \ket{u(x,y,t)}        \big)  -  \frac{1}{(\Delta y)^2} \big(   A^{\dagger}_y \ket{u(x,y,t)}  - 2 \ket{u(x,y,t)} \\ + A_y \ket{u(x,y,t)}      \big) = 0  \text{, }\\
              \end{align*}
              
\noindent from which multiplying terms scaled in the increment of spatial derivatives produces the following sequence of rearrangements,

\begin{align*}
\frac{1}{(\Delta x)^3} \big( A_x \ket{u(x,y,t)} -     \ket{u(x,y,t)}  \big)    \big(     A^{\dagger}_x \ket{u(x,y,t)} - 2 \ket{u(x,y,t)}    + A_x  \ket{u(x,y,t)}        \big)    \\ 
 \Updownarrow \\ 
 \frac{1}{(\Delta x)^3} \big( A_x \ket{u(x,y,t)}   A^{\dagger}_x \ket{u(x,y,t)} - A_x \ket{u(x,y,t)} 2 \ket{u(x,y,t)}  + \textbf{1} - \ket{u(x,y,t)}  \\ \times A^{\dagger}_x \ket{u(x,y,t)} + 2 \\ - \ket{u(x,y,t)} A_x \ket{u(x,y,t)} \big)  + 
\frac{\tau \Delta x}{2} \Bigg\{  \frac{2}{\tau \Delta x} A_x \ket{u(x,y,t+\tau)}  + 
\frac{1}{(\Delta x)^3} \\ \times \big( A_x \ket{u(x,y,t)} A^{\dagger}_x \ket{u(x,y,t)} \\ - A_x \ket{u(x,y,t)} 2 \ket{u(x,y,t)}  + \textbf{1}  - \ket{u(x,y,t)}  A^{\dagger}_x \ket{u(x,y,t)} \\ + 2 - \ket{u(x,y,t)} A_x \ket{u(x,y,t)} \big)        \\ -  \frac{1}{(\Delta y)^2} \big(   A^{\dagger}_y \ket{u(x,y,t)} - 2 \ket{u(x,y,t)}  +  A_y \ket{u(x,y,t)}      \big) \Bigg\}  \text{. }        
\end{align*}

\noindent The resulting time evolved variational state isolated above yields the cost function,

    \begin{align*} \bigg|\bigg|     A_x \ket{u(x,y,t+\tau)}   + \frac{\tau}{(\Delta x)^2} \big( A_x \ket{u(x,y,t)} \text{  }\Big(  A^{\dagger}_x   - 2  \Big)  \ket{u(x,y,t)}  +  \textbf{1} - \ket{u(x,y,t)} \Big(   A^{\dagger}_x \ket{u(x,y,t)}\\  -   A_x  \ket{u(x,y,t)}  \Big)  + 2  \text{     }  \big)  - 
    \frac{\tau \Delta x}{2  (\Delta y)^2} \big( A^{\dagger}_y \ket{u(x,y,t)} - 2 \ket{u(x,y,t)} + A_y \ket{u(x,y,t)} \big)   \bigg| \bigg|^2    \text{. }
    \end{align*}

    \noindent We collect terms from squaring the norm below,


    \begin{align*} 
     \big(  \frac{\tau}{(\Delta x)^2} \big)^2 \bigg\{   \big(    A_x \ket{u(x,y,t)} \Big(   A^{\dagger}_x   -              2  \Big)  \ket{u(x,y,t)} +  \textbf{1} - \ket{u(x,y,t)} A^{\dagger}_x  \ket{u(x,y,t)}  - \ket{u(x,y,t)} \\ \times A_x \ket{u(x,y,t)} + 2   - \frac{1}{(\Delta y)^2} \big( A^{\dagger}_y \ket{u(x,y,t)} - 2 \ket{u(x,y,t)}  + A_y \ket{u(x,y,t)} \big)  \bigg\}^2        \big( \frac{\tau}{(\Delta x)^2}  \big)^2 \text{         }\\ \times  \Bigg\{   \text{         }   \bigg[   A_x \ket{u(x,y,t)} \Big( A^{\dagger}_x   - 2  \Big)  \ket{u(x,y,t)}     \text{          } \bigg] ^2   + 2   \Bigg\{    A_x \ket{u(x,y,t)}  \Big(   A^{\dagger}_x   - 2 \Big) \ket{u(x,y,t)} \Bigg\} \big(    \textbf{1} \\ - \ket{u(x,y,t)} A^{\dagger}_x \ket{u(x,y,t)} - \ket{u(x,y,t)} A_x \ket{u(x,y,t)} +    2  \bigg\}   +  \text{          }
                     \big(     \textbf{1} - \ket{u(x,y,t)} \text{            }        A^{\dagger}_x \ket{u(x,y,t)} \\ - \ket{u(x,y,t)} A_x \ket{u(x,y,t)} +   2 - \frac{1}{(\Delta y)^2} \big( A^{\dagger}_y \ket{u(x,y,t)}  \text{         }  - 
                        2 \ket{u(x,y,t)}  + A_y \ket{u(x,y,t)} \big)          \bigg)^2   \text{. } 
 \end{align*}

  \noindent To obtain the desired cost function, we determine all terms in the quantum superposition resulting from the expansion of terms above. Rearrangements imply,
  
  \begin{align*}
            3 - \ket{u(x,y,t)} A^{\dagger}_x \ket{u(x,y,t)}    - \ket{u(x,y,t)} A_x \ket{u(x,y,t)} - \frac{A^{\dagger}_y}{(\Delta y)^2} \text{         }   \ket{u(x,y,t)} +  \frac{2}{(\Delta y)^2} \text{        }         \ket{u(x,y,t)} \\ - \frac{A_y}{(\Delta y)^2} \text{          }  \ket{u(x,y,t)}    
            \\ \Updownarrow  \\  3 - \bigg[   \ket{u(x,y,t)} A^{\dagger}_x - \ket{u(x,y,t)} A_x - \frac{1}{(\Delta y)^2} \big(    A^{\dagger}_y - 2 \textbf{1} + A_y          \big)      \bigg]   \ket{u(x,y,t)} \text{.   }  
\end{align*}

\noindent Squaring the last term of the expansion for the cost function above gives,

\begin{align*}
                      \Big(  3 - \big(  \ket{u(x,y,t)} A^{\dagger}_x - \ket{u(x,y,t)} A_x - \frac{1}{(\Delta y)^2} \big(    A^{\dagger}_y - 2 \textbf{1} + A_y      \big) \text{         }          \big)    \text{          }   \ket{u(x,y,t)}  \Big)^2  \text{      }                                   
                     \\ \Updownarrow  \\ \Big(  3 - \big(   \ket{u(x,y,t)} A^{\dagger}_x - \ket{u(x,y,t)} A_x - \frac{1}{(\Delta y)^2} \big(     A^{\dagger}_y - 2 \textbf{1} + A_y      \big) \text{         }          \big)    \text{          }   \ket{u(x,y,t)} \Big) 
                \Big(  3 - \bra{u(x,y,t)} \\ \times \big(   \ket{u(x,y,t)} A^{\dagger}_x - \ket{u(x,y,t)} A_x - \frac{1}{(\Delta y)^2}  \big(      A^{\dagger}_y - 2 \textbf{1} + A_y     \big) \text{         }          \big)^{\dagger}    \text{          }    \Big)   \text{, }   
                \end{align*}
                
  \noindent   which is further rearranged as,

    \begin{align*}
                  \text{              }    9 -      \bigg[  3  \bigg[ \bra{u(x,y,t)} \bigg[  \ket{u(x,y,t)} A^{\dagger}_x - \ket{u(x,y,t)} A_z - \frac{1}{(\Delta y)^2}   \big(  A^{\dagger}_y - 2 \textbf{1} + A_y  \big)  \bigg] ^{\dagger} \text{  } \bigg]   \bigg]  \\  -        3   \bigg[  \ket{u(x,y,t)}    \text{         } A^{\dagger}_x  - \ket{u(x,y,t)}     A_x \text{   }  - \frac{1}{(\Delta y)^2}       \big(  A^{\dagger}_y  - 2 \textbf{1} \text{         } + A_y \text{      } \big)  \bigg]   \text{    } \ket{u(x,y,t)}  \\ + 
                 \bigg[  \ket{u(x,y,t)} A^{\dagger}_x - \ket{u(x,y,t)}        A_x   - \frac{1}{(\Delta y)^2} \big(  A^{\dagger}_y - 2 \textbf{1} \text{      }   + A_y \big) \text{   } \bigg]  \text{             } \ket{u(x,y,t)} \\ \times \text{             } 
      \bra{u(x,y,t)}  \bigg[  \ket{u(x,y,t)}     A^{\dagger}_x   - \ket{u(x,y,t)} A_x - \frac{1}{(\Delta y)^2} \big(  A^{\dagger}_y  - 2 \textbf{1} \text{     } + \text{          } A_y   \text{                      }      \big)      \bigg] ^{\dagger} \text{. } 
      \end{align*}
      
      \noindent In order from the superposition above, we distribute terms amongst the superposition, and take the Hermitian conjugate of the last term,
      
      \begin{align*}
       9 - \bigg[  3    \bigg[      \text{          }     \bra{u(x,y,t)} A^{\dagger}_x \ket{u(x,y,t)} \text{       }    -  \text{             }  \bra{u(x,y,t)} A_x \ket{u(x,y,t)} - \frac{A^{\dagger}_y }{(\Delta y)^2} \ket{u(x,y,t)}   -   2 \ket{u(x,y,t)}     \\  + A_y \ket{u(x,y,t)}               \bigg]   - 
                3 \text{  } \bigg[   \bra{u(x,y,t)} A_x \ket{u(x,y,t)}  - 
                        \bra{u(x,y,t)} A^{\dagger}_x \ket{u(x,y,t)} - 
     \frac{A^{\dagger}_y}{(\Delta y)^2} \ket{u(x,y,t)} \text{  } \\  - \frac{2}{(\Delta y)^2} \ket{u(x,y,t)}  -  \text{       }  \frac{A_y}{(\Delta y)^2} \ket{u(x,y,t)}        \bigg]    + 
 \bigg[  \bra{u(x,y,t)} A_x \ket{u(x,y,t)}   - \bra{u(x,y,t)} A^{\dagger}_x \ket{u(x,y,t)} \text{        } \\ - \frac{A^{\dagger}_y}{(\Delta y)^2} \ket{u(x,y,t)} \text{       }   \bigg]  \bigg[ \text{           }     \ket{u(x,y,t)} A^{\dagger}_x - \ket{u(x,y,t)} A_x - \frac{1}{(\Delta y)^2} \big(   A^{\dagger}_y - 2 \textbf{1} + A_y \big)        \text{           }          \bigg]   \ket{u(x,y,t)} \\ \times 
                    \bra{u(x,y,t)} A_x \ket{u(x,y,t)}  \bra{u(x,y,t)} \text{     } A_x \ket{u(x,y,t)}  - 
                    \bra{u(x,y,t)} \text{         }   \\ \times   A_x \ket{u(x,y,t)}   \bra{u(x,y,t)} A^{\dagger}_x \ket{u(x,y,t)}   - \bra{u(x,y,t)} A_x  \ket{u(x,y,t)}  \frac{A^{\dagger}_y}{(\Delta y)^2} \ket{u(x,y,t)} \\ +  \bra{u(x,y,t)} A_x \ket{u(x,y,t)}       \text{    }      \frac{2}{(\Delta y)^2} \ket{u(x,y,t)}       \text{    }        - \text{    } \bra{u(x,y,t)} A_x \ket{u(x,y,t)} \\ \times     \frac{A_y}{(\Delta y)^2} \ket{u(x,y,t)} \text{        }     -     \text{         }   \bra{u(x,y,t)} A^{\dagger}_x \ket{u(x,y,t)} \text{        } \bra{u(x,y,t)} A_x \ket{u(x,y,t)}  \text{        } \\ -  \bra{u(x,y,t)} A^{\dagger}_x \ket{u(x,y,t)}     
                       \bra{u(x,y,t)} A^{\dagger}_x \ket{u(x,y,t)}       \text{.      }         
                  \end{align*}     
                       
          \noindent Besides these rearrangements, additional expectation terms in the superposition are of the following form,     
                  
                  \begin{align*}     \bra{u(x,y,t)} A^{\dagger}_x \ket{u(x,y,t)}  \text{         }     \frac{A^{\dagger}_y}{(\Delta y)^2} \ket{u(x,y,t)}          -       \text{       }   \bra{u(x,y,t)}    A^{\dagger}_x \ket{u(x,y,t)}   \frac{2}{(\Delta y)^2}      
                                        \ket{u(x,y,t)}
  \\ \times   \bra{u(x,y,t)}  A^{\dagger}_x   \ket{u(x,y,t)}  \frac{A_y}{(\Delta y)^2} \ket{u(x,y,t)}     \text{        }  -    \frac{A^{\dagger}_y}{(\Delta y)^2} \ket{u(x,y,t)}     \\ \times        \bra{u(x,y,t)}   A_x \ket{u(x,y,t)} \text{       } 
                                        -          \frac{A^{\dagger}_y}{(\Delta y)^2}  \ket{u(x,y,t)} \text{        }       \bra{u(x,y,t)} A^{\dagger}_x \\ \times  \ket{u(x,y,t)}                  +  \frac{A^{\dagger}_y}{(\Delta y)^2} \ket{u(x,y,t)} \frac{A^{\dagger}_y}{(\Delta y)^2} \ket{u(x,y,t)}           -  
   \frac{A^{\dagger}_y}{(\Delta y)^2}  \ket{u(x,y,t)}         \\ \times \frac{2}{(\Delta y)^2} \ket{u(x,y,t)} 
                  +  \frac{A^{\dagger}_y}{(\Delta y)^2} \ket{u(x,y,t)}      \text{       }     \frac{A_y}{(\Delta y)^2} \ket{u(x,y,t)}         
                   \bra{u(x,y,t)} A_x   \\ \times  \text{     } A_x \ket{u(x,y,t)}   - \bra{u(x,y,t)} \text{         }     A_x    A^{\dagger}_x \ket{u(x,y,t)}  - 
                   \bra{u(x,y,t)}   A_x  \frac{A_y}{(\Delta y)^2} \text{      } \\ + \bra{u(x,y,t)}   A_x        \text{    }      \frac{2}{(\Delta y)^2}   \text{   }     -  \text{         } \bra{u(x,y,t)}  A_x  
                   \frac{A^{\dagger}_y}{(\Delta y)^2}  \text{      }     -        \bra{u(x,y,t)} \\ \times  A^{\dagger}_x  \text{        }  A_x \ket{u(x,y,t)}  \text{        }   - \bra{u(x,y,t)} A^{\dagger}_x         A^{\dagger}_x \ket{u(x,y,t)}       \text{         }      
       + \bra{u(x,y,t)}\\ \times  A^{\dagger}_x   \text{         }    \frac{A_y}{(\Delta y)^2}          -        \text{       }   \bra{u(x,y,t)} A^{\dagger}_x        \frac{2}{(\Delta y)^2}         - 
       \text{        }    A^{\dagger}_x    \ket{u(x,y,t)}   -    \frac{A^{\dagger}_y}{(\Delta y)^2}    \\ \times      A_x \ket{u(x,y,t)} \text{       }      -         \frac{A^{\dagger}_y}{(\Delta y)^2} \text{        }            A^{\dagger}_x \ket{u(x,y,t)}    \text{  }               + \text{     }  \bra{u(x,y,t)} \frac{1}{(\Delta y)^2}     \frac{1}{(\Delta y)^2}  \\ \times  \ket{u(x,y,t)}           -    \bra{u(x,y,t)}       \frac{2}{(\Delta y)^2}   \frac{A^{\dagger}_y}{(\Delta y)^2} \ket{u(x,y,t)}  +         \text{                        }        \frac{1}{(\Delta y)^2}        \frac{1}{(\Delta y)^2} \text{. }    
               \end{align*}        
                       
     \noindent Continuing, we further expand terms after distributing states from the expansion of the norm to determine the cost function whose ground state we investigate,                  
     
                   \begin{align*}    
                       \text{            }   \big(  \bra{u(x,y,t)} \text{              }        A_x A_x - \bra{u(x,y,t)} A_x A^{\dagger}_x - \bra{u(x,y,t)} A^{\dagger}_x A_x - \bra{u(x,y,t)} A^{\dagger}_x A^{\dagger}_x \text{      } - A^{\dagger}_x \\ -  \frac{A^{\dagger}_y}{(\Delta x)^2} A_x \text{         } - \frac{A^{\dagger}_y}{(\Delta y)^2}       A^{\dagger}_x  +  \bra{u(x,y,t)} \text{              }                \frac{1}{(\Delta y)^4 }          \text{         } - \bra{u(x,y,t)} \frac{2}{(\Delta y)^2 } \times \frac{A^{\dagger}_x}{(\Delta y)^2 }          - 3 \bra{u(x,y,t)}         A_x \\ - \text{         } 3 \bra{u(x,y,t)}   A^{\dagger}_x \text{               }     - \frac{A^{\dagger}_y}{(\Delta y)^2} - \frac{2}{(\Delta y)^2}  -   \frac{A_y}{(\Delta y)^2}    +  A^{\dagger}_x \frac{A^{\dagger}_y }{(\Delta y)^2 }   \text{          }  +   2 \frac{A^{\dagger}_x}{(\Delta y)^2 } - A^{\dagger}_x \frac{A_y}{(\Delta y)^2 }           \end{align*}

                       \begin{align*} +   A_x \frac{A^{\dagger}_y}{(\Delta y)^2 }   - A_x \frac{2}{(\Delta y)^2 }    \big) \ket{u(x,y,t)} 
       +    9 -  \big(  3 \bra{u(x,y,t)} A^{\dagger}_x - 3 \bra{u(x,y,t)} A_x - 3  \frac{A^{\dagger}_y}{(\Delta y)^2}   - 2 + A_y\\ + 
          \bra{u(x,y,t)} \text{              }        A_x A_x - \bra{u(x,y,t)} A_x A^{\dagger}_x - \bra{u(x,y,t)} A^{\dagger}_x A_x - \bra{u(x,y,t)} A^{\dagger}_x A^{\dagger}_x \text{      } - A^{\dagger}_x \\ -  \frac{A^{\dagger}_y}{(\Delta x)^2} A_x \text{         } - \frac{A^{\dagger}_y}{(\Delta y)^2}       A^{\dagger}_x  +   \bra{u(x,y,t)} \text{              }                \frac{1}{(\Delta y)^4 }          \text{         } - \bra{u(x,y,t)} \frac{2}{(\Delta y)^2 } \frac{A^{\dagger}_x}{(\Delta y)^2 }          - 3 \bra{u(x,y,t)}         A_x \\ -  3 \bra{u(x,y,t)}   A^{\dagger}_x \text{               }     - \frac{A^{\dagger}_y}{(\Delta y)^2} - \frac{2}{(\Delta y)^2} -   \frac{A_y}{(\Delta y)^2}    +  A^{\dagger}_x \frac{A^{\dagger}_y }{(\Delta y)^2 }   \text{          }  +   2 \frac{A^{\dagger}_x}{(\Delta y)^2 } - A^{\dagger}_x \frac{A_y}{(\Delta y)^2 }         +   A_x \frac{A^{\dagger}_y}{(\Delta y)^2 }  \\  - A_x \frac{2}{(\Delta y)^2 }    \big) \ket{u(x,y,t)} \big)  \text{               }      \\ \Updownarrow \\ 
                                           \Big(     A_x \ket{u(x,y,t)} \Big( A^{\dagger}_x - \text{            }  2 \Big) \ket{u(x,y,t)}     \text{                   } \Big)   \text{       }  
                       \Big(          \bra{u(x,y,t)} A^{\dagger}_x \big(      \text{      }            A^{\dagger}_x - 2            \text{      }   \big)^{\dagger}   \ket{u(x,y,t)}                      \Big)        \text{. }    \end{align*} 
                     
  \noindent We continue rearranging terms from the superposition obtained in previous steps above, 
                     
                     \begin{align*}    \text{             }      \Big(      \text{         }            A_x \ket{u(x,y,t)} \Big( \text{         } A^{\dagger}_x - 2   \text{         } \Big)         \text{         }   \ket{u(x,y,t)} \bra{u(x,y,t)} A^{\dagger}_x \text{          }  \text{      }    \text{              }       \big(   A_x - 2      \big)        \text{      }   \bra{u(x,y,t)} \text{         }    \Big)                              \text{             }  \text{      }   \\  \Updownarrow \\    \Big(        \text{      }    A_x \ket{u(x,y,t)}   \Big(  A^{\dagger}_x A^{\dagger}_x - 2 A^{\dagger}_x \Big) \text{         }      \Big(  A_x - 2 \Big) \bra{u(x,y,t)}    \text{        }     \Big)      \text{          }             \\
    \Updownarrow   \\
    \text{          } \Big(         A_x \ket{u(x,y,t)}  \Big(  A^{\dagger}_x  - \text{     }  2 A^{\dagger}_x A^{\dagger}_x - 2 \textbf{1}  + 4 A^{\dagger}_x \Big)  \text{     }   \bra{u(x,y,t)} \text{           }      \Big) \text{          }\\
        \Updownarrow         \\     \Big(      A_x \ket{u(x,y,t)} A^{\dagger}_x  \bra{u(x,y,t)}     - 2   A_x \ket{u(x,y,t)} A^{\dagger}_x A^{\dagger}_x \bra{u(x,y,t)} - 2 A_x        \ket{u(t)} \text{          }    \ket{u(t+\tau)} \\ +  4 \text{   }  A_x \ket{u(x,y,t)} A^{\dagger}_x \ket{u(x,y,t)}        \Big) -        \big(  \bra{u(x,y,t)} A_x  A_x \ket{u(x,y,t)} -  \text{     } \bra{u(x,y,t)}  2 A_x \ket{u(x,y,t)} \big)  \\ \times  \big(  3 -         \bra{u(x,y,t)} A_x \ket{u(x,y,t)} - \bra{u(x,y,t)}   A^{\dagger}_x \ket{u(x,y,t)} \big)     \text{.      }   
         \end{align*}

        \noindent Collecting expectation values from the multiplication of quantum states yields,

        \begin{align*}
                     6 \bra{u(x,y,t)} A_x A_x \ket{u(x,y,t)} \bra{u(x,y,t)} \text{      } A_x \ket{u(x,y,t)}      -  2 \bra{u(x,y,t)} A_x A_x \ket{u(x,y,t)} \bra{u(x,y,t)} \\ \times \ A^{\dagger}_x \ket{u(x,y,t)} -      6\bra{u(x,y,t)} 2 A_x \ket{u(x,y,t)} \bra{u(x,y,t)} A_x \ket{u(x, y,t)}      \\ - 2 \bra{u(x,y,t)} \text{           }     A_x A_x \ket{u(x,y,t)} \bra{u(x,y,t)} A^{\dagger}_x \ket{u(x,y,t)} \\ +   2 \bra{u(x,y,t)} 2 A_x \ket{u(x,y,t)} \bra{u(x,y,t)} A^{\dagger}_x \ket{u(x,y,t)} \text{       }   \\ \Updownarrow  \\ 
      6 \bra{u(x,y,t)} A_x A_x \lambda^2 A_x \ket{u(x,y,t)} \text{      }      - 2 \bra{u(x,y,t)} A_x A_x  \lambda^2 \text{         }              A^{\dagger}_x \ket{u(x,y,t)} \text{      }  \\ -          6 \bra{u(x,y,t)} 2 A_x \lambda^2 \text{         }  \ket{u(x,y,t)} \text{          }       - 2 \bra{u(x,y,t)} \text{      }      A_x A_x  \lambda^2 A^{\dagger}_x \ket{u(x,y,t)} \\ +     \text{   }          2 \bra{u(x,y,t)} 2 A_x \lambda^2 A^{\dagger}_x \ket{u(x,y,t)}   \text{. }  
      \end{align*}
      
 \noindent      Previous rearrangements imply, 
      
                \begin{align*}
                       \big(   6 \bra{u(x,y,t)} A_x A_x \lambda^2 A_x - 2 \bra{u(x,y,t)} A_x A_x \lambda^2 A^{\dagger}_x  -     
                         6 \bra{u(x,y,t)} 2 A_x \lambda^2 \text{        }     - 2 \bra{u(x,y,t)} A_x A_x \\ \times \lambda^2 A^{\dagger}_x  \\  + 2 \bra{u(x,y,t)} 2 A_x \lambda^2 A^{\dagger}_x  + 
         \bra{u(x,y,t)} A_x A_x - \bra{u(x,y,t)} A_x A^{\dagger}_x - \bra{u(x,y,t)} A^{\dagger}_x\\ \times  A_x - \bra{u(x,y,t)} A^{\dagger}_x   A^{\dagger}_x \\
         -   \frac{A^{\dagger}_y}{(\Delta x)^2} A_x - \frac{A^{\dagger}_y}{(\Delta y)^2} A^{\dagger}_x \text{           }         + \bra{u(x,y,t)}   \frac{1}{(\Delta y)^4 \text{       } }      - \bra{u(x,y,t)} \frac{2}{(\Delta y)^2} \text{      }  \frac{A^{\dagger}_x}{(\Delta y)^2}  -      3 \bra{u(x,y,t)} \text{                   }       A_x \\ - 3 \bra{u(x,y,t)} A^{\dagger}_x  - \frac{A^{\dagger}_y}{(\Delta y)^2}         - \frac{2}{(\Delta y)^2}   -         \frac{A_y}{(\Delta y)^2} \text{          }           + A^{\dagger}_x \frac{A^{\dagger}_y}{(\Delta y)^2} \text{             }    + 2 \text{            } \frac{A^{\dagger}_x}{(\Delta y)^2}            - A^{\dagger}_x \frac{A_y}{(\Delta y)^2} \\ + A_x \frac{A^{\dagger}_y}{(\Delta y)^2} - A_x \frac{2}{(\Delta y)^2}           \big) \ket{u(x,y,t)} \text{              }     \text{, }        
                       \end{align*}
                       
          \noindent from which rearrangements of the final collection of terms yields, 
          
           
           \begin{align*}
             \frac{\tau}{(\Delta x)^2} \Bigg\{         A_x \ket{u(x,y,t+\tau)}     A_x \ket{u(x,y,t)} \big( A^{\dagger}_x - 2 \big)  \ket{u(x,y,t)}  + 
                 A_x \ket{u(x,y,t+\tau)}   \\  - A_x \ket{u(x,y,t+\tau)}  \ket{u(x,y,t)} \big(     A^{\dagger}_x \ket{u(x,y,t)}  - A_x \ket{u(x,y,t)}     \big)   + 2                 \Bigg\}   -       \frac{\tau}{(\Delta x)^2}   \\ \times    \frac{A_x \ket{u(x,y,t+\tau)}}{(\Delta y)^2} \big( A^{\dagger}_y \ket{u(x,y,t)}  - 2 \ket{u(x,y,t)} + \text{                             }   \end{align*}

                    \begin{align*}  \times      A_y \ket{u(x,y,t)} \big)  \text{                        } \\     
                    \Updownarrow \\ 
                    \frac{\tau}{(\Delta x)^2} \Bigg\{  \text{                          }          A_x \ket{u(x,y,t+\tau)} A_x \ket{u(x,y,\tau)} A^{\dagger}_x \ket{u(x,y,t)}      -   A_x \ket{u(x,y,t+\tau)} A_x \ket{u(x,y,t)}  \\ \times  2 \ket{u(x,y,t)}    \text{           } \text{             }   + 
                                         \text{      }    A_x \ket{u(x,y,t+\tau)}   - \text{        }    A_x \ket{u(x,y, t+\tau)}   \ket{u(x,y,t)} A^{\dagger}_x  \ket{u(x,y,t)} \\   - 
                 \text{              }    A_x \ket{u(x,y,t+\tau)  }    A_x \ket{u(x,y,t)} \ket{u(x,y,t)} \bra{u(x,y,t)} A^{\dagger}_x  + 2 \Bigg\}    \\   - 
          \frac{\tau}{(\Delta x)^2 (\Delta y)^2 } \big( A_x \ket{u(x,y,t)} A^{\dagger}_x \ket{u(x,y,t)} \\   - A_x \ket{u(x,y,t+\tau)} 2 \ket{u(x,y,t)} \big) \text{ . } \\
            \end{align*}

\newpage

\subsection{Camassa-Holm equation}

\subsubsection{Statement}

\begin{align*}
   \underline{ \mathcal{C}^{\mathrm{C-H}} (\lambda,\lambda^{\mathrm{C-H}}_0 ) } \approx    \big(\lambda^{\mathrm{C-H}}_0\big)^2 +    \mathrm{Re }   \bigg\{ \frac{4 \lambda^2}{(\Delta x)^3} \frac{\tau}{ \big(                   1 + \frac{2}{(\Delta x)^3 \tau} \big)} \bra{u} A_x A_x \ket{u}     - \frac{2 \lambda^2}{(\Delta x)^3} \frac{\tau}{ \big(              1 + \frac{2}{(\Delta x)^3 \tau} \big)} \bra{u} A_x \ket{u} \end{align*}
   
   \begin{align*} -   \frac{8 }{(\Delta x)^3}  \frac{ \tau^2}{\big(     1 + \frac{1}{(\Delta x)^3 \tau}     \big)^2 }  A_x \ket{u}   - 
    \frac{ 8  \lambda^2}{(\Delta x)^3} \frac{\tau}{ \big(                    1 + \frac{2}{(\Delta x)^3 \tau} \big)} A_x \ket{u}          + \frac{ 4   \lambda^2}{(\Delta x)^4} \frac{\tau}{ \big(                   1 + \frac{2}{(\Delta x)^3 \tau} \big)} A^{\dagger}_x \ket{\widetilde{u}}   \\ +   \frac{ 8   \lambda^2}{(\Delta x)^6} \frac{\tau}{ \big(                     1 + \frac{2}{(\Delta x)^3 \tau} \big)}  \ket{\widetilde{u}}   +  \frac{ 8   \lambda^2}{(\Delta x)^6} \frac{\tau}{ \big(                  1 + \frac{2}{(\Delta x)^3 \tau} \big)}  A_x \ket{u} +   \frac{ 8   \lambda^2}{(\Delta x)^6} \frac{\tau}{ \big(                     1 + \frac{2}{(\Delta x)^3 \tau} \big)}  A^{\dagger}_x \ket{u}      \\   -     \frac{ 4  \lambda^2}{(\Delta x)^6} \frac{\tau}{ \big(         1 + \frac{2}{(\Delta x)^3 \tau} \big)}   A^{\dagger}_x \ket{u}    -  \frac{ 8 \lambda^2}{(\Delta x)^6} \frac{\tau}{ \big(                   1 + \frac{2}{(\Delta x)^3 \tau} \big)}    \ket{\widetilde{u}}           -    \frac{ 4 \lambda^2}{(\Delta x)^6} \frac{\tau}{ \big(                   1 + \frac{2}{(\Delta x)^3 \tau} \big)}    A_x \ket{u}  \\  -           \frac{ 4 \lambda^2}{(\Delta x)^6} \frac{\tau}{ \big(                     1 + \frac{2}{(\Delta x)^3 \tau} \big)}           -   \frac{ 12 \lambda^2}{(\Delta x)^4} \frac{\tau}{ \big(                    1 + \frac{2}{(\Delta x)^3 \tau} \big)} \bra{u} A^{\dagger}_x \ket{u}         +  \frac{ 12 \lambda^2}{(\Delta x)^4} \frac{\tau}{ \big(                  1 + \frac{2}{(\Delta x)^3 \tau} \big)} \bra{u} \ket{u}  \\  +            \frac{ 8 \lambda^2}{(\Delta x)^4} \frac{\tau}{ \big(                    1 + \frac{2}{(\Delta x)^3 \tau} \big)} A_x  \ket{u}  + \frac{ 4  \lambda^2}{(\Delta x)^4} \frac{\tau}{ \big(                   1 + \frac{2}{(\Delta x)^3 \tau} \big)}  \ket{u}  + \frac{ 8  \lambda^2}{(\Delta x)^7}   \frac{\tau}{ \big(                 1 + \frac{2}{(\Delta x)^3 \tau} \big)} A^{\dagger}_x \ket{\widetilde{u}}          \\   -       \frac{ 8  \lambda^2}{(\Delta x)^8}   \frac{\tau}{ \big(                 1 + \frac{2}{(\Delta x)^3 \tau} \big)}  \ket{\widetilde{u}}     +  \frac{ 16  \lambda^2}{(\Delta x)^7}  \frac{\tau}{ \big(                 1 + \frac{2}{(\Delta x)^3 \tau} \big)}  A_x \ket{\widetilde{u}}  - 
   \frac{ 8  \lambda^2}{(\Delta x)^7}   \frac{\tau}{ \big(              1 + \frac{2}{(\Delta x)^3 \tau} \big)} A^{\dagger}_x \ket{u}    \\  +   \frac{ 8  \lambda^2}{(\Delta x)^7}  \frac{\tau}{ \big(                     1 + \frac{2}{(\Delta x)^3 \tau} \big)}  \ket{u}   -   \frac{ 8  \lambda^2}{(\Delta x)^7}   \frac{\tau}{ \big(               1 + \frac{2}{(\Delta x)^3 \tau} \big)} A_x \ket{u}     + \frac{1}{(\Delta x)^6}   \frac{8 \tau^2}{\big(     1 + \frac{2}{(\Delta x)^3 \tau}     \big)^2 }              \bra{u} A_x A_x \ket{u}   \\  -    \frac{1}{(\Delta x)^6}     \frac{8 \tau^2}{\big(     1 + \frac{2}{(\Delta x)^3 \tau}     \big)^2 }       \bra{u} A_x A^{\dagger}_x \ket{u}      +   \frac{1}{(\Delta x)^6}        \frac{4 \tau^2}{\big(     1 + \frac{2}{(\Delta x)^3 \tau}     \big)^2 }         \bra{u} A_x A^{\dagger}_x \ket{u}         \\  +   \frac{1}{(\Delta x)^6}    \frac{4 \tau^2}{\big(     1 + \frac{2}{(\Delta x)^3 \tau}     \big)^2 }   \bra{u} A^{\dagger}_x A_x \ket{u}    -  \frac{2}{(\Delta x)^6} \frac{4 \tau^2}{\big(     1 + \frac{2}{(\Delta x)^3 \tau}     \big)^2 }      \bra{\widetilde{u}} A_x \ket{u}     \\   +  \frac{2}{(\Delta x)^6} \frac{4 \tau^2}{\big(     1 + \frac{1}{(\Delta x)^3 \tau}     \big)^2 } \bra{u} A_x \ket{u}    +    \frac{8}{(\Delta x)^9}  \frac{4 \tau^2}{\big(     1 + \frac{1}{(\Delta x)^3 \tau}     \big)^2 }   \bra{u} A^{\dagger}_x A_x \ket{u}     \\   - \frac{16}{(\Delta x)^9} 
   \frac{4 \tau^2}{\big(     1 + \frac{1}{(\Delta x)^3 \tau}     \big)^2 }    \bra{u} A_x A_x \ket{u}      +  \frac{8}{(\Delta x)^9}    \frac{4 \tau^2}{\big(     1 + \frac{1}{(\Delta x)^3 \tau}     \big)^2 }            \bra{u} A_x \ket{u}       \\  -   \frac{8}{(\Delta x)^9}  \frac{4 \tau^2}{\big(     1 + \frac{1}{(\Delta x)^3 \tau}     \big)^2 }  \bra{u} A^{\dagger}_x A_x \ket{\widetilde{u}}             -    \frac{16}{(\Delta x)^9}    \frac{4 \tau^2}{\big(     1 + \frac{1}{(\Delta x)^3 \tau}     \big)^2 }           \bra{u} A_x \ket{\widetilde{u}}               \end{align*}

   \begin{align*}     + \frac{4}{(\Delta x)^3}   \frac{ \tau^2}{\big(     1 + \frac{1}{(\Delta x)^3 \tau}     \big)^2 }   \bra{u} A_x A_x \ket{u} -       \frac{4}{(\Delta x)^3}   \frac{ \tau^2}{\big(     1 + \frac{1}{(\Delta x)^3 \tau}     \big)^2 }   \bra{u} A_x \ket{u}    \\  -  \frac{8}{(\Delta x)^9}  \frac{4 \tau^2}{\big(     1 + \frac{1}{(\Delta x)^3 \tau}     \big)^2 }           \bra{u} A_x \ket{u}  +   \frac{2  \tau^2}{(\Delta x)^3 \big(     1 + \frac{1}{ (\Delta x)^3 \tau}     \big)^2 }    \bra{\widetilde{u}} A_x \ket{u}        +     \frac{ 4 \tau^2}{(\Delta x)^3 \big(     1 + \frac{1}{ (\Delta x)^3 \tau}     \big)^2 }    \bra{\widetilde{u}}   \ket{u}  \\  +     \frac{ 2 \tau^2}{(\Delta x)^3 \big(     1 + \frac{1}{ (\Delta x)^3 \tau}     \big)^2 }     \bra{u}  A^{\dagger}_x \ket{u}   +  \frac{ 2 \tau^2}{(\Delta x)^3 \big(     1 + \frac{1}{ (\Delta x)^3 \tau}     \big)^2 }   \bra{u} A^{\dagger}_x \ket{u}  +   \frac{ 9  \tau^2}{ \Delta x \big(     1 + \frac{2}{ (\Delta x)^2 \tau}     \big)^2 } \\ \times     \bra{u} A^{\dagger}_x A_x \ket{u}       -     \frac{  6  \tau^2}{\Delta x\big(     1 + \frac{1}{ (\Delta x)^3 \tau}     \big)^2 }     \bra{u} A^{\dagger}_x A_x \ket{u}       -  \frac{ 9 \tau^2}{\Delta x\big(     1 + \frac{1}{ (\Delta x)^3 \tau}     \big)^2 }         \\ \times   \bra{u} A^{\dagger}_x A_x \ket{u}   - 12     \bra{u} A^{\dagger}_x  A_x \ket{u}     - 12  \bra{u}  A_x \ket{u} +   12   \bra{u}  A_x \ket{u}   +  4 \bra{u} A_x \ket{u} \\ -   8 \bra{u} A_x \ket{u}    +   8 \bra{u} \ket{u}  \bigg\}  \text{.}  \end{align*}

\subsubsection{Derivation}

\noindent For $\kappa\equiv 1$, the equation after introducing the standard quantum state transformation for evolution of the variational state for the solution is equivalent to,

\begin{align*}
            u_t = 2 u_x u_{xx} + u u_{xxx} -   ( 3 u   + 2 \kappa  ) u_x    + u_{xxt}      \\  
               \text{              } \Updownarrow      \text{                       }        \\
               \frac{\ket{u(x,t+\tau)} - \ket{u(x,t)}}{\tau} =     2 \big(     \frac{A_x -  \textbf{1}}{\Delta x}    \big)   \ket{u(x,t)}      \big(       \frac{A^{\dagger}_x - 2 \textbf{1} + A_x}{(\Delta x)^2}            \big)   \ket{u(x,t)}       +    \frac{\ket{u(x,t)}}{\Delta x} \\ \times   \bigg[ \big(     \frac{A^{\dagger}_x - 2 \textbf{1} + A_x}{(\Delta x)^2} \big)            \ket{u(x,t+\tau)}  -   \big(     \frac{A^{\dagger}_x - 2 \textbf{1} + A_x}{(\Delta x)^2}        \big)  \ket{u(x,t)} \bigg]   -   \big(    3 \ket{u(x,t)} + 2 \big) \\ \times  \big(  \frac{A_x - \textbf{1}}{\Delta x}     \big)      \ket{u(x,t)}  +   \big(   \frac{ A^{\dagger}_x - 2 \textbf{1} + A_x }{(\Delta x)^2}   \ket{u(x,t)}   \big)_t \text{ . }\\
               \end{align*}
                              
                          \noindent We evaluate the first order time derivative and additional terms of the superposition,

                              \begin{align*}
                              \frac{\big( \frac{A^{\dagger}_x - 2 \textbf{1} + A_x}{(\Delta x)^2} \big)    \ket{u(x,t+\tau)} -    \big( \frac{A^{\dagger}_x - 2 \textbf{1} + A_x}{(\Delta x)^2} \big) \ket{u(x,t)}}{    (      \Delta x) \tau }                     \\
                              \text{              } \Updownarrow      \text{                       }        \\
                                                     \big(    3 \ket{u(x,t)} + 2 \big) \big(  \frac{A_x - \textbf{1}}{\Delta x}     \big)  \ket{u(x,t)}   =   \frac{1}{\Delta x}    \big(    3 \ket{u(x,t)} +     2       \big)  \big( A_x \ket{u(x,t)} - \ket{u(x,t)} \big)     \\ =       \frac{1}{\Delta x}  \big( \{ 3 \ket{u(x,t)} \}   \{ A_x \ket{u(x,t)} \}  - \{ 3 \ket{u(x,t)} \} \{ \ket{u(x,t)} \} \\ +  2 A_x \ket{u(x,t)} - 2 \ket{u(x,t)} \big)        \\
                           \text{              } \Updownarrow      \text{                       }        \\
                         \big(     \frac{A^{\dagger}_x - 2 \textbf{1} + A_x}{(\Delta x)^2}        \big)  \ket{u(x,t)}  =     \frac{1}{(\Delta x)^2}  \big( A^{\dagger}_x \ket{u(x,t)} - 2 \ket{u(x,t)} + A_x \ket{u(x,t)}       \big)    \end{align*}

                         \begin{align*}
                                 \text{              } \Updownarrow    \text{                       }         \\
                                  \big(     \frac{A^{\dagger}_x - 2 \textbf{1} + A_x}{(\Delta x)^2} \big)          \ket{u(x,t+\tau)}      =   \frac{1}{(\Delta x)^2}  \big(   A^{\dagger}_x \ket{u(x,t+\tau)}  - 2 \ket{u(x,t+\tau)} + A_x \ket{u(x,t)}  \big)     \\
 \text{              } \Updownarrow   \\ 
                           \big(     \frac{A_x - \textbf{1}}{\Delta x}    \big)   \ket{u(x,t)}    \big(       \frac{A^{\dagger}_x - 2 \textbf{1} + A_x}{(\Delta x)^2}            \big) \ket{u(x,t)}  \\   \Updownarrow  \\      
                  \frac{1}{(\Delta x)^3}  \bigg[ \{ A_x - \textbf{1} \} \ket{u(x,t)}  A^{\dagger}_x - 2 \{ A_x - \textbf{1} \} \ket{u(x,t)} + \{ A_x -   \textbf{1} \} \ket{u(x,t)}  A_x     \bigg] \ket{u(x,t)} \text{, } \end{align*}
                  
                 \noindent for terms on the right hand side of the equation, and, 
                  
                  \begin{align*}
                  \frac{1}{(\Delta x)^3} \bigg[              A_x \ket{u(x,t)} A^{\dagger}_x - \ket{u(x,t)} A^{\dagger}_x -   2 A_x \ket{u(x,t)} -  \ket{u(x,t)} + A_x \ket{u(x,t)} A_x - \ket{u(x,t)} A_x \bigg] \\ \times \ket{u(x,t)} \text{             }   \\  \Updownarrow   \\ \frac{1}{(\Delta x)^3} \bigg[    A_x \ket{u(x,t)} A^{\dagger}_x \ket{u(x,t)} - \ket{u(x,t)} A^{\dagger}_x \ket{u(x,t)}    - 2 A_x                     \bigg]          \text{, }         \\
                    \end{align*}
                    
       \noindent for terms on the left hand side of the equation. As a consequence of previous manipulations, the time evolved variational state corresponding to the solution takes the form,

             \begin{align*}
                \ket{u(x,t+\tau)} =  \tau   \bigg\{ \frac{2 }{(\Delta x)^3} \big( A_x \ket{u(x,t)} A^{\dagger}_x \ket{u(x,t)} - \ket{u(x,t)} A^{\dagger}_x \ket{u(x,t)} - 2 A_x \big)  + 
                     \frac{\ket{u(x,t)}}{(\Delta x)^3}   \big(   A^{\dagger}_x \\ \times  \ket{u(x,t+\tau)} - 2 \ket{u(x,t+\tau)} + A_x \ket{u(x,t)}            \big)    -    \frac{\ket{u(x,t)}}{(\Delta x)^3} \bigg[ A^{\dagger}_x \ket{u(x,t)} - 2 \ket{u(x,t)} \\ + A_x \ket{u(x,t)} \bigg]   -     \frac{1}{\Delta x} \big( { 3 \ket{u(x,t)} } { A_x \ket{u(x,t)}} - { 3 \ket{u(x,t)} }{ \ket{u(x,t)}} + 2 A_x \ket{u(x,t)} \\ - 2 \ket{u(x,t)}    \big) +                \frac{\big( \frac{A^{\dagger}_x - 2 \textbf{1} + A_x}{(\Delta x)^2} \big)    \ket{u(x,t+\tau)} -    \big( \frac{A^{\dagger}_x - 2 \textbf{1} + A_x}{(\Delta x)^2} \big)  \ket{u(x,t)}}{    (      \Delta x) \tau }     \bigg\} 
                     \end{align*}

               \noindent In turn, the superposition above yields the cost function that can be identified with the expansion of the norm below, after consolidating terms with the time evolved variational state of the solution,      
                     
                     \begin{align*}
                     \big(                     1 + \frac{2}{(\Delta x)^3 \tau}     \big)  \ket{u(x,t+\tau)} =  \tau  \bigg\{   \frac{2 }{(\Delta x)^3} \big( A_x \ket{u(x,t)} A^{\dagger}_x \ket{u(x,t)} - \ket{u(x,t)} A^{\dagger}_x \ket{u(x,t)} - 2 A_x \big) \\ + 
                     \frac{\ket{u(x,t)}}{(\Delta x)^3}  \bigg[   A^{\dagger}_x \ket{u(x,t+\tau)} - 2 \ket{u(x,t+\tau)} + A_x \ket{u(x,t)}            \bigg]     - \frac{\ket{u(x,t)}}{(\Delta x)^3} \big( A^{\dagger}_x \ket{u(x,t)} \\ - 2 \ket{u(x,t)} + A_x \ket{u(x,t)} \big)  -     \frac{1}{\Delta x} \bigg[ { 3 \ket{u(x,t)} } { A_x \ket{u(x,t)}} - { 3 \ket{u(x,t)} }{ \ket{u(x,t)}} \\ + 2 A_x \ket{u(x,t)} - 2 \ket{u(x,t)}    \bigg]  +     \frac{\big( \frac{A^{\dagger}_x -2 \textbf{1} +  A_x}{(\Delta x)^2} \big)    \ket{u(x,t+\tau)} -    \big( \frac{A^{\dagger}_x - 2 \textbf{1} + A_x}{(\Delta x)^2} \big) \ket{u(x,t)}}{    (      \Delta x) \tau }     \bigg\} \text{. } \end{align*}

                     \noindent The above superposition equals,
                     
                     \begin{align*}
                      \ket{u(x,t+\tau)} =  \frac{\tau}{ \big(                 1 + \frac{2}{(\Delta x)^3 \tau}    \big)}   \bigg\{   \frac{2 }{(\Delta x)^3} \bigg[ A_x \ket{u(x,t)} A^{\dagger}_x \ket{u(x,t)} - \ket{u(x,t)} A^{\dagger}_x \ket{u(x,t)} - 2 A_x \bigg] \\  +     
                     \frac{\ket{u(x,t)}}{(\Delta x)^3}  \bigg[   A^{\dagger}_x \ket{u(x,t+\tau)} - 2 \ket{u(x,t+\tau)} + A_x \ket{u(x,t)}            \bigg]     -    \frac{\ket{u(x,t)}}{(\Delta x)^3} \bigg[ A^{\dagger}_x \ket{u(x,t)}\\  - 2 \ket{u(x,t)} + A_x \ket{u(x,t)} \bigg]    -     \frac{1}{\Delta x} \bigg[ { 3 \ket{u(x,t)} } { A_x \ket{u(x,t)}}  - { 3 \ket{u(x,t)} }{ \ket{u(x,t)}} \end{align*}

                     \begin{align*} + 2 A_x \ket{u(x,t)} - 2 \ket{u(x,t)}    \bigg]  +                \frac{\big( \frac{A^{\dagger}_x - 2 \textbf{1}  +  A_x}{(\Delta x)^2} \big)    \ket{u(x,t+\tau)} -    \big( \frac{A^{\dagger}_x - 2 \textbf{1} + A_x}{(\Delta x)^2} \big) \ket{u(x,t)}}{    (      \Delta x) \tau }     \bigg\} \text{ . }   \end{align*}    
                     
                     \noindent Expanding the cost function, $ \mathcal{C}^{\mathrm{C-H} } = \big| \big| \ket{u(x,t+\tau)} - \ket{u(x,t)} \big| \big|^2 $, yields,
                     
                     \begin{align*} 
                  \bigg| \bigg|   \frac{\tau}{ \big(                     1 + \frac{2}{(\Delta x)^3 \tau}     \big)}   \bigg\{   \frac{2 }{(\Delta x)^3} \big( A_x \ket{u(x,t)} A^{\dagger}_x \ket{u(x,t)} - \ket{u(x,t)} A^{\dagger}_x \ket{u(x,t)} - 2 A_x \big) \\  + 
                     \frac{\ket{u(x,t)}}{(\Delta x)^3}  \bigg[   A^{\dagger}_x \ket{u(x,t+\tau)}  - 2 \ket{u(x,t+\tau)} + A_x \ket{u(x,t)}            \bigg]     -    \frac{\ket{u(x,t)}}{(\Delta x)^3} \bigg[ A^{\dagger}_x \ket{u(x,t)} \\ - 2 \ket{u(x,t)} + A_x \ket{u(x,t)} \bigg]   -    \frac{1}{\Delta x} \bigg[ { 3 \ket{u(x,t)} } { A_x \ket{u(x,t)}} - { 3 \ket{u(x,t)} }{ \ket{u(x,t)}} + 2 A_x \ket{u(x,t)} \\ - 2 \ket{u(x,t)}    \bigg]    + 
                      \frac{\big( \frac{A^{\dagger}_x - 2 \textbf{1} + A_x}{(\Delta x)^2} \big)    \ket{u(x,t+\tau)} -    \big( \frac{A^{\dagger}_x - 2 \textbf{1}  +  A_x}{(\Delta x)^2} \big) }{    (      \Delta x) \tau }   \ket{u(x,t)}   \bigg\}   - \ket{u(x,t)} \bigg| \bigg|^2  \text{. }    \end{align*}  
                     
                 \noindent As with derivations of previous cost functions, the first component of the superposition is solely dependent upon the initial choice of variational parameter $\lambda$, while remaining terms are inversely proportional with respect to either the time step $\tau$ with which the solution state is evolved, or the grid resolution $\Delta x$. The superposition is of the form, 
                     
                     \begin{align*}      \lambda^2 +  \bigg\{   \frac{\tau}{ \big(                     1 + \frac{2}{(\Delta x)^3 \tau}     \big)}   \bigg\{   \frac{2 }{(\Delta x)^3} \big( A_x \ket{u(x,t)} A^{\dagger}_x \ket{u(x,t)} - \ket{u(x,t)} A^{\dagger}_x \ket{u(x,t)} - 2 A_x \big) \\ + 
         \frac{\ket{u(x,t)}}{(\Delta x)^3}  \big(   A^{\dagger}_x \ket{u(x,t+\tau)} - 2 \ket{u(x,t+\tau)} + A_x \ket{u(x,t)}            \big)    -    \frac{\ket{u(x,t)}}{(\Delta x)^3} \big( A^{\dagger}_x \ket{u(x,t)} - 2 \ket{u(x,t)} \\ + A_x \ket{u(x,t)} \big)  -    \frac{1}{\Delta x} \big( { 3 \ket{u(x,t)} } { A_x \ket{u(x,t)}} - { 3 \ket{u(x,t)} }{ \ket{u(x,t)}}  + 2 A_x \ket{u(x,t)} - 2 \ket{u(x,t)}    \big) \end{align*} \begin{align*}  +             \frac{\big( \frac{A^{\dagger}_x - 2 \textbf{1} + A_x}{(\Delta x)^2} \big)    \ket{u(x,t+\tau)} -    \big( \frac{A^{\dagger}_x - 2 \textbf{1} + A_x}{(\Delta x)^2} \big) \ket{u(x,t)}}{    (      \Delta x) \tau }     \bigg\} \bigg\}^2 + 
                     2 \lambda^2   \frac{\tau}{ \big(                     1 + \frac{2}{(\Delta x)^3 \tau}     \big)} \\ \times   \bigg\{   \frac{2 }{(\Delta x)^3} \big( A_x \ket{u(x,t)} A^{\dagger}_x \ket{u(x,t)}  - \ket{u(x,t)} A^{\dagger}_x \ket{u(x,t)} - 2 A_x \big)  \\ +  \frac{\ket{u(x,t)}}{(\Delta x)^3}  \big(   A^{\dagger}_x \ket{u(x,t+\tau)} - 2 \ket{u(x,t+\tau)} + A_x \ket{u(x,t)}            \big)   \\  - 
                   \frac{\ket{u(x,t)}}{(\Delta x)^3} \big( A^{\dagger}_x \ket{u(x,t)} - 2 \ket{u(x,t)}  + A_x \ket{u(x,t)} \big)  -     \frac{1}{\Delta x} \big( { 3 \ket{u(x,t)} } \\ \times { A_x \ket{u(x,t)}}  - { 3 \ket{u(x,t)} }{ \ket{u(x,t)}}  + 2 A_x \ket{u(x,t)} \\ - 2 \ket{u(x,t)}    \big) +                \frac{\big( \frac{A^{\dagger}_x - 2 \textbf{1} + A_x}{(\Delta x)^2} \big)    \ket{u(x,t+\tau)}}{ (      \Delta x) \tau }  -    \frac{\big( \frac{A^{\dagger}_x - 2 \textbf{1} + A_x}{(\Delta x)^2} \big) \ket{u(x,t)}}{    (      \Delta x) \tau }     \bigg\} \text{. }      
\end{align*}

\noindent Expanding all terms in the order that they appear in the superposition following $\lambda^2$ yields,

\begin{align*}
      \bigg\{  \frac{\tau}{ \big(                     1 + \frac{2}{(\Delta x)^3 \tau}     \big)}   \bigg\{   \frac{2 }{(\Delta x)^3} \bigg[ A_x \ket{u(x,t)} A^{\dagger}_x \ket{u(x,t)} - \ket{u(x,t)} A^{\dagger}_x \ket{u(x,t)} - 2 A_x \bigg]  \\  + 
                     \frac{\ket{u(x,t)}}{(\Delta x)^3}  \bigg[   A^{\dagger}_x \ket{u(x,t+\tau)}  - 2 \ket{u(x,t+\tau)} + A_x \ket{u(x,t)}            \bigg]     -    \frac{\ket{u(x,t)}}{(\Delta x)^3} \bigg[ A^{\dagger}_x \ket{u(x,t)} \\  - 2 \ket{u(x,t)} + A_x \ket{u(x,t)} \bigg]   -     \frac{1}{\Delta x} \bigg[ { 3 \ket{u(x,t)} } { A_x \ket{u(x,t)}}  - { 3 \ket{u(x,t)} }{ \ket{u(x,t)}} + 2 A_x \ket{u(x,t)} \\ - 2 \ket{u(x,t)}    \bigg]  +                \frac{\big( \frac{A^{\dagger}_x - 2 \textbf{1} + A_x}{(\Delta x)^2} \big)    \ket{u(x,t+\tau)}}{    (      \Delta x) \tau }  -    \frac{\big( \frac{A^{\dagger}_x - 2 \textbf{1} + A_x}{(\Delta x)^2} \big) \ket{u(x,t)}}{    (      \Delta x) \tau }     \bigg\} \bigg\}^2 \text{. } \end{align*}  
                     
                  \noindent Squaring the terms included in the superposition above implies,   
                     
                     \begin{align*}   \bigg\{  \frac{\tau}{ \big(                     1 + \frac{2}{(\Delta x)^3 \tau}     \big)}   \bigg\{  \mathscr{T}_1 \bigg\}^2    +      \bigg\{               \mathscr{T}_2    \bigg\}^2 + \mathscr{T}_3    
                     \bigg\{  \frac{\tau}{ \big(                     1 + \frac{2}{(\Delta x)^3 \tau}     \big)}   \mathscr{T}_4    -     \frac{1}{\Delta x}  \mathscr{T}_5 \bigg\}  \bigg\}^2 \text{, }
\end{align*}

\noindent where,

\begin{align*}
  \mathscr{T}_1 \equiv  \frac{2 }{(\Delta x)^3} \big( A_x \ket{u(x,t)} A^{\dagger}_x \ket{u(x,t)} - \ket{u(x,t)} A^{\dagger}_x \ket{u(x,t)} - 2 A_x \big)  + 
                     \frac{\ket{u(x,t)}}{(\Delta x)^3}  \big(   A^{\dagger}_x \ket{u(x,t+\tau)} \\ - 2 \ket{u(x,t+\tau)} + A_x \ket{u(x,t)}            \big)    -    \frac{\ket{u(x,t)}}{(\Delta x)^3} \big( A^{\dagger}_x \ket{u(x,t)} - 2 \ket{u(x,t)}  + A_x \ket{u(x,t)} \big) \\  -     \frac{1}{\Delta x} \big( { 3 \ket{u(x,t)} } { A_x \ket{u(x,t)}} - { 3 \ket{u(x,t)} }{ \ket{u(x,t)}} + 2 A_x \ket{u(x,t)} \\ - 2 \ket{u(x,t)}    \big)   \text{,} \\ \\   \mathscr{T}_2 \equiv    \frac{\big( \frac{A^{\dagger}_x - 2 \textbf{1} + A_x}{(\Delta x)^2} \big)    \ket{u(x,t+\tau)} -    \big( \frac{A^{\dagger}_x - 2 \textbf{1} + A_x}{(\Delta x)^2} \big) \ket{u(x,t)}}{    (      \Delta x) \tau }     \text{, } \end{align*}
                     
                     \begin{align*} \mathscr{T}_3 \equiv   2 \mathscr{T}_2 \equiv     \frac{2 \big( \frac{A^{\dagger}_x - 2 \textbf{1} + A_x}{(\Delta x)^2} \big)    \ket{u(x,t+\tau)} -    2 \big(  \frac{A^{\dagger}_x - 2 \textbf{1} + A_x}{(\Delta x)^2} \big) \ket{u(x,t)}}{    (      \Delta x) \tau }          \text{, } \\ \\ \mathscr{T}_4 \equiv  \frac{2 }{(\Delta x)^3} \big( A_x \ket{u(x,t)} A^{\dagger}_x \ket{u(x,t)}  - \ket{u(x,t)} A^{\dagger}_x \ket{u(x,t)} - 2 A_x \big)  + 
                     \frac{\ket{u(x,t)}}{(\Delta x)^3}  \big(   A^{\dagger}_x \ket{u(x,t+\tau)} \\ - 2 \ket{u(x,t+\tau)}  + A_x \ket{u(x,t)}            \big)    -    \frac{\ket{u(x,t)}}{(\Delta x)^3} \big( A^{\dagger}_x \ket{u(x,t)} - 2 \ket{u(x,t)} + A_x \ket{u(x,t)}  \text{, } \\ \\    \mathscr{T}_5 \equiv        { 3 \ket{u(x,t)} } { A_x \ket{u(x,t)}} - { 3 \ket{u(x,t)} }{ \ket{u(x,t)}} + 2 A_x \ket{u(x,t)} - 2 \ket{u(x,t)}    \big)               \text{. } 
\end{align*}

\noindent Furthermore, additional rearrangements of the expansion above imply,

\begin{align*}
            \bigg\{  \frac{\tau}{ \big(                     1 + \frac{2}{(\Delta x)^3 \tau}     \big)}   \bigg\{   \frac{2 }{(\Delta x)^3} \bigg[ A_x \ket{u(x,t)} A^{\dagger}_x \ket{u(x,t)} - \ket{u(x,t)} A^{\dagger}_x \ket{u(x,t)} - 2 A_x \bigg]  \\ + 
                     \frac{\ket{u(x,t)}}{(\Delta x)^3}  \bigg[   A^{\dagger}_x \ket{u(x,t+\tau)} - 2 \ket{u(x,t+\tau)} + A_x \ket{u(x,t)}            \bigg]     -    \frac{\ket{u(x,t)}}{(\Delta x)^3} \bigg[ A^{\dagger}_x \ket{u(x,t)} - 2 \ket{u(x,t)} \\ + A_x \ket{u(x,t)} \bigg]   -     \frac{1}{\Delta x} \bigg[ { 3 \ket{u(x,t)} } { A_x \ket{u(x,t)}} - { 3 \ket{u(x,t)} }{ \ket{u(x,t)}} + 2 A_x \ket{u(x,t)} - 2 \ket{u(x,t)}    \bigg]  \bigg\}^2  \text{ . } \end{align*}

             \noindent  Distributing further,        
                     
                     \begin{align*}
                     \frac{\tau^2}{ \big(                     1 + \frac{2}{(\Delta x)^3 \tau}     \big)^2}  \bigg\{               \frac{2 }{(\Delta x)^3} \bigg[ A_x \ket{u(x,t)} A^{\dagger}_x \ket{u(x,t)} - \ket{u(x,t)} A^{\dagger}_x \ket{u(x,t)} - 2 A_x \bigg]  \\ +  
                     \frac{\ket{u(x,t)}}{(\Delta x)^3}  \bigg[   A^{\dagger}_x \ket{u(x,t+\tau)}  - 2 \ket{u(x,t+\tau)} + A_x \ket{u(x,t)}            \bigg]     -    \frac{\ket{u(x,t)}}{(\Delta x)^3} \bigg[ A^{\dagger}_x \ket{u(x,t)} \\ - 2 \ket{u(x,t)} + A_x \ket{u(x,t)} \bigg]                     \bigg\}^2  - \frac{\tau^2}{  \Delta x     \big(                     1 + \frac{2}{(\Delta x)^3 \tau}     \big)^2}   \bigg\{ { 3 \ket{u(x,t)} } { A_x \ket{u(x,t)}}  \\ -    { 3 \ket{u(x,t)} }{ \ket{u(x,t)}}  + 2 A_x \ket{u(x,t)} - 2 \ket{u(x,t)}    \bigg\}^2 \text{ . } \end{align*}  
                     
                     \noindent From the resulting terms above, the terms of the superposition following the prefactor dependent upon $\tau^2$ is,

                     \begin{align*}  \frac{\tau^2}{ \big(                     1 + \frac{2}{(\Delta x)^3 \tau}     \big)^2}  \bigg\{   \bigg\{ \frac{2 }{(\Delta x)^3} \bigg[ A_x \ket{u(x,t)} A^{\dagger}_x \ket{u(x,t)} - \ket{u(x,t)} A^{\dagger}_x \ket{u(x,t)} - 2 A_x \bigg] \\ + \frac{\ket{u(x,t)}}{(\Delta x)^3}  \bigg[   A^{\dagger}_x \ket{u(x,t+\tau)} - 2 \ket{u(x,t+\tau)} + A_x \ket{u(x,t)}            \bigg]                                            \bigg\}^2  \\ +    2  \bigg[ \frac{2 }{(\Delta x)^3} \bigg[ A_x \ket{u(x,t)} A^{\dagger}_x \ket{u(x,t)} - \ket{u(x,t)} A^{\dagger}_x \ket{u(x,t)} - 2 A_x \bigg]   + 
                     \frac{\ket{u(x,t)}}{(\Delta x)^3}  \bigg[   A^{\dagger}_x \ket{u(x,t+\tau)} \\ - 2 \ket{u(x,t+\tau)} + A_x \ket{u(x,t)}            \bigg]                  \bigg]   +  \frac{\ket{u(x,t)}}{(\Delta x)^3} \bigg[ A^{\dagger}_x \ket{u(x,t)} - 2 \ket{u(x,t)} + A_x \ket{u(x,t)} \bigg]        \bigg\} \\ - 
            \frac{\tau^2}{  \Delta x     \big(                     1 + \frac{2}{(\Delta x)^3 \tau}     \big)^2}   \bigg[ \bigg[  3 \ket{u(x,t)} A_x \ket{u(x,t)} -        3 \ket{u(x,t)} A_x \ket{u(x,t)}    \bigg] ^2   -  2 \bigg[   3 \ket{u(x,t)} A_x \ket{u(x,t)} \\ -         3 \text{    }\ket{u(x,t)} A_x \ket{u(x,t)}      \bigg]  \bigg[  2 A_x \ket{u(x,t)} - 2 \ket{u(x,t)}      \bigg]         +       \big(  2 A_x \ket{u(x,t)} - 2 \ket{u(x,t)}      \big)^2 \bigg]   \\   \Updownarrow \\     \frac{\tau^2}{ \big(                     1 + \frac{2}{(\Delta x)^3 \tau}     \big)^2}  \bigg\{    \bigg[ \frac{4 }{(\Delta x)^6} \bigg[ A_x \ket{u(x,t)} A^{\dagger}_x \ket{u(x,t)} - \ket{u(x,t)} A^{\dagger}_x \ket{u(x,t)} - 2 A_x \bigg] ^2\\ +  \frac{1}{(\Delta x)^6}  \bigg[ \bra{u(x,t+\tau)} A_x \ket{u(x,t)}  - 2 \bra{u(x,t+\tau)}  \ket{u(x,t)} +  \bra{u(x,t)}  A^{\dagger}_x \ket{u(x,t)}          \bigg] ^2 \text{, }   \end{align*} 
            
                              \noindent corresponding to terms of the superposition from squaring expectation values, as shown above. The remaining terms are,

            \begin{align*}   \frac{ 8}{(\Delta x)^9}    T_1    T_2     +    2  \bigg[ \frac{2 }{(\Delta x)^3} T_3   + \frac{\ket{u(x,t)}}{(\Delta x)^3}  T_4   \bigg]      
            + \frac{\ket{u(x,t)}}{(\Delta x)^3} T_5      - \frac{\tau^2}{  \Delta x     \big(                     1 + \frac{2}{(\Delta x)^3 \tau}     \big)^2} T_6      -  2 T_7  \\ +  2  T_8    +  T_9       \text{, }          
\end{align*}

\noindent where,

\begin{align*}
   T_1 \equiv   A_x \ket{u(x,t)} A^{\dagger}_x \ket{u(x,t)} - \ket{u(x,t)} A^{\dagger}_x \ket{u(x,t)} - 2 A_x   \text{, } \\ \\          T_2 \equiv    \bra{u(x,t+\tau)} A_x \ket{u(x,t)}   - 2 \bra{u(x,t+\tau)}  \ket{u(x,t)} +  \bra{u(x,t)}  A^{\dagger}_x \ket{u(x,t)}           \text{, }\\ \\     T_3 \equiv A_x \ket{u(x,t)} A^{\dagger}_x \ket{u(x,t)}  - \ket{u(x,t)} A^{\dagger}_x \ket{u(x,t)} - 2 A_x           \text{, } \\ \\ T_4 \equiv   A^{\dagger}_x \ket{u(x,t+\tau)} - 2 \ket{u(x,t+\tau)} + A_x \ket{u(x,t)}           \text{, } \\ \\ T_5 \equiv     A^{\dagger}_x \ket{u(x,t)} - 2 \ket{u(x,t)} + A_x \ket{u(x,t)}          \text{,} \\ \\ T_6 \equiv      \bra{u(x,t)} A^{\dagger}_x   \bra{u(x,t)}     9 \ket{u(x,t)} A_x \ket{u(x,t)}   -  6   \bra{u(x,t)} A^{\dagger}_x \bra{u(x,t)} \ket{u(x,t)} A_x \ket{u(x,t)}    \\ + 9 \bra{u(x,t)} A^{\dagger}_x \bra{u(x,t)} \ket{u(x,t)} A_x \ket{u(x,t)}                   \text{, } \\ \\ T_7 \equiv   3 \ket{u(x,t)} A_x \ket{u(x,t)} -  3 \ket{u(x,t)} A_x \ket{u(x,t)}     \text{,} \\ \\ T_8 \equiv        2 A_x \ket{u(x,t)} - 2 \ket{u(x,t)} \text{, } \\ \\ T_9 \equiv   \bra{u(x,t)}      4 A_x \ket{u(x,t)} - \bra{u(x,t)} 8 A_x  \ket{u(x,t)}     + \bra{u(x,t)} 8 \ket{u(x,t)}          \text{. }
\end{align*}

\noindent Expanding the squares of expectation values in the above superposition yields,

\begin{align*}
 \frac{\tau^2}{ \big(                     1 + \frac{2}{(\Delta x)^3 \tau}     \big)^2}  \bigg\{     \frac{4 }{(\Delta x)^6}  \mathcal{T}_1      + \frac{1}{(\Delta x)^6}  \mathcal{T}_2   + 
 \frac{ 8}{(\Delta x)^9} \mathcal{T}_3    +    2  \bigg[ \frac{2 }{(\Delta x)^3} \mathcal{T}_4  + \frac{\ket{u(x,t)}}{(\Delta x)^3} \mathcal{T}_5                 \bigg]   +  \frac{\ket{u(x,t)}}{(\Delta x)^3}   \mathcal{T}_6    \bigg\} \text{,}
 \end{align*}

\noindent where,

\begin{align*}
   \mathcal{T}_1 \equiv \bra{u(x,t)} A_x \bra{u(x,t)} A^{\dagger}_x  -   A_x \ket{u(x,t)} A^{\dagger}_x \ket{u(x,t)}  - 2 \bra{u(x,t)} A_x \bra{u(x,t)} \\ \times A_x \ket{u(x,t)} A^{\dagger}_x \ket{u(x,t)}  + \bra{u(x,t)} A_x \bra{u(x,t)} \ket{u(x,t)} A^{\dagger}_x \ket{u(x,t)}   \text{, } \\  \\     \mathcal{T}_2 \equiv \bra{u(x,t)} A^{\dagger}_x  \times  \ket{u(x,t+\tau)} \bra{u(x,t+\tau)} A_x \ket{u(x,t)}   - 2 \bra{u(x,t)} \ket{u(x,t+\tau)} \\ \times  \bra{u(x,t+\tau)} A_x \ket{u(x,t)}   + \bra{u(x,t)} A_x \ket{u(x,t)} \bra{u(x,t+\tau)} A_x \ket{u(x,t)}  \text{, } \\ \\  \mathcal{T}_3 \equiv     \bra{u(x,t)} A^{\dagger}_x \bra{u,x,t+\tau)}  A_x \ket{u(x,t)}  A^{\dagger}_x \ket{u(x,t)}    - 2 \bra{u(x,t)} \ket{u(x,t+\tau)} \\ \times A_x \ket{u(x,t)} A^{\dagger}_x \ket{(x,t)}   + \bra{u(x,t)} A_x \ket{u(x,t)}  -  \bra{u(x,t)} A^{\dagger}_x \ket{u(x,t+\tau)} \\ \times \ket{u(x,t)} A^{\dagger}_x \ket{u(x,t)}   - 2 \bra{u(x,t)} \ket{u(x,t+\tau)}  \ket{u(x,t)} A^{\dagger}_x \ket{u(x,t)}  -  \bra{u(x,t)} A_x \\ \times  \ket{u(x,t)} A_x \ket{u(x,t)} A^{\dagger}_x \ket{u(x,t)}    \text{, } \\ \\  \mathcal{T}_4 \equiv    A_x \ket{u(x,t)} A^{\dagger}_x \ket{u(x,t)}  - \ket{u(x,t)} A^{\dagger}_x \ket{u(x,t)} - 2 A_x    \text{, } \\ \\ \mathcal{T}_5 \equiv A^{\dagger}_x \ket{u(x,t+\tau)}  - 2 \ket{u(x,t+\tau)} + A_x \ket{u(x,t)}     \text{,} \\ \\ \mathcal{T}_6 \equiv    A^{\dagger}_x \ket{u(x,t)} - 2 \ket{u(x,t)} + A_x \ket{u(x,t)}    \text{. }    
\end{align*}

 \noindent The cross term resulting in the superposition with prefactor $\frac{\tau^2}{\Delta x \big( 1 + \frac{2}{(\Delta x)^2 \tau }  \big)^2} $ takes the form,
 
 \begin{align*}
\frac{2 \tau^2}{  \Delta x     \big(               1 + \frac{2}{(\Delta x)^3 \tau}    \big)^2} \bigg[  6 \ket{u(x,t)} A_x \ket{u(x,t)} A_x \ket{u(x,t)}  - 6 \bra{u(x,t)} \ket{u(x,t)} A_x \ket{u(x,t)}  - 6 \ket{u(x,t)} \\ \times A_x \ket{u(x,t)} A_x \ket{u(x,t)}   + 6 \bra{u(x,t)} \ket{u(x,t)} A_x \ket{u(x,t)} \bigg]  \text{. } 
\end{align*}

\noindent The superposition with prefactor $\frac{\tau^2}{\Delta x \big( 1 + \frac{2}{(\Delta x)^2 \tau }  \big)^2} $ takes the form,

\begin{align*}
\frac{\tau^2}{\Delta x \big( 1 + \frac{2}{(\Delta x)^2 \tau }  \big)^2} \bigg[  \big(           \bra{u(x,t)} A^{\dagger}_x \bra{u(x,t)}     9 \ket{u(x,t)} A_x \ket{u(x,t)}  -  6   \bra{u(x,t)} A^{\dagger}_x \bra{u(x,t)} \\ \times \ket{u(x,t)}  A_x \ket{u(x,t)}       + 9 \bra{u(x,t)} A^{\dagger}_x \bra{u(x,t)} \ket{u(x,t)} A_x \ket{u(x,t)}                \bigg]   -  2 \bigg[   6 \ket{u(x,t)} \\ \times A_x \ket{u(x,t)} A_x \ket{u(x,t)}   -   6 \bra{u(x,t)} \ket{u(x,t)}  A_x \ket{u(x,t)}  - 6 \ket{u(x,t)} \\ \times A_x \ket{u(x,t)} A_x \ket{u(x,t)}  + 6 \bra{u(x,t)} \ket{u(x,t)} A_x \ket{u(x,t)}     \bigg]        \\    +     \bra{u(x,t)}      4 A_x \ket{u(x,t)}  - \bra{u(x,t)} 8 A_x  \ket{u(x,t)}   + \bra{u(x,t)} 8 \ket{u(x,t)}    \big)   \text{, }   
\end{align*}

\noindent implying that the finalized cost function the quantum state whose ground state we approximate takes the desired form.

\section{Acknowledgements}

The author would like to thank Peter McMahon for suggesting a direction for the project, in addition to Abhi Sarma for extremely informative discussions surrounding implementation of ansatzae for the initial state of solutions, assisting with troubleshooting time evolution with the Qsim package, along with possible future research directions with quantum machine learning, and Yilian Liu for discussions of applications of the VQA with other papers in the literature.





\section{Declarations}

\subsection{Ethics approval and consent to participate}

The author consents to participate in the peer review process.

\subsection{Consent for publication}

The author consents to submit the following work for publication.

\subsection{Availability of data and materials}

Not applicable

\subsection{Conflict of interest}

The author declares no competing interests.

\subsection{Funding}

Not applicable


\begin{thebibliography}{00}


 

  \bibitem[Frahm, Gehrmann, Nepomechie, Retore(2023)]{frahm2023}
  [1] Albert, V., Covey, J., Preskill, J.: Robust encoding of a qubit in a molecule. Phys
Rev 10(031050) (2020) https://doi.org/10.1103/PhysRevX.10.031050



  \bibitem[Frahm, Gerhmann(2022)]{Frahm2022}
 [2] Brandao, F., al.: Models of quantum complexity growth. PRX Quantum
2(030316) (2021) https://doi.org/10.1103/PRXQuantum.2.030316


 \bibitem[Frahm, Gerhmann(2022)]{Frahm2022}
 [3] Cerezo, M., Arrasmith, A., Babbush, R. et al. Variational quantum algorithms. Nat Rev Phys 3, 625–644 (2021). https://doi.org/10.1038/s42254-021-00348-9

     \bibitem[Frahm, Gerhmann(2022)]{frahm2022}
   [4] Chowdhury, A., Low, G.H., Wiebe, N.: A variational quantum algorithm for
preparing quantum gibbs states. arXiv 2002.00055 (2020) https://doi.org/10.
48550/arXiv.2002.00055

   \bibitem[Jimbo(1986)]{jimbo1986}
 [5] Eremenko, A.: Spectral theorems for hermitian and unitary matrices. Lecture Notes (2016).



   \bibitem[Nepomechie, Retore(2021)]{nepomechie2021}
 [6] Faist, P., al.: Continuous symmetries and approximate quantum error correction.
Phys. Rev. X 10(041018) (2020) https://doi.org/10.1103/PhysRevX.10.041018



   \bibitem[Nepomechie, Pimenta, Retore (2017)]{nepomechie2017}
 [7] Gidney, C., Ekera, M.: How to factor 2048 bit rsa integers in 8 hours using
20 million noisy qubits. Quantum 5(433) (2021) https://doi.org/10.22331/q-2021-04-15-433







   \bibitem[Nepomechie, Pimenta, Retore (2019)]{nepomechie2019}
[8] Hadfield, S., Wang, Z., O’Gorman, B., Rieffel, E., Venturelli, D., Biswas, R.: From
the quantum approximate optoimization algorithm to a quantum alternating
operator ansatz. Quantum Optimization Theory, Algorithms, and Applications
34(12.2) (2019) https://doi.org/10.3390/a12020034

   \bibitem[Robertson, Pawelkiewicz, Jacobsen, Saleur (2020)]{Robertson2020}
[9] Iverson, J., Preskill, J.: Coherence in logical quantum channels. New Journal of
Physics 22(073066) (2020) https://doi.org/10.1088/1367-2630/ab8e5c







   \bibitem[Robertson, Jacobsen, Saleur (2019)]{Robertson2019}
[10] Kandala, A., al.: Harware-efficient variational quantum eigensolver for small
molecules and quantum magnets. Nature 549(242) (2017) https://doi.org/10.1038/nature23879

\bibitem[Robertson, Jacobsen, Saleur (2019)2]{Robertson20192}
[11] Kim, I., Swingle, B.: Robust entanglement renormalization on a noisy quantum
computer. arXiv 1711.07500 (2017) https://doi.org/10.48550/arXiv.1711.07500


  \bibitem[Robertson, Jacobsen, Saleur (2019)2]{Robertson20192}
[12] Lubasch, M., al.: Variational quantum algorithms for nonlinear problems. Phys
Rev 101(010301) (2020) https://doi.org/10.1103/PhysRevA.101.010301




  \bibitem[Robertson, Jacobsen, Saleur (2019)2]{Robertson20192}
[13] McClean, J.R., al.: The theory of variational hybrid quantum-classical algorithms. New J. Phys. 18(023023) (2016) https://doi.org/10.1088/1367-2630/18/2/023023

  \bibitem[Robertson, Jacobsen, Saleur (2019)2]{Robertson20192}
[14] Molina, P.G., Mediavilla, J.R., Garcia-Ripoll, JJ. Quantum Fourier analysis for multivariate functions and applications to a class of Schrödinger-type partial differential equations. Phys. Rev. A 105, 012433 (2022). https://doi.org/10.1103/PhysRevA.105.012433

  \bibitem[Robertson, Jacobsen, Saleur (2019)2]{Robertson20192}
[15] Orus, R.: A practical introduction to tensor networks: Matrix product states and
projected entangled pair states. Annals of Physics 349, 117–158 (2014) https://doi.org/10.1016/j.aop.2014.06.013
  
  \bibitem[Robertson, Jacobsen, Saleur (2019)2]{Robertson20192}
[16] Perez-Garcia, D., Verstraete, F., Wolf, M.M., Cirac, J.I.: Matrix product state
representations. Quantum Inf. Comput. 7(401) (2007) https://doi.org/10.26421/QIC7.5-6-1






  \bibitem[Robertson, Jacobsen, Saleur (2019)2]{Robertson20192}
[17] Peruzzo, A., al.: A variational eigenvalue solver on a quantum processor. Nature
Communications 5(4213) (2014) https://doi.org/10.1038/ncomms5213

  \bibitem[Robertson, Jacobsen, Saleur (2019)2]{Robertson20192}
[18] J., P.: Quantum computing in the nisq era and beyond. Quantum 2(79) (2018)
https://doi.org/10.22331/q-2018-08-06-79


 \bibitem[Robertson, Jacobsen, Saleur (2019)2]{Robertson20192}
[19] Romero, J.M., Montoya-González, E., Guillermo Cruz, G., Romero, R.C. A novel quantum circuit for the quantum Fourier transform. https://doi.org/10.48550/arXiv.2507.08699 (2025).


  \bibitem[Robertson, Jacobsen, Saleur (2019)2]{Robertson20192}
[20] Sim, S., Johnson, P., Aspurur-Guzik, A.: Expressibility and entangling capability of parametrized quantum circuits for hybrid quantum-classical algorithms.
Advanced Quantum Technologies 00070(12) (2019) https://doi.org/10.1002/qute.201900070


  \bibitem[Robertson, Jacobsen, Saleur (2019)2]{Robertson20192}
[21] Sung, K., al.: Using models to improve optimizers for variational quantum algorithms. Quantum Science and Technology 5(4) (2020) https://doi.org/10.1088/2058-9565/abb6d9

    \bibitem[Robertson, Jacobsen, Saleur (2019)2]{Robertson20192}
[22] Vedral, V., Barenco, A., Ekert, A.: Quantum networks for elementary arithmetic
operations. Phys. Rev. A 54(147) (1996) https://doi.org/10.1103/PhysRevA.54.147

      \bibitem[Robertson, Jacobsen, Saleur (2019)2]{Robertson20192}
[23] Xu, X., al.: Variational algorithms for linear algebra. Science Bulletin 66(21),
2181–2188 (2021) https://doi.org/10.1016/j.scib.2021.06.023

        \bibitem[Robertson, Jacobsen, Saleur (2019)2]{Robertson20192}
[24] Zhang, D.B., Zhan-Hao, Y., Yin, T.: Variational quantum eigensolvers by variance minimization. Chin.Phys.B 31(120301) (2022) https://doi.org/10.1088/1674-1056/ac8a8d

          \bibitem[Robertson, Jacobsen, Saleur (2019)2]{Robertson20192}
[25] Thapliyal, H., Munoz-Coreas, E., Varun, T.S.S., Humble, T.S.: Quantum circuit
designs of integer division optimizing t-count and t-depth. arXiv (1809.09732)
(2018) https://doi.org/10.48550/arXiv.1809.09732

            \bibitem[Robertson, Jacobsen, Saleur (2019)2]{Robertson20192}

[26] Zhuang, Q., Preskill, J., Jiang, L.: Distributed quantum sensing enhanced by
continuous-variable error correction. New Journal of Physics 22 (2020) https://doi.org/10.1088/1367-2630/ab7257

            
\end{thebibliography}
\end{document}